\documentclass[fleqn,usenatbib,useAMS]{mnras}
\usepackage{newtxtext,newtxmath}
\usepackage[flushleft]{threeparttable}
\usepackage[caption=false]{subfig}
\usepackage{lscape}
\usepackage[T1]{fontenc}
\usepackage{Alegreya}
\usepackage{csquotes}
\usepackage[table,xcdraw]{xcolor}
\DeclareRobustCommand{\VAN}[3]{#2}
\let\VANthebibliography\thebibliography
\def\thebibliography{\DeclareRobustCommand{\VAN}[3]{##3}\VANthebibliography}

\usepackage{hyperref}
\hypersetup{linkcolor=blue,citecolor=blue,filecolor=cyan,urlcolor=magenta}

\DeclareRobustCommand{\ion}[2]{%
\relax\ifmmode
\ifx\testbx\f@series
{\mathbf{#1\,\mathsc{#2}}}\else
{\mathrm{#1\,\mathsc{#2}}}\fi
\else\textup{#1\,{\mdseries\textsc{#2}}}%
\fi}
\usepackage{graphicx}	 
\usepackage{amsmath}

\usepackage{newtxmath} 
\usepackage{float}
\usepackage[flushleft]{threeparttable}
\usepackage{multirow}
\usepackage{adjustbox}

\newcommand{\ebmvgas}{E(B-V)_{\rm neb}}

\newcommand{\ha}{\text{H$\alpha$}}
\newcommand{\hb}{\text{H$\beta$}}

\newcommand{\sfrha}{\text{SFR[H$\alpha$]}}
\newcommand{\sfrsed}{\text{SFR[SED]}}

\newcommand{\angstrom}{\text{\normalfont\AA}}
\usepackage{amsmath,amssymb}

\let\originallesssim\lesssim
\let\originalgtrsim\gtrsim

\DeclareRobustCommand{\lesssim}{%
  \mathrel{\mathpalette\lowersim\originallesssim}%
}
\DeclareRobustCommand{\gtrsim}{%
  \mathrel{\mathpalette\lowersim\originalgtrsim}%
}

\makeatletter
\newcommand{\lowersim}[2]{%
  \sbox\z@{$#1<$}%
  \raisebox{-\dimexpr\height-\ht\z@}{$\m@th#1#2$}%
}
\title[$L(\ha)/L(\rm UV)$]{Exploring the Correlation between $\rm{H}\alpha$-to-UV Ratio and Burstiness for Typical Star-forming Galaxies at $z\sim2$}
\author[S. Rezaee et al.]{
Saeed Rezaee$^{1}$\thanks{E-mail: saeed.rezaee@email.ucr.edu},
Naveen A. Reddy$^{1}$,
Michael W. Topping$^{2}$,
Irene Shivaei$^{2,3}$,
Alice E. Shapley$^{4}$,
Tara Fetherolf$^{1}$,
\newauthor
Mariska Kriek$^{5}$,
Alison Coil$^{6}$,
Bahram Mobasher$^{1}$,
Brian Siana$^{1}$,
Xinnan Du$^{7}$,
Ali Ahmad Khostovan$^{8,9}$,
\newauthor
Andrew Weldon$^{1}$,
Najmeh Emami$^{10}$, and
Nima Chartab$^{11}$
\\
$^{1}$Department of Physics and Astronomy, University of California, Riverside, CA 92521, USA\\
$^{2}$Steward Observatory, University of Arizona, 933 North Cherry Avenue, Tucson, AZ 85721, USA\\
$^{3}$Nasa Hubble fellow\\
$^{4}$Department of Physics and Astronomy, University of California, Los Angeles, CA 90095, USA\\
$^{5}$Astronomy Department, University of California, Berkeley, CA 94720, USA\\
$^{6}$Center for Astrophysics and Space Sciences, Department of Physics, University of California, San Diego, 9500 Gilman Drive, La Jolla, CA 92093, USA\\
$^{7}$ Kavli Institute for Particle Astrophysics $\&$ Cosmology, P. O. Box 2450, Stanford University, Stanford, CA 94305, USA\\
$^{8}$Astrophysics Division, NASA Goddard Space Flight Center, Greenbelt, MD 20771, USA\\
$^{9}$NASA Postdoctoral Program Fellow\\
$^{10}$Minnesota Institute for Astrophysics, University of Minnesota, 116 Church St SE, Minneapolis, MN 55455, USA\\
$^{11}$The Observatories of the Carnegie Institution for Science, 813 Santa Barbara St., Pasadena, CA 91101, USA\\
}
\date{Accepted XXX. Received YYY; in original form ZZZ}
\pubyear{2022}
\begin{document}
\label{firstpage}
\pagerange{\pageref{firstpage}--\pageref{lastpage}}
\maketitle
%We use multi-waveband CANDELS/3D-HST photometry
\begin{abstract}

    The $\rm{H}\alpha$-to-UV luminosity ratio ($L(\text{H}\alpha)/L(\rm UV)$) is often used to probe bursty star-formation histories (SFHs) of star-forming galaxies and it is important to validate it against other proxies for burstiness. To address this issue, we present a statistical analysis of the resolved distribution of star-formation-rate surface density ($\Sigma_{\rm{SFR}}$) as well as stellar age and their correlations with the globally measured $L(\text{H}\alpha)/L(\rm UV)$ for a sample of 310 star-forming galaxies in two redshift bins of $1.37 < z < 1.70$ and $ 2.09 < z < 2.61$ observed by the MOSFIRE Deep Evolution Field (MOSDEF) survey. We use the multi-waveband CANDELS/3D-HST imaging of MOSDEF galaxies to construct $\Sigma_{\rm{SFR}}$ and stellar age maps. We analyze the composite rest-frame far-UV spectra of a subsample of MOSDEF targets obtained by the Keck Low Resolution Imager and Spectrometer (LRIS), which includes 124 star-forming galaxies (MOSDEF-LRIS) at redshifts $1.4 < z < 2.6$, to examine the average stellar population properties, and the strength of age-sensitive FUV spectral features in bins of $L(\text{H}\alpha)/L(\rm UV)$. Our results show no significant evidence that individual galaxies with higher $L(\text{H}\alpha)/L(\rm UV)$ are undergoing a burst of star formation based on the resolved distribution of $\Sigma_{\rm{SFR}}$ of individual star-forming galaxies. We segregate the sample into subsets with low and high $L(\text{H}\alpha)/L(\rm UV)$. The high -$L(\text{H}\alpha)/L(\rm UV)$ subset exhibits, on average, an age of $\log[\rm{Age/yr}]$ = 8.0, compared to $\log[\rm{Age/yr}]$ = 8.4 for the low-$L(\text{H}\alpha)/L(\rm UV)$ galaxies, though the difference in age is significant at only the $2\sigma$ level. Furthermore, we find no variation in the strengths of \ion{Si}{iv} $\lambda\lambda1393, 1402$ and \ion{C}{iv} $\lambda\lambda1548, 1550$ P-Cygni features from massive stars between the two subsamples, suggesting that the high-$L(\text{H}\alpha)/L(\rm UV)$ galaxies are not preferentially undergoing a burst compared to galaxies with lower $L(\text{H}\alpha)/L(\rm UV)$. On the other hand, we find that the high-$L(\text{H}\alpha)/L(\rm UV)$ galaxies exhibit, on average, more intense \ion{He}{ii} $\lambda1640$ emission, which may possibly suggest the presence of a higher abundance of high-mass X-ray binaries.

\end{abstract}


\section{Introduction}
%\citep{Somerville_1999,Springel_2000,Springel_2005,Kere_2009,Governato_2010,Hopkins_2014,Hayward_2017,Fujimoto_2019}.
While most galaxies follow a tight sequence in star-formation rate (\text{SFR}) versus stellar mass ($M_{\ast}$), there are some that are significantly offset above this relation at any given redshift, suggestive of a recent burst of star formation \citep{Schmidt_1959,Kennicutt_1998,Somerville_1999,Springel_2000,Springel_2005,Noeske_2007,Kere_2009,Knapen_2009,Dobbs_2009,Genzel_2010,Governato_2010,Reddy_2012,Rodighiero_2014,Hopkins_2014,Shivaei_2015,Hayward_2017,Fujimoto_2019,Wang_2020}. For example, the apparent increase in scatter of the relationship between SFR and $M_{\ast}$ at low stellar masses suggests that such galaxies are characterized by bursty star-formation histories \citep{Noeske_2007,Hopkins_2014,Shen_2014,Guo_2016,Asquith_2018,Dickey_2021,Atek2022}. In addition, simulations with resolved scaling comparable to the star-forming clouds suggest that the burst amplitude and frequency increase with redshift (e.g., \citealt{Feldmann_2017,Sparre_2017,Ma_2018}). Given that bursty SFHs are inferred to be the likely mode of galaxy growth for at least lower mass galaxies at high-redshift (e.g., \citealt{Atek2022} found evidence of bursty SFHs for lower mass galaxies with $M_{\ast}<10^{9}\,M_{\odot}$ at $z\sim1.1$), it is important to determine the effectiveness of commonly-used proxies for burstiness.

A key method that has been used to infer the burstiness\footnote{In the context of this study, when we discuss 'burstiness,' we primarily refer to the incidence rate of starbursts within galaxies. We acknowledge that the term 'burstiness' has also been used in the literature to describe broader variations in SFR \citep{Knapen_2009,Sparre_2016}.} of star-forming galaxies is to compare SFR indicators that are sensitive to star formation on different timescales. Two of the widely used SFR indicators are derived from the $\ha$ nebular recombination line ($\lambda = 6564.60\, \angstrom$) and far-ultraviolet (FUV) continuum  ($1300\, \angstrom < \lambda <2000\, \angstrom$). The $\ha$ emission line originates from the recombination of the ionized gas around young massive stars ($M_{\ast}\gtrsim20\,M_{\odot}$) and traces SFR over a timescales of $\sim10$\,Myr \citep{Kennicutt_2012}. The UV continuum is sensitive to the same stars that are responsible for $\ha$, as well as lower-mass stars (B stars, and A stars at wavelengths redder than 1700 A) with lifetimes of $\sim 100$\,Myr and $M_{\ast}\gtrsim\,3M_{\odot}$. As a result, when compared to the $\ha$ emission line, the FUV continuum traces SFRs averaged over a longer timescale. Therefore, variations in the dust-corrected $\ha$-to-UV luminosity ratio ($L(\ha)/L(\rm UV)$) may reveal information about recent burst activity \citep{Glazebrook_1999,Iglesias_2004,Lee_2009,Meurer_2009,Hunter_2010,Fumagalli_2011,Lee_2011,Weisz_2012,dasilva_2012,dasilva_2014,Dominguez,Emami_2019,Caplar_2019,Faisst_2019}.

For a constant star-formation history (SFH), the $\ha$-to-UV luminosity ratio will reach to its equilibrium after a few tens of Myr (e.g., \citealt{Reddy_2012}). However, variations in the inferred integrated $\ha$-to-UV ratio may result from a number of effects, including variations in the IMF \citep{Leitherer_1995,Elmegreen_2006,Pflamm_2007,Meurer_2009,Pflamm_2009,Hoversten_2008,Boselli_2009,Mas_Ribas_2016}, nebular and stellar dust reddening \citep{Kewley_2002,Lee_2009,Reddy_2012,Reddy_2015,Shivaei_2015,Shivaei_2018,Theios_2019}, ionizing escape fraction \citep{Steidel_2001,Shapley_2006,Siana_2007}, and binary stellar evolution \citep{Eldridge_2012,Eldridge_2017,Choi_2017}.
In addition, comparing the mock HST and JWST galaxy catalogs with 3D-HST observations of $z\sim1$ galaxies, \cite{Broussard_2019} finds that the average $\ha$-to-UV ratio is not impacted significantly by variations in the high-mass slope of the IMF, and metallicity. Similar studies also show that the average $\ha$-to-UV is not a good indicator of business but rather a probe of the average SFH or dust law uncertainties \citep{Broussard_2019,Broussard_2022}. Given these possibilities, any interpretation about the burstiness of galaxies based on the variations in $L(\ha)/L(\rm UV)$ must be approached with caution.

The MOSFIRE Deep Evolution Field (MOSDEF) survey is ideally suited to examine the extent to which variations in $L(\ha)/L(\rm UV)$ trace burstiness. MOSDEF probes galaxies at $z\sim2$, which marks a key epoch for galaxy growth when the cosmic star-formation density reaches its maximum \citep{Madau_1996,Hopkins_2006,Madau_2014}. Additionally, the deep Hubble Space Telescope (HST) imaging of the MOSDEF galaxies obtained by CANDELS \citep{Grogin_2011,Koekemoer_2011} enables the construction of stellar population maps that can be used to assess burstiness on smaller (resolved) spatial scales (e.g., \citealt{Wuyts_2011,Wuyts_2012,Hemmati_2014,Jafariyazani_2019,TARA_Voro}). Moreover, the availability of follow-up Keck/LRIS rest-FUV spectra of a subset of 259 MOSDEF galaxies \citep{Topping,Reddy_2022} allows us to investigate the relationship between the $L(\ha)/L(\rm UV)$ ratio and age-sensitive FUV spectral features.

The goal of this study is to determine whether the dust-corrected $\ha$-to-UV luminosity ratio is a reliable tracer of a bursty SFH at $z\sim2$. We address this question by examining the correlations between the differences in properties of the stellar populations and the $L(\ha)/L(\rm UV)$ ratio. The structure of this paper is as follows. In Section~\ref{sec:sample}, we introduce the samples used in this work, and outline the sample selection criteria and data reduction. In Section~\ref{sec:morph}, we describe the method used for constructing the stellar population maps, and the result of the morphology analysis. Our approach for constructing rest-FUV composite spectra is described in Section~\ref{sec:compnorm}. Our results on variations of the average physical properties of galaxies, and the strength of age-sensitive FUV spectral features in bins of $L(\ha)/L(\rm{UV})$ are presented in Section~\ref{Sec:lrisbins}. Finally, the conclusions are summarized in Section~\ref{sec:Sum}. Wavelengths are in the vacuum frame. We adopt a flat cosmology with $H_{0}=70\,\rm{km \,s^{-1}}$, $\Omega_{\Lambda}=0.7$, and $\Omega_{m}=0.3$. A \cite{Chabrier} IMF is assumed throughout this work.

\section{Sample}
\label{sec:sample}
\begin{figure*}
	\includegraphics[width=0.95\textwidth]{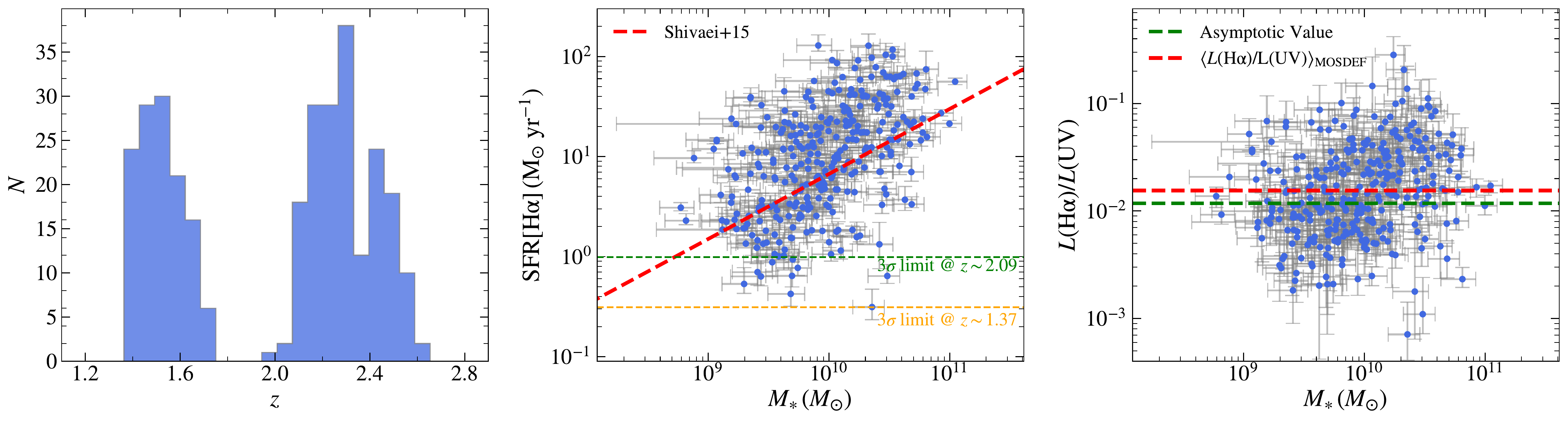}
    \caption{Physical properties of 310 star-forming galaxies in the MOSDEF/MORPH sample used in this work. \textit{Left}: The histogram indicates the MOSDEF spectroscopic redshift distribution in two bins with the average redshifts of $z\sim1.5$ and $z\sim2.3$. \textit{Middle}: $\sfrha$ vs. $M_{\ast}$ relationship. $\sfrha$ is computed using the dust-corrected $\ha$  luminosity. The conversion factor between the $\ha$ luminosity and $\sfrha$, as well as stellar mass are derived using the SED modeling. The dashed red line shows \citet{Shivaei_2015} relationship between $\sfrha$ and $M_{\ast}$, which has been adjusted to represent the assumptions used in this work, based on the first two years of MOSDEF (including galaxies with undetected \hb). The horizontal dashed lines represent the $3\sigma$ detection limits of the $\sfrha$ determined for the two redshift bins ($1.37<z<1.70$, and $2.09<z<2.61$ ) using H and K band line sensitivities \citep{Kriek_2015}. \textit{Right}: The distribution of dust-corrected $L(\ha)/L(\rm UV)$ with respect to the stellar mass where $L(\ha)$ and $L(\rm{UV})$ are dust-corrected using the \citet{Cardelli} and SMC extinction curves, respectively. The red dashed line indicates the average dust-corrected $L(\ha)/L(\rm UV)$ of all the galaxies in the MOSDEF parent sample that have coverage of $\ha$ and $\hb$ emission lines with $S/N\geq3$. The green dashed line indicates the asymptotic value of $L(\ha)/L(\rm UV)$ for a constant SFH using BPASS SED models (Section~\ref{sec:sedmodels}).}
    \label{fig:sample}
\end{figure*}

\subsection{Rest-Frame Optical MOSDEF Spectroscopy}
\label{sec:mosdef}

The MOSDEF survey \citep{Kriek_2015} used the Keck/MOSFIRE spectrograph \citep{Mclean} to obtain rest-frame optical spectra of $\sim1500$ $H$-band-selected star-forming galaxies and active galactic nuclei (AGNs). The five extragalactic legacy fields (GOODS-S, GOODS-N, COSMOS, UDS, AEGIS) covered by the CANDELS survey \citep{Grogin_2011,Koekemoer_2011} were targeted. The targets were chosen to lie in three redshift bins: $1.37 < z < 1.70$, $2.09 < z < 2.61$, and $2.95 < z < 3.80$ where the strong rest-frame optical emission lines ($\text{[\ion{O}{ii}]}\lambda3727,3730$, $\hb$, $\text{[\ion{O}{iii}]}\lambda\lambda4960,5008$, $\ha$, $\text{[\ion{N}{ii}]}\lambda\lambda6550,6585$, and $\text{[\ion{S}{ii}]}\lambda6718,6733$ ) are redshifted into the YJH, JHK, and HK transmission windows, respectively. Further details of the survey and MOSFIRE spectroscopic data reduction are provided in \cite{Kriek_2015}.

We use the spectroscopic redshifts and emission lines measured by the MOSDEF survey. The spectroscopic redshift for each target was measured from the observed wavelength centroid of the highest signal-to-noise emission line in each spectrum. Emission line fluxes were measured from the 1D-spectra of the individual objects by fitting Gaussian functions along with a linear continuum. The $\ha$ was fit simultaneously with the $\text{[\ion{N}{II}]}$ doublet using three Gaussian functions. The $\ha$ emission line flux was corrected for the underlying Balmer absorption, which was measured from the best-fit stellar population model (Section~\ref{sec:sedmodels}). Line flux uncertainties were calculated by perturbing the observed spectra according to their error spectra and remeasuring the line fluxes $1000$ times. The 68th percentile of the distribution obtained from these iterations was adopted to represent the upper and lower flux uncertainties (e.g., \citealt{Reddy_2015,Freeman_2019}).    
\subsection{CANDELS/3D-HST Imaging}
\label{sec:candels}
Resolved broad-band photometry of the MOSDEF galaxies was obtained by CANDELS using $HST/$ACS in the $F435W$ ($B_{435}$), $F606W$ ($V_{606}$), $F775W$ ($i_{775}$), $F813W$ ($I_{814}$), and $F850LP$ ($z_{850}$) filters and $HST/$WFC3 in the $F125W$ ($J_{125}$), $F140W$ ($JH_{140}$), and $F160W$ ($H_{160}$) filters. CANDELS imaging covered $\sim 960\,$arcmin$^{2}$ up to a $90\%$ completeness in the $H_{160}$ filter at a magnitude of $25\,$mag. To construct stellar population maps for the sample galaxies, we use the processed CANDELS images provided by the $3$D-HST grism survey team \citep{Brammer_2012,Skelton_2014,Momcheva2016} along with the publicly available\footnote{\url{https://archive.stsci.edu/prepds/3d-hst/}} photometric catalogs with coverage from $0.3\,\mu$m to $0.8\,\mu$m. The HST images provided by the 3D/HST team were drizzled to a $0.06\,$arcsec pixel$^{-1}$ scale and smoothed to produce the same spatial resolution as the $H_{160}$ images ($0.18\,$arcsec).    

The final sample used in this work contains 310 typical star-forming galaxies at $1.36<z<2.66$, all meeting the following criteria. They all have spectroscopic redshifts from the MOSDEF survey and detections of $\ha$ and $\hb$ emission lines with $S/N\geq3$. AGNs were identified and excluded from the sample based on the IR properties, X-ray luminosities, or \text{[\ion{N}{II}]}$\lambda6584/\ha$ line ratio criteria as described in \citet{Coil-2015}, \citet{Azadi-2017,Azadi_2018}, and \citet{Leung-2019}. Additional $S/N$ and resolution constraints were applied to the HST photometry as a result of our approach of grouping pixels which will be discussed in Section~\ref{sec:morph}.

The final sample described above is used to analyze the morphological information of the MOSDEF galaxies in the first part of this work (i.e., Section~\ref{sec:morph}), and is referred to as the MOSDEF/MORPH sample throughout this work. This sample is based on that used by \cite{Tara_2022}. The MOSDEF/MORPH sample covers a range of stellar mass of $8.77<\log[M_{\ast}/M_{\odot}]<11.04$, and a $\sfrha$ range of $0.40<(\sfrha/M_{\odot}\rm{yr}^{-1})<130$. As shown in the middle panel of Figure~\ref{fig:sample}, the MOSDEF/MORPH sample galaxies lie systematically above the mean main-sequence relation found by \cite{Shivaei_2015} based on the first two years of MOSDEF. This is due to the $S/N$ and resolution criteria (Section~\ref{sec:morph}) imposed on the HST photometry of MOSDEF galaxies. Using these requirements results in a sample that is biased against low-mass and compact galaxies \citep{TARA_Voro}. The $S/N$ requirement for $\hb$ emission line is necessary to obtain a more reliable Balmer decrement measurement for each galaxy and does not introduce a significant bias in the sample \citep{Shivaei_2015,Reddy_2015,Sanders_2018,Tara_2021a}. As evidenced in the middle panel of Figure~\ref{fig:sample}, our sample galaxies do not display any substantial bias relative to the mean main-sequence relation determined by \cite{Shivaei_2015}, which was derived irrespective of $\hb$ detection. The MOSDEF/MORPH sample galaxies exhibit a similar range of $L(\ha)/L(\rm{UV})$ to the MOSDEF parent sample galaxies that have coverage of $\ha$ and $\hb$ with significant detections ($S/N\geq3$) and include galaxies with $L(\ha)/L(\rm{UV})$ that lie at least $5\sigma$ below the mean ratio for the MOSDEF parent sample. The $S/N$ requirement for $\hb$ emission does not significantly impact the average $L(\ha)/L(\rm UV)$ ratio. If we consider  those galaxies where $\hb$ is not detected at the  $S/N\geq3$ but still covered in the spectra, the average $L(\ha)/L(\rm UV)$  decreases by approximately $31\%$, which falls within the measurement uncertainty when considering the $S/N\geq3$ requirement. Regardless of the $\hb$ detection requirement, the average $L(\ha)/L(\rm UV)$ values and the asymptotic $L(\ha)/L(\rm UV)$ are consistent within the measurement uncertainties. As mentioned earlier in this section, $L(\ha)$ used here is obtained by the MOSDEF survey, and is corrected for the effect of dust using an MW extinction curve \citep{Cardelli} which is shown to best represent the nebular attenuation curve for both high-redshift and local galaxies \citep{Reddy_2020,Rezaee_2021}. UV luminosity ($L(\rm UV)$) is estimated by using the best-fit SED models at $\lambda=1500\,\angstrom$. A more detailed discussion on calculating the dust-corrected $L(\rm UV)$ is presented in Section~\ref{sec:sedmodels}.

\subsection{MOSDEF/LRIS Rest-FUV Spectroscopy}
\label{sec:lris}
A subset of 259 objects from the MOSDEF parent sample were selected for deep rest-FUV spectroscopic follow-up observations with the Keck I/Low Resolution Imager and Spectrometer (LRIS; \citealt{Oke,Steidel_2004}). We refer the reader to \citet{Topping} and \citet{Reddy_2022} for further details about the MOSDEF/LRIS survey data collection and reduction procedures. In brief, targets were prioritized based on $S/N\geq3$ detection of the four emission lines (\text{[\ion{O}{III}]}, \hb, \text{[\ion{N}{II}]}$\lambda6584$, and \ha) measured by the MOSDEF survey. Objects with available \ha, \hb, and [O III] as well as an upper limit on \text{[\ion{N}{II}]} were accorded the next highest priority. The objects with available spectroscopic redshifts from the MOSDEF survey, as well as those without a secure redshift measurements, were also included. The lowest priority was assigned to the objects that were not included in the MOSDEF survey, but had photometric redshifts and apparent magnitudes from the 3D-HST catalogs that were within the MOSDEF survey redshift ranges.

Rest-FUV LRIS spectra were obtained within 9 multi-object slit masks with $1\farcs2$ slits in four extragalactic legacy fields: GOODS-S, GOODS-N, AEGIS, COSMOS. The d500 dichroic was used to split the incoming beam at $\simeq5000\,\angstrom$ were used to obtain the LRIS spectra. The blue and red-side channels of LRIS were observed with the $400$ line/mm grism blazed at $4300 \ \angstrom$, and the $600$ line/mm grating blazed at $5000 \ \angstrom$, respectively. This configuration yielded a continuous wavelength range from the atmospheric cutoff at $3100\, \angstrom$ to $\sim 7650\, \angstrom$ (the red wavelength cutoff depends on the location of the slit in the spectroscopic field of view) with a resolution of $R \sim 800$ on the blue side and $R \sim 1300$ on the red side.   
The final MOSDEF/LRIS sample used in the second part of this work (i.e., Section~\ref{Sec:lrisbins}) includes 124 star-forming galaxies at $1.42<z<2.58$, all meeting the same $S/N$ and redshift measurement requirements as those mentioned in Section~\ref{sec:mosdef}.

\subsection{SED Modeling}
\label{sec:sedmodels}

We use the Binary Population and Spectral Synthesis (\texttt{BPASS}) version $2.2.1$ models\footnote{\url{https://bpass.auckland.ac.nz/}} \citep{Eldridge_2017,Stanway_2018} to infer UV luminosity ($L(\rm UV)$), stellar continuum reddening ($E(B-V)_{\rm cont}$), stellar ages, conversion factors between luminosities and SFRs, as well as stellar masses ($M_{\ast}$). The effect of binary stellar evolution is included in the \texttt{BPASS} SED models, which has been found to be an important assumption in modeling the spectra of high redshift galaxies \citep{Steidel_2016,Eldridge_2017,Reddy_2022}. These models are characterized by three sets of parameters, stellar metallicity ($Z_{\ast}$) ranging from $10^{-5}$ to $0.040$ in terms of mass fraction of metals where solar metallicity ($Z_{\odot}$) is equal to 0.0142 \citep{Asplund_2009}, the upper-mass cutoff of the IMF ($M_{\rm cutoff}=\{100M_{\odot}, 300M_{\odot}\}$), and the choice of including binary stellar evolution. These parameters divide the models into four sets of model assumptions with various $M_{\rm cutoff}$ and whether or not the binary effects are included. Throughout, we refer to these model combinations as \enquote{\rm 100bin}, \enquote{\rm 300bin}, \enquote{\rm 100sin}, and \enquote{\rm 300sin} where the initial number indicate the $M_{\rm cutoff}$ of the IMF and \enquote{\rm bin} (\enquote{\rm sin}) indicates that the binary evolution is (or is not) included \citep{Reddy_2022}. 

Stellar population synthesis (SPS) models are constructed by adding the original instantaneous-burst \texttt{BPASS} models for ages ranging from $10^{7}\,$-$10^{10}\,$yr while adopting a constant star-formation history\footnote{According to \cite{Reddy_2012}, if the stellar ages are constrained to be older than the typical dynamical timescale, SED models with either constant or exponentially rising star formation histories (SFHs) are best at reproducing the star formation rate (SFR) of galaxies at $z\sim2$. The study also found that assuming exponentially rising SFHs leads to stellar population age estimates that are on average $30\%$ older than those obtained under the assumption of constant SFHs. Our sample of galaxies, assuming both exponentially rising and declining SFHs, exhibit SFRs that are typically within 0.03 dex of those obtained under constant SFHs, which is within the usual measurement uncertainties of SED-derived SFRs.}. The choice of constant SFH over instantaneous burst models is based on the fact that the latter are better suited for the individual massive star clusters that are more age-sensitive than the entire high-redshift star-forming galaxies, which have dynamical times that are typically far greater than a few Myr \citep{Shapley_2001,Papovich_2001,Reddy_2012}. The reddening of the stellar continuum is added to the models assuming the following attenuation curves: the SMC \citep{Gordon}, \cite{Reddy_2015}, and \cite{Calzetti_2000}, with stellar continuum reddening in range of $E(B-V)_{\rm cont}=0.0-0.60$. Based on earlier studies, these curves are shown to best represent the shape of the dust attenuation curves for the majority of high-redshift galaxies (e.g., \citealt{Reddy_2018,Fudamoto_2020,Shivaei_2020}).

When fitting the broadband photometry, the stellar metallicity is held fixed at $\left<Z_{\ast}\right>=0.001$ as this value was found to best fit the rest-FUV spectra of galaxies in the MOSDEF/LRIS sample \citep{Topping,Reddy_2022}. The stellar population ages of the models are permitted to range between $\sim10\,$Myr and the age of the universe at the redshift of each galaxy. Unless mentioned otherwise, the \texttt{BPASS} model with binary stellar evolution, an upper-mass cutoff of $100\,M_{\odot}$ (\enquote{\rm 100bin}), and the SMC  extinction curve are adopted for this analysis. Previous studies (e.g., \citealt{Reddy_2022}) have shown that using the SMC dust attenuation curve results in better agreement between $\ha$ and SED-derived SFRs. Assuming the $Z_{\ast}=0.001$ 100bin \texttt{BPASS} SPS models in fitting the broadband photometry yields a conversion factor of $2.12 \times 10^{-42}\,M_{\odot} \rm{yr}^{-1} \rm{erg}^{-1} s$ between the dust-corrected $\ha$ luminosity and $\sfrha$. The dust-corrected $L(\rm UV)$ is determined using the best-fit model at $\lambda=1500\,\angstrom$ and the best-fit stellar continuum reddening.

The best-fit SED model is chosen by fitting the aforementioned models to broadband photometry. The parameters of the model with the lowest $\chi^{2}$ relative to the photometry are considered to be the best-fit values. The errors in the parameters are calculated by fitting the models to many perturbed realizations of the photometry according to the photometric errors. The resulting standard deviations in the best-fit model values give the uncertainties in these values.

\section{Morphology Analysis}
\label{sec:morph}
In this section, we present a methodology to construct resolved stellar population maps that may unveil galaxies undergoing bursts of star formation on smaller ($\sim$10\,kpc) spatial scales. We also examine the correlation between the resolved stellar population properties and $L(\ha)/L(\rm{UV})$. 
\subsection{Pixel Binning}
\label{sec:pix}

\begin{figure}
\centering
   \subfloat{%
      \includegraphics[width=\columnwidth]{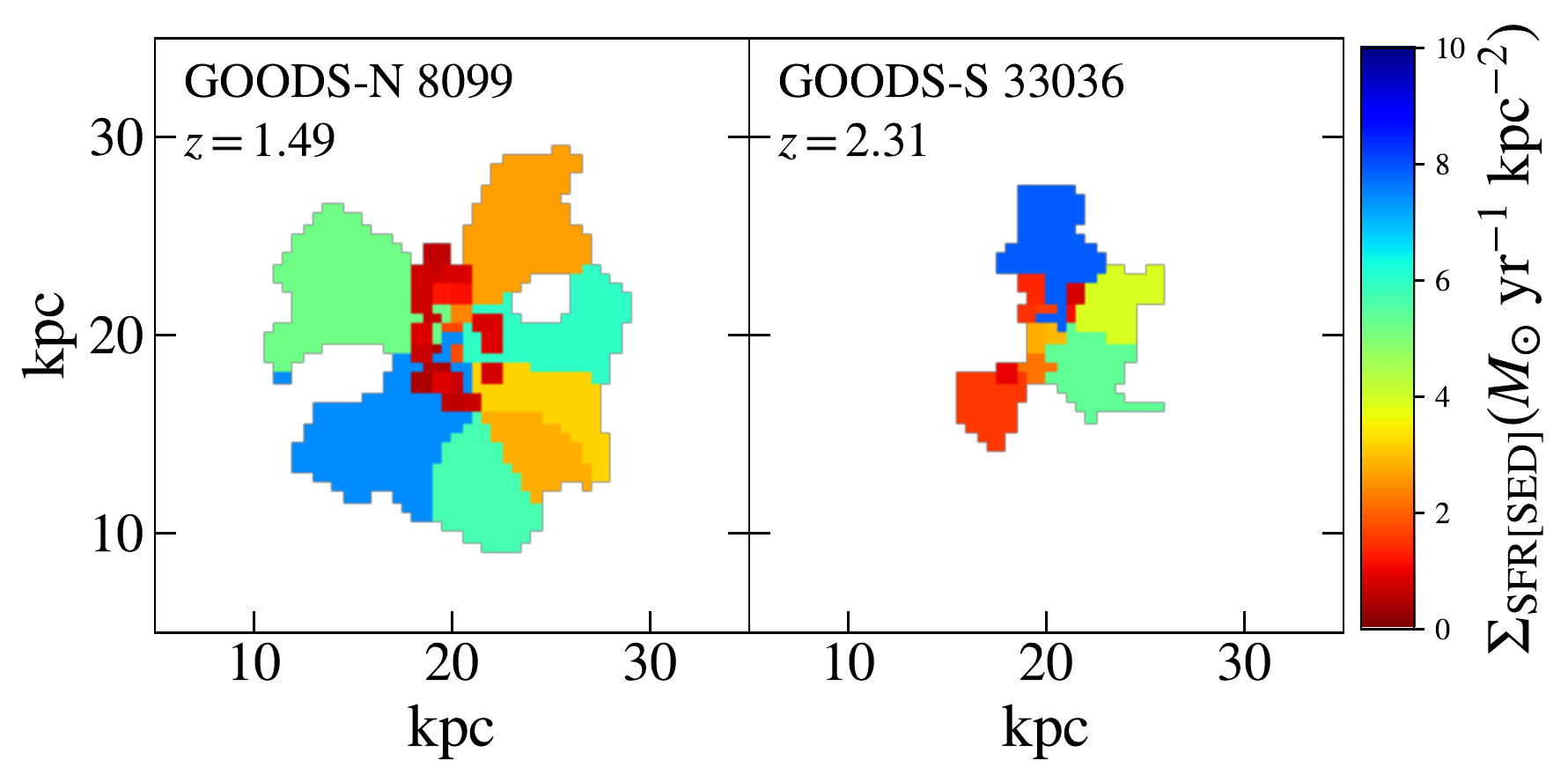}}
    \caption{Examples of star-formation-rate surface density ($\Sigma_\text{SFR[SED]}$) maps using Voronoi bins. The field name and 3D-HST Version 4.0 ID of each galaxy, as well as their redshifts, are indicated in the top left corner of each panel.}
    \label{fig:patchmap}
\end{figure}

Rather than studying the individual images pixel by pixel, we group pixels using the two-dimensional Voronoi binning technique introduced by \cite{Cappellari} and further modified by \cite{TARA_Voro}. The point spread function of the CANDELS imaging is larger than the individual pixels (0\farcs18), such that we apply a Voronoi binning technique to the imaging in order to avoid correlated noise between individual analyzed elements. In brief, each of the 3D-HST images (Section~\ref{sec:candels}) is divided into sub-images 80 pixels on a side. We use the SExtractor \citep{Bertin} segmentation map to mask pixels in each sub-image that are not associated with the galaxy. The pixels are grouped following the algorithm presented in \cite{Cappellari} to attain $S/N\geq5$ in at least five different filters (e.g., see \citealt{TARA_Voro}). 
Alongside CANDELS imaging, we use unresolved $Spitzer$/IRAC photometry to cover the rest-frame near-infrared part of the spectrum. As the HST and Spitzer/IRAC photometry have different spatial resolutions, we assign IRAC fluxes to each of the Voronoi bins proportionally according to the $H_{160}$ flux (see \citealt{TARA_Voro} for further details). The stellar population properties for each Voronoi bin are inferred using the SED models (see Section~\ref{sec:sedmodels}) that best fit the resolved 3D-HST photometry \citep{Wuyts_2011,Wuyts_2012,Wuyts_2013,Hemmati_2014,Lang_2014}. We calculate star-formation-rate surface density ($\Sigma_{\sfrsed}$) for each Voronoi bin by dividing the SFR determined from the best-fit resolved SED model by the area of each Voronoi bin. Figure~\ref{fig:patchmap} shows examples of the Voronoi bins and stellar population maps for two galaxies in the sample, one in each targeted redshift range.

\begin{figure}
    \centering
    \includegraphics[width=\columnwidth]{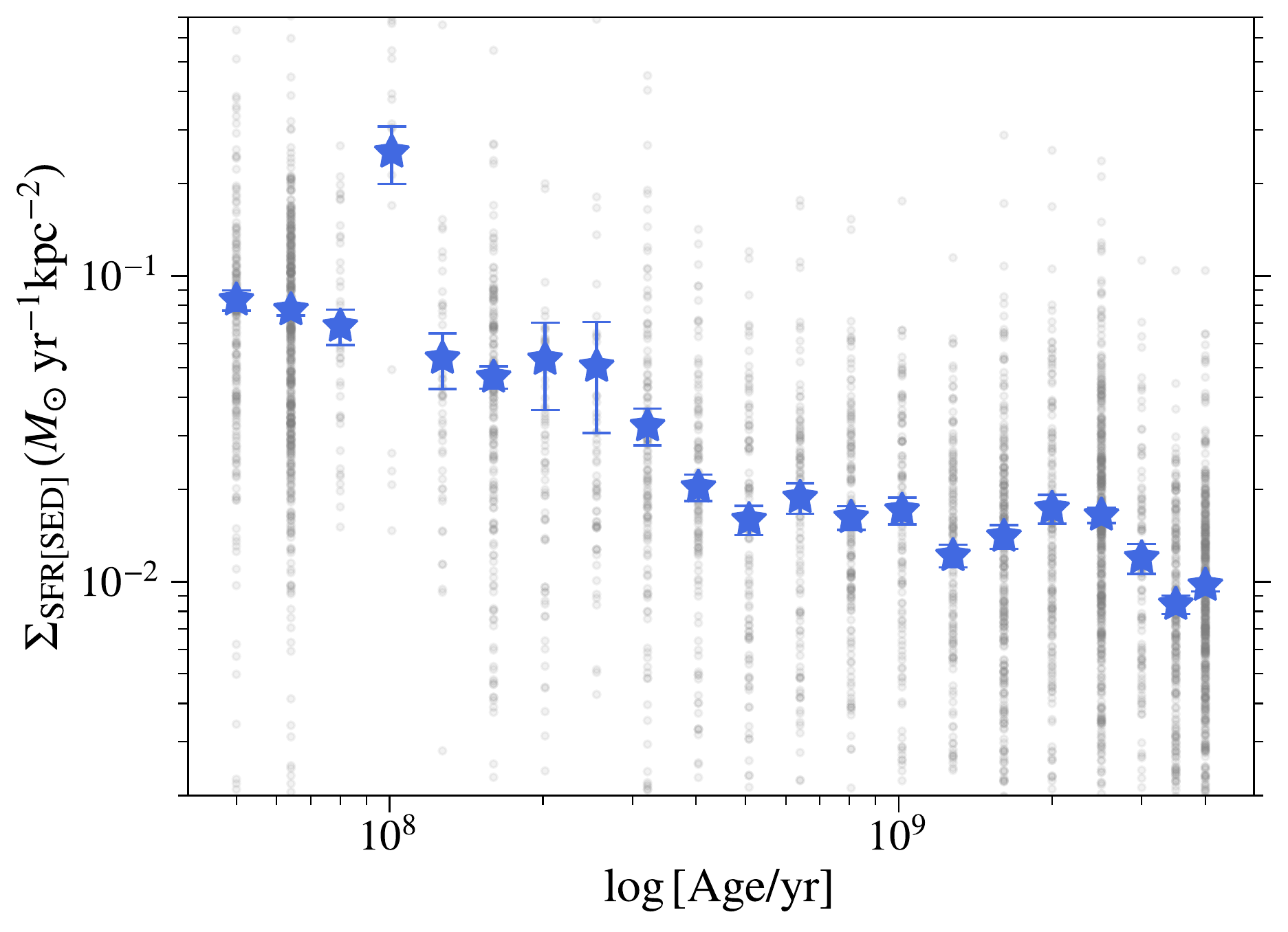}
    \caption{Star-formation-rate surface density vs. stellar age of Voronoi bins constructed for all the galaxies in the MOSDEF/MORPH sample (gray). Average values of $\Sigma_\text{SFR[SED]}$ in bins of stellar age $\log[\rm Age/yr]$ are shown by the blue stars. A total of 26 Voronoi bins are estimated at a stellar age of $10^{8}\,$yr, with 12 bins exhibiting $\Sigma_\text{SFR[SED]} \geq 0.1$, accounting for the noticeable outlier. }
    \label{fig:agevssfr}
\end{figure}

\begin{figure*}
\centering
    \includegraphics[width=0.96\textwidth]{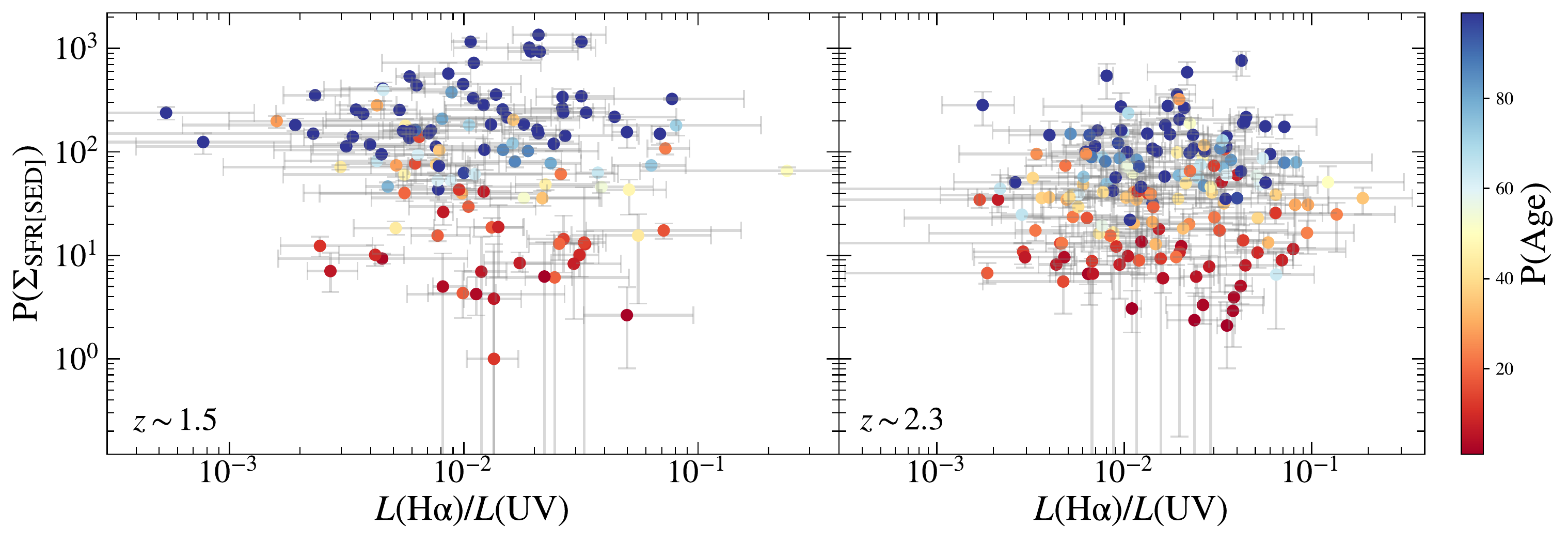}
    \caption{The MOSDEF/MORPH sample: $P(\Sigma_{\sfrsed})$ versus dust-corrected $L(\ha)/L(\rm UV)$ for two redshift bins centered at $z\sim1.5$ ($\it{left}$) and $z\sim2.3$ ($\it{right}$). The points are colored by patchiness of the stellar age. No significant correlations is found between $L(\ha)/L(\rm UV)$ and $P(\Sigma_{\sfrsed})$, or between $L(\ha)/L(\rm UV)$ and $P(\rm Age)$. The Spearman correlation properties for the relations shown in this figure are reported in Table~\ref{tab:Spearman}.}
    \label{fig:patchvar}
\end{figure*} 

\subsection{Patchiness}
\label{patch}
Patchiness ($P$) is a recently introduced morphology metric \citep{Tara_2022} that evaluates the Gaussian likelihood that each of the distinct components of a distribution are equal to the weighted average of the distribution. In this analysis, individual elements are values of a parameter measured for each of the resolved Voronoi bins. The area-weighted average of the parameter $X$ measured from individual Voronoi bins is defined by
\begin{eqnarray}
\left<X\right> = \frac{ \sum_{i=1}^{ N_{\rm{bins}}} n_{\rm {pix},\it {i}} X_{i}}{\sum_{i=1}^{N_{\rm bins}} n_{\rm pix,\it{i}}},
\label{eq:ave}
\end{eqnarray}
where $X_{i}$ are the values measured for the parameter $X$ inside each of the Voronoi bins with  uncertainty $\sigma_{i}$, $N_{\rm bins}$ is the total number of Vornoi bins in a galaxy photometry, and $n_{\rm pix}$ is the total number of pixels inside a single Voronoi bin (area). The patchiness, $P(X)$, can be calculate by Equation 2 in \cite{Tara_2022} as:
\begin{eqnarray}
P(X) = -\ln \Bigg\{\prod_{i=1}^{ N_{bins}} \frac{1}{\sqrt{2\pi}\sigma_{i}} \exp{\Bigg[-\frac{(X_{i}-\left<X\right>)^{2}}{2\sigma_{i}^2}\Bigg]}\Bigg\}.
\label{eq:patchiness}
\end{eqnarray}

\begin{table}
\centering
\begin{threeparttable}[b]
\caption{Results of Spearman correlation tests between $L(\ha)/L(\rm UV)$ and $P(\rm Age)$, as well as $L(\ha)/L(\rm UV)$ and $P(\Sigma_{\sfrsed})$. }
\label{tab:Spearman}
\def\arraystretch{1.3}%
\begin{tabular*}{\columnwidth}{@{\extracolsep{\fill}} lccc}
\hline
Redshift bins&&$\rho_{s}$\tnote{a}& $P_{n}$\tnote{b}\\
\hline
\multirow{2}{*}{$z\sim1.5$}&{$P(\rm Age)$}\tnote{c}& -0.04& 0.62\\
&{$P(\Sigma_{\sfrsed})$}\tnote{d}&-0.09 & 0.30 \\
\hline
\multirow{2}{*}{$z\sim2.3$}&{$P(\rm Age)$}& 0.06 & 0.43\\
&{$P(\Sigma_{\sfrsed})$}& 0.07 & 0.36 \\
\hline
\end{tabular*}
\begin{tablenotes}
\item [a] The Spearman correlation coefficient between $L(\ha)/L(\rm UV)$ and each of the listed parameters using the MOSDEF/MORPH sample.
\item [b] The probability of null correlation between $L(\ha)/L(\rm UV)$ and each of the listed parameters using the MOSDEF/MORPH sample.
\item [c] Patchiness of the stellar population age.
\item [d] Patchiness of the star-formation-rate surface density.
\end{tablenotes}
\end{threeparttable}
\end{table}
 
A detailed discussion and evaluation of the patchiness metric properties are presented in \cite{Tara_2022}. In brief, patchiness can be compared most reliably between galaxies with similar redshifts. Thus, we divide galaxies into two bins of redshift and analyze the patchiness separately for galaxies in each bin. Moreover, patchiness can be used on parameters with large dynamic ranges or parameters with values close to zero. We study patchiness of $\Sigma_{\sfrsed}$ which traces the concentration of star formation within the Voronoi bins, and exhibits a large dynamic range among our sample galaxies. A single Voronoi bin has, on average, a size of $\sim4.5\,\rm{kpc^{2}}$, and a median size of $\sim1.5\,\rm{kpc^{2}}$. The estimated typical dynamical timescale for a Voronoi bin is $\sim10\,$Myr, based on the velocity dispersion assumption inferred from studies on $z\sim2$ galaxies with comparable size measurements to the Voronoi bins used in this work \citep{Erb_2006,Law_2009,Law_2012,Reddy_2012,TARA_Voro}. A physical example of how patchiness can be used is presented in \cite{Tara_2022}, where higher patchiness values of stellar reddening indicate a more complex dust distribution.

A burst of star formation on top of an underlying constant SFH can result in an increase in $\Sigma_{\rm SFR}$. An element of a resolved population containing a burst of star formation has a higher $\Sigma_{\rm SFR}$ and a younger stellar age compared to other elements, resulting in a larger $P(\Sigma_{\sfrsed})$, and $P(\rm{Age})$. Therefore, large $P(\Sigma_{\sfrsed})$ may suggest the presence of bursts in localized (Voronoi) regions of galaxies. Figure~\ref{fig:agevssfr} indicates the relationship between the stellar age and $\Sigma_{\rm SFR[SED]}$ derived for Voronoi bins constructed for all the galaxies in the MOSDEF/MORPH sample. The figure indicates that younger stellar populations are found in regions with higher SFR surface densities.

\subsection{Patchiness of $\Sigma_{\sfrsed}$ vs. $L(\ha)/L(\rm UV)$}
\label{Sec:patchvslratio}
This section presents our results on the correlation between $P(\Sigma_{\sfrsed})$ and $L(\ha)/L(\rm UV)$. Given that star-formation mode varies in strength, duration, or a combination of both factors in different regions of a galaxy \citep{Reddy_2012,Dale_2016,Dale_2020,Smith_2021}, and patchiness is sensitive to outliers below and above the average, we expect $P(\Sigma_{\sfrsed})$ to be large for galaxies that are undergoing a burst of star formation that could be detected on resolved scales. 

Due to surface brightness dimming, higher-redshift objects on average have fewer and larger Voronoi bins. To control for this effect, we divide the MOSDEF sample into two subsamples in the redshift ranges of $ 1.37 < z < 1.70$ ($z\sim1.5$) and $ 2.09 < z < 2.61$ ($z\sim2.3$). Figure~\ref{fig:patchvar} shows the relationship between $P(\Sigma_{\rm SFR[SED]})$ and $L(\ha)/L(\rm UV)$ for galaxies in each redshift bin. Based on a Spearman correlation test, we find no significant correlation between the two for both the $z\sim1.5$ and $z\sim2.3$ subsamples, with probabilities of $P_{\rm n}=0.30$ and 0.36, respectively, of a null correlation. As shown by the stellar age color-coded points, a higher $P(\Sigma_{\rm SFR[SED]})$ corresponds to a higher $P(\rm Age)$, which is expected given that stellar age and star-formation-rate surface density are correlated for a given SPS model. There is also a lack of correlation between $L(\ha)/L(\rm UV)$ and $P(\rm Age)$ with correlation properties reported in Table~\ref{tab:Spearman}.

One possible cause for the absence of correlation is the large uncertainties in $L(\ha)/L(\rm{UV})$, $P(\Sigma_{\rm SFR[SED]})$, and $P(\text{Age})$\footnote{The SED parameters, such as SFR, stellar age, etc., are determined for individual Voronoi bins by performing SED fitting on resolved scales, as explained in Section~\ref{sec:sedmodels}. However, determining the uncertainty or noise in each SED parameter is a time-consuming process due to the large number of Voronoi bins and galaxies involved \citep{Tara_2022}. To address this challenge, we chose 50 galaxies with stellar population parameters similar to the overall sample and perturbed the resolved photometric fluxes based on their respective errors. We conducted SED fitting on these perturbed values, and the standard deviation of the SED parameters from the models with the lowest chi-squared were treated as the 1$\sigma$ uncertainty in the SED parameters.}. Using $L(\ha)/L(\rm{UV})$ as a tracer of stochastic star formation may be complicated by uncertainties in dust corrections and aperture mismatches between the Ha and UV measurements (e.g., \citealt{Brinchmann_2004,Kewley_2005,Salim_2007,Richards_2016,green_2017,Tara_2022}). These issues are discussed in more detail below.

Although there is a consensus that the \cite{Cardelli} curve is an adequate description for the dust reddening of nebular recombination lines such as $\ha$ \citep{Reddy_2020,Rezaee_2021}, a variety of different stellar attenuation curves have been found for high redshift galaxies, depending on their physical properties. For example, several studies have found that more massive galaxies ($M_{\ast}>10^{10.4}\,M_{\odot}$) tend to have a slope of the attenuation curve that is similar to the \cite{Calzetti_2000}, while the SMC extinction curve has been shown to be applicable for less massive galaxies \citep{Reddy_2015,Du_2018,Shivaei_2020}. We obtain the same lack of correlation between $P(\Sigma_{\sfrsed})$ and $L(\ha)/L(\rm UV)$ when the \cite{Reddy_2015} and the metallicity-dependent \cite{Shivaei_2020} curves are used to dust-correct $L(\rm UV)$. We find that the degree by which the variation in the attenuation curves affects the $P(\Sigma_{\sfrsed})$ and $L(\ha)/L(\rm UV)$ correlation is insignificant as long as a fixed curve is assumed to dust-correct $L(\rm UV)$. However, a correlation may still be washed out if the attenuation varies from galaxy to galaxy systematically as a function of $L(\ha)/L(\rm UV)$ ratio. 

Another factor that might cause the $\ha$-to-UV luminosity ratios of high redshift galaxies to deviate from their true values is  aperture mismatch. $L(\rm UV)$ is measured using broadband photometry, while $\ha$ luminosity is measured using slit spectroscopy. However, \cite{Tara_2021a} conducted an aperture-matched analysis utilizing a MOSDEF sample comparable to the one used in this study and found that the variations between $\ha$ and UV SFRs are not caused by the aperture mismatches. Another possible reason for the absence of correlation is that the variations in SFH may be occurring in regions that are still spatially unresolved with the HST imaging (i.e., on scales smaller than a few kpc). Additionally, the lack of correlation could be expected if variations in the SFH are occurring on even shorter timescales than the typical dynamical timescale of the spatial region probed by a Voronoi bin ($\sim 10\,$Myr). In this case, such short and localized bursts of star formation may only affect the $\ha$-to-UV ratio on similar spatial scales.

\section{Rest-FUV Composite Spectra Construction, And Model-Predicted $L(\ha)/L(\rm UV)$ }
Aside from patchiness, there are several key FUV spectral features that are age-sensitive and can potentially be used to probe bursty SFHs. In this section, we outline a stacking analysis methodology that allows us to measure the average strength of FUV features in bins of $L(\ha)/L(\rm{UV})$.
\label{sec:compnorm}
\subsection{Rest-FUV Composite Spectra Construction}
\label{sec:compfuv}
Rest-FUV spectra are averaged together to produce high $S/N$ composite spectra. Individual LRIS spectra have limited $S/N$ to make measurements on the FUV spectral features. Using the stacked spectra, we measure the average physical properties of galaxies contributing to each composite, as well as measuring FUV spectral features associated with massive stellar populations. We use the procedures that are outlined in \citet{Reddy_2016,Reddy_2022} to construct the composites. In brief, the science and error spectrum of sample galaxies are shifted to the rest-frame based on the MOSDEF spectroscopic redshift (Section~\ref{sec:mosdef}), converted to luminosity density, and interpolated to a grid with wavelength steps $\Delta \lambda = 0.5\,\angstrom$. The composite spectrum at each wavelength point is calculated as the average luminosity density after rejecting $3\sigma$ outliers. The error in the composite spectrum is calculated by perturbing the individual spectra according to their error, and using bootstrap resampling to construct the stacked spectrum for those perturbed spectra 100 times. The standard deviation of the luminosity densities at each wavelength point gives the error in the composite spectrum.
\subsection{Continuum Normalization}
\label{sec:norm}
Rest-FUV composite spectra must be continuum-normalized in order to accurately measure the average strength of the FUV stellar features. We use the \texttt{SPS+Neb} models discussed in \cite{Reddy_2022} to aid in the normalization process. \texttt{SPS+Neb} models consist of  the \texttt{BPASS} SPS models described in Section~\ref{sec:sedmodels} as the stellar component. Each \texttt{BPASS} SPS model is used as an input to the \texttt{Cloudy}\footnote{\url{https://gitlab.nublado.org/cloudy/cloudy/-/wikis/home}} version 17.02 radiative transfer code \citep{Ferland17} to compute the nebular continuum. The final \texttt{SPS+Neb} models are then built by combining the stellar and nebular components. We refer the reader to \cite{Reddy_2022} for more details. In brief, all the \texttt{BPASS} SPS models with a range of stellar ages of $\log [\rm Age/\rm yr]= \{7.0, 7.3, 7.5, 7.6, 7.8, 8.0, 8.5, 9.0\}$ are interpolated to construct models with stellar metallicities comparable to the values expected for $z\sim2$ galaxies \citep{Steidel_2016} rather than the original metallicity values of \texttt{BPASS} models described in Section~\ref{sec:sedmodels}. This results in a grid of models with stellar metallicities ranging from $Z_{\ast}=10^{-4}$ to $3\times10^{-3}$ spaced by $2\times10^{-4}$. Our assumptions for the ionization parameter ($U$) and gas-phase oxygen abundance (i.e., nebular metallicity; $Z_{\rm neb}$) match the average values for the MOSDEF/LRIS sample where $\log[Z_{\rm neb}/Z_{\odot}] = -0.4 $ and $\log U=-3.0$ \citep{Topping,Reddy_2022}. 

We fit the composite spectra with \texttt{SPS+Neb} models to model the continuum. The \texttt{SPS+Neb} models are normalized for a constant SFR of $1\,M_{\odot}/\rm{yr}$. To re-normalize the models to the observed spectra, these models are forced to have the same median luminosity as the composites in the \cite{Steidel_2016} \enquote{Mask 1} wavelength windows. These wavelength windows are chosen to include regions of the spectrum that are not affected by interstellar absorption and emission features. We smooth the \texttt{SPS+Neb} models for wavelengths below $1500\angstrom$ to have the same rest-frame resolution as the MOSDEF/LRIS spectra. To identify the best-fit \texttt{SPS+Neb} model for a composite spectrum, the $\chi^{2}$ between the models and the composite are computed. The model that yields the smallest $\chi^{2}$ is taken as the best-fit model. Using the median luminosity densities defined in the \cite{Rix_2004} wavelength windows, a spline function is fitted to the best-fit model. Finally, the composite spectrum is divided by that spline function to produce a continuum-normalized spectrum. 

\begin{figure}
\centering

    \includegraphics[width=1\columnwidth]{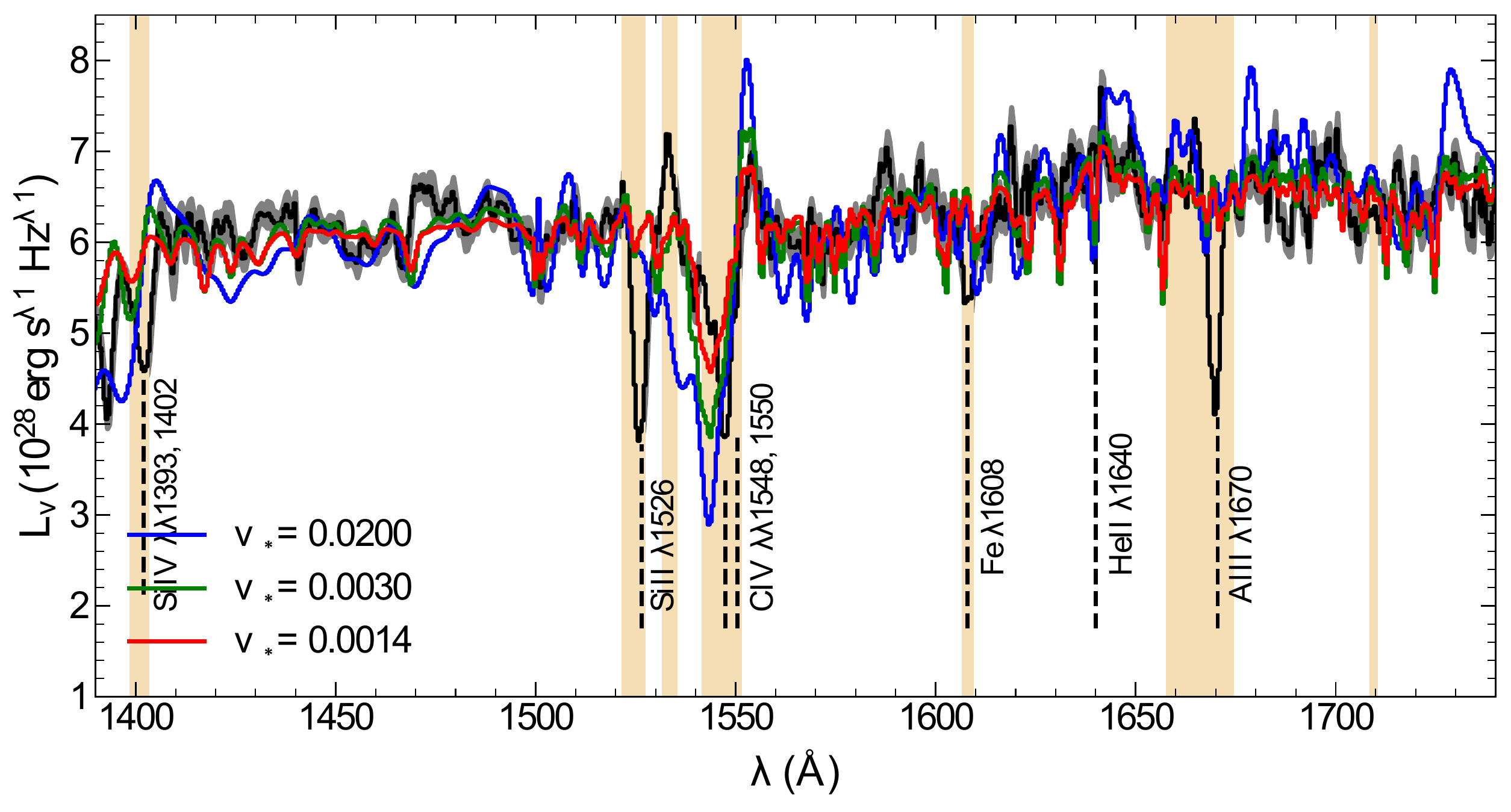}
    \caption{Composite spectrum constructed for the 124 galaxies in the MOSDEF/LRIS sample (black) with $1\sigma$ uncertainty (gray). The \texttt{SPS+Neb} models with fixed stellar age of $\log[\rm{Age/yr}]=8.0$ and various stellar metallicities are shown alongside. Some of the prominent FUV spectral features are labeled. Regions that are not included in the fitting process are shaded in orange. The wavelengths that have been excluded are the ones that are impacted by interstellar absorption and emission features.}
    \label{fig:comp}
\end{figure}

Any line measurements derived from the continuum-normalized spectra are affected by uncertainties in the normalization of the composite spectra. In order to compute this uncertainty, the normalization process outlined above is applied to $100$ realizations of the composite spectrum constructed by bootstrap resampling, and fitting the \texttt{SPS+Neb} models to those realizations. The standard deviation of the best-fit models gives the uncertainty in the continuum normalization at each wavelength point. In addition, all of the model parameters and their uncertainties, including stellar age, metallicity, continuum reddening, and $\sfrsed$ are derived using the mean and standard deviation of the best-fit values when fitting those realizations, respectively. Figure~\ref{fig:comp} shows an example of the comparison between the composite spectrum computed for all the galaxies in the MOSDEF/LRIS sample along with \texttt{SPS+Neb} models of different stellar metallicities. Models with lower metallicities are more consistent with the observed composite spectrum of $z\sim2$ galaxies \citep{Steidel_2016,Reddy_2022}.

\subsection{$L(\ha)/L(\rm UV)$ Predicted by the \texttt{SPS+Neb} Models versus Physical Properties and Model Assumptions}
\label{Sec4}

\begin{figure*}
\centering
   %\subfloat{%
      %\includegraphics[width=0.9\textwidth]{images/bpasslratiooriginal.pdf}}
     
    \subfloat{%
      \includegraphics[width=0.5\textwidth]{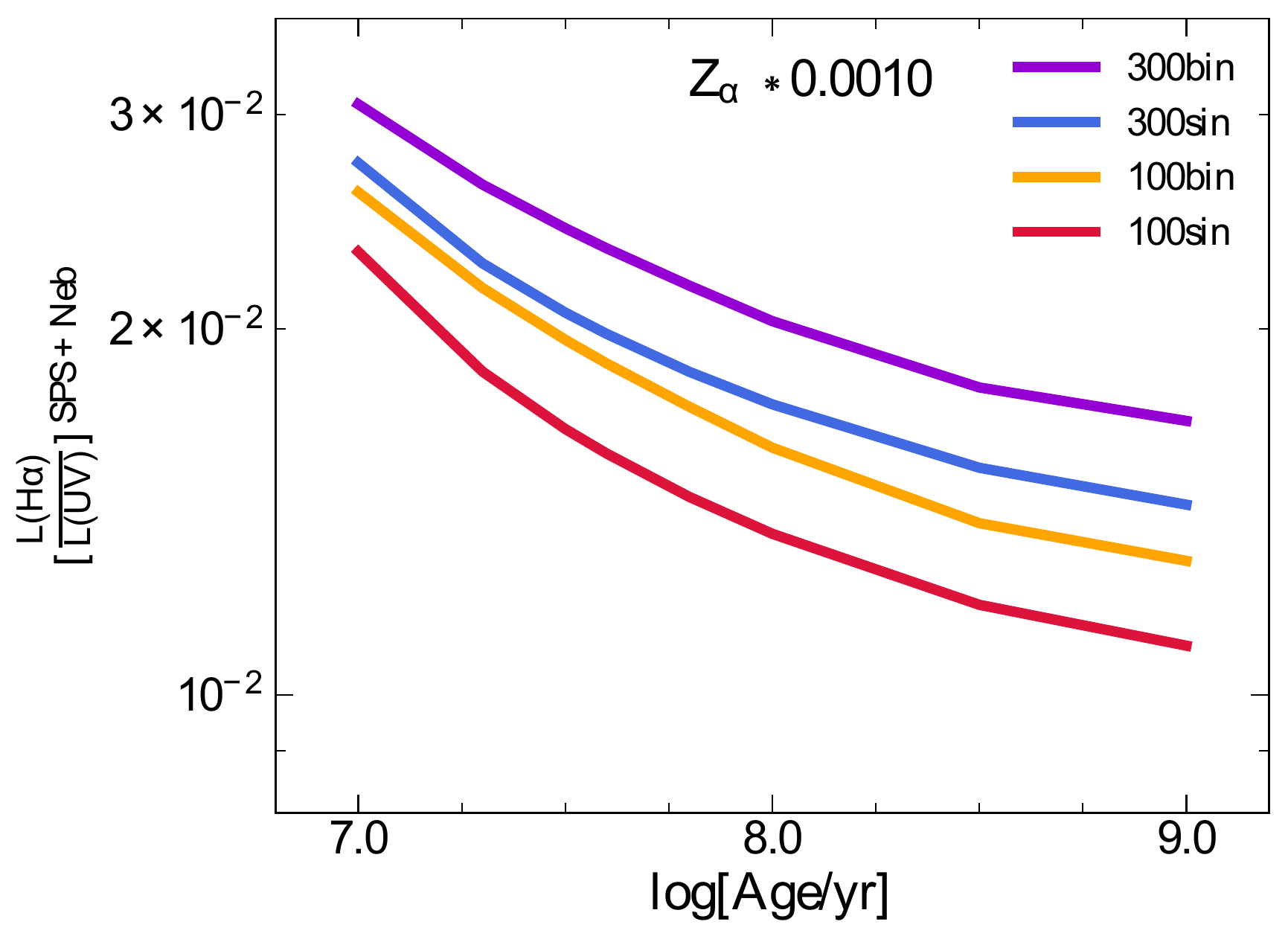}}
    \subfloat{%
      \includegraphics[width=0.5\textwidth]{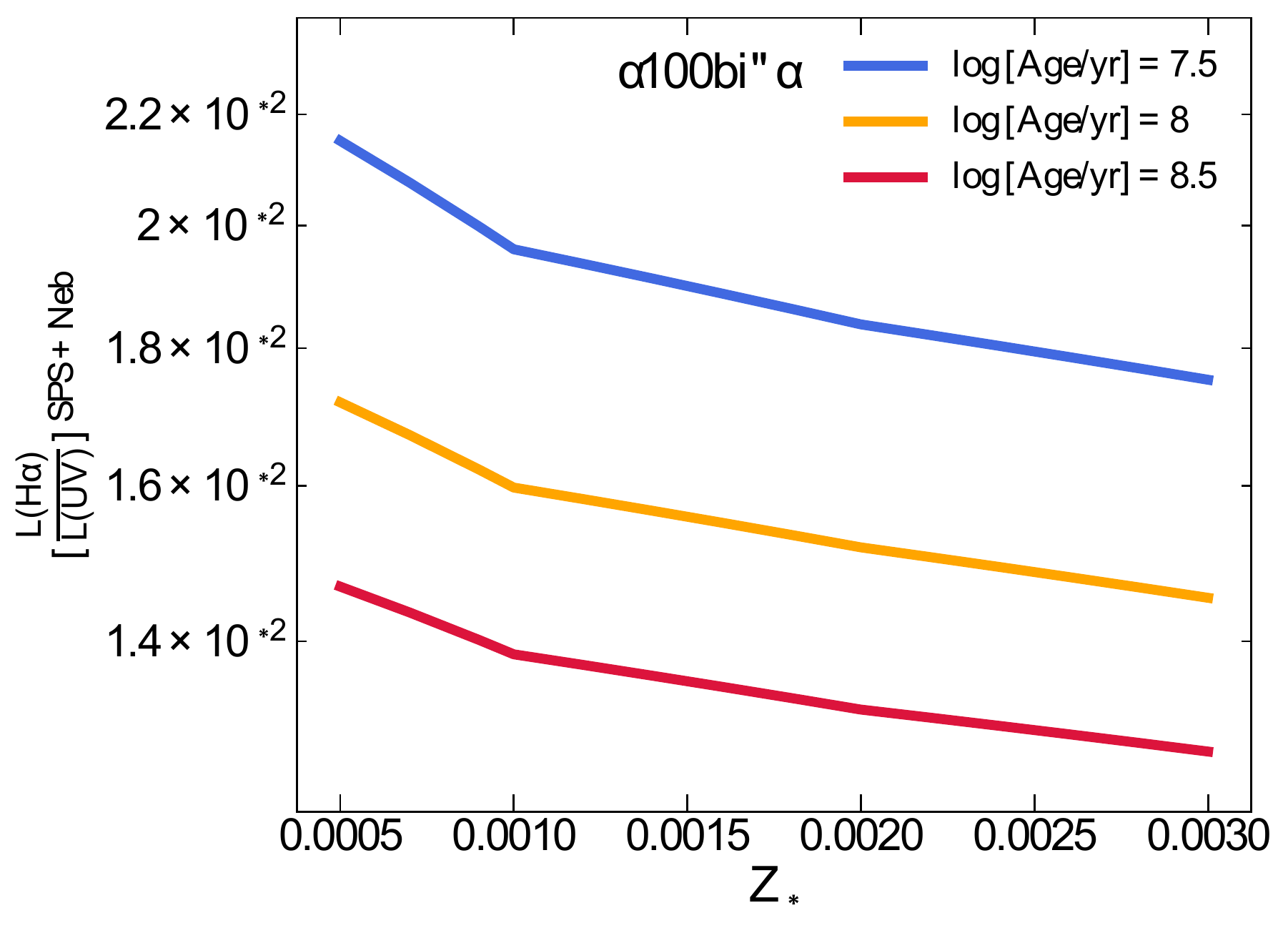}}\\
\caption{Variation of the $\ha$-to-UV luminosity ratio derived from the \texttt{SPS+Neb} models with physical properties including stellar age, stellar metallicity, inclusion of binary stellar evolution, and upper-mass cutoff of the IMF. The \texttt{SPS+Neb} models are constructed using a constant SFH assumption.}
\label{fig:bpass}
\end{figure*}
In this section, we examine how the $\ha$-to-UV ratio varies with stellar population properties, including stellar age, metallicity, inclusion of binaries, and $M_{\rm cutoff}$ of the IMF using the \texttt{SPS+Neb} models. These relations are shown in Figure \ref{fig:bpass} and are used to study the systematic variation observed in $L(\ha)/L(\rm UV)$ of the MOSDEF/LRIS galaxies in Section~\ref{Sec:lrisbins}. We calculate the $\ha$ luminosity of each model using the relation: 
\begin{eqnarray}
L(\ha)\,[\rm erg\, s^{\rm -1}]=1.37 \times 10^{-12}\, \it N(\rm H^{\rm 0})\,[\rm s^{\rm -1}]
\end{eqnarray}
where $N(H^{0})$ is the hydrogen ionizing photon rate. We calculate $N(H^{0})$ by integrating the model spectrum below $912 \angstrom$. $L(\rm UV)$ is calculated using the \texttt{SPS+Neb} model at $\lambda=1500\, \angstrom$.

The left panel of Figure~\ref{fig:bpass} indicates that the ratio predicted by the constant SFH models ($\left[L(\ha)/L(\rm UV)\right]_{\rm \texttt{SPS+Neb}}$) at a fixed stellar metallicity is influenced by both the choice of upper-mass cutoff of the IMF and inclusion of binary stellar evolution. The $\ha$ luminosity increases by the presence of extremely massive stars with masses greater than $100 M_{\odot}$ and inclusion of energetic binary systems. For example, at $\log[\rm{Age/yr}]=8.0$, the $L(\ha)/L(\rm UV)$ ratio grows by a factor of 1.2, and 1.3, respectively, from \enquote{100sin} to  the \enquote{100bin} and \enquote{300bin} models. The number of ionizing photons (and thus $L(\ha)$) will decrease as massive O-stars evolve off the main sequence, whereas less massive stars will still contribute significantly to the non-ionizing UV luminosity. As a result, the $L(\ha)/L(\rm UV)$ ratio decreases with increasing age as shown in the left panel of Figure~\ref{fig:bpass}.

The right panel of Figure~\ref{fig:bpass} shows the sensitivity of $\left[L(\ha)/L(\rm UV)\right]_{\rm \texttt{SPS+Neb}}$ of "100bin" model assumption to the stellar metallicity at various stellar population ages of the models. At a fixed stellar age, decreasing stellar metallicity increases the \ha-to-UV luminosity ratio. For example, $\left[L(\ha)/L(\rm UV)\right]_{\rm \texttt{SPS+Neb}}$ grows by a factor of $\sim1.1$ from $Z_{\ast}=0.0020$ to $Z_{\ast}=0.0010$ models, at $\log[\rm Age/yr]=8.0$.  This relationship is expected given that lower-metallicity stellar atmospheres (less opaque) result in higher effective temperatures and therefore harder ionizing spectra \citep{Bicker_2005}. 

\section{Variations of the average Physical properties of galaxies with $L(\ha)/L(\rm UV)$}
\label{Sec:lrisbins}

In addition to variations in the physical properties of galaxies such as stellar age and metallicity, variations in the strength of age-sensitive FUV spectral features with $L(\ha)/L(\rm UV)$ may contain important information on burstiness. To investigate the above-mentioned variations, we divide the MOSDEF/LRIS sample into two $L(\ha)/L(\rm UV)$ subsamples (hereafter referring to as \textit{low-} and \textit{high-}$L(\ha)/L(\rm UV)$ bin) with an equal number of galaxies in each. When binning the galaxies, we are using the $\ha$-to-UV luminosity ratio rather than the $\sfrha$-to-$\rm SFR[\rm{UV}]$ ratio because the latter requires some assumptions of the SFH to convert luminosity to SFR, and when trying to probe the SFH (i.e., whether a galaxy has a bursty or constant SFH), it is useful to use a probe which is independent of such assumptions. The results of the measurements on the two subsamples are presented in the following sections.

\subsection{Physical Properties of Galaxies vs. $L(\ha)/L(\rm UV)$}
The bestfit \texttt{SPS+Neb} models to the rest-FUV composites are used to derive the average stellar age, metallicity, and continuum reddening of galaxies in each of the  $L(\ha)/L(\rm UV)$ bins (Table~\ref{tab:eqpcygni}). In order for the \texttt{SPS+Neb} models to self-consistently explain all the observations, we checked that the $L(\ha)/L(\rm UV)$ predicted by the best-fit \texttt{SPS+Neb} model to each composite is in agreement with the mean ratio of all individual galaxies contributing to the composite as well as the average ratio directly measured from the rest-FUV and optical composite spectra~\footnote{The same procedure outlined in Section~\ref{sec:compfuv} is applied to construct the optical composite spectrum (e.g., \citealt{Shivaei_2018,Reddy_2020,Rezaee_2021}). The \texttt{Python} code presented in \url{https://github.com/IreneShivaei/specline/} is used in constructing the optical composite spectra here.}.

Table~\ref{tab:eqpcygni} reports the average physical properties of galaxies in each $L(\ha)/L(\rm UV)$ bin. 
The \textit{high-}$L(\ha)/L(\rm UV)$ subset exhibits, on average, an stellar population age of $\log[\rm{Age/yr}]=8.0$, compared to $\log[\rm{Age/yr}]=8.4$ for the \textit{low-}$L(\ha)/L(\rm UV)$ galaxies, though the difference in age is significant at only the $2\sigma$ level. The stellar population age of the \textit{high-}$L(\ha)/L(\rm UV)$ galaxies is $100\,$Myr, longer than the dynamical timescale of a few tens of Myr, implying that the \textit{high-}$L(\ha)/L(\rm UV)$ galaxies are not necessarily undergoing a burst of star formation.
However, this conclusion comes with the caveat that the minimum SED-fitting age would be equivalent to the dynamical timescale for a single initial burst of star formation. Using the \texttt{SPS+Neb} models, $L(\ha)/L(\rm UV)$ increases by a factor of $\sim1.1$ from $\log[\rm{Age/yr}] = 8.4$ to $\log[\rm{Age/yr}] = 8.0$ for a fixed stellar metallicity (Figure~\ref{fig:bpass}). The \textit{high-}$L(\ha)/L(\rm UV)$ subset exhibits an average $L(\ha)/L(\rm UV)$ which is $\sim5$ times larger than that of the \textit{low-}$L(\ha)/L(\rm UV)$ subset. This implies that the difference in the average $L(\ha)/L(\rm UV)$ ratio of the subsets cannot be solely attributed to the variation in the stellar age of those subsets. It is essential to examine other burst indicators, such as the strength of the FUV P-Cygni features in both bins, to find whether there is any strong evidence that the \textit{high-}$L(\ha)/L(\rm UV)$ subset traces recent starbursts. This is further discussed in the next section.

The effective radius ($R_{\rm{e}}$) of each galaxy is taken from \cite{van_der_Wel2014}, and is defined as the radius that contains half of the total $HST/F160W$ light. The star-formation-rate surface density ($\Sigma_{\sfrha}$) of individual galaxies is then computed as:

\begin{equation}
\Sigma_{\sfrha}=\frac{\rm{\sfrha}}{2 \pi R_{\rm{e}}^{2}}.
\end{equation}
For an ensemble of galaxies, $\left<{\sfrha}\right>$ is computed by multiplying the dust-corrected $\left<L(\ha)\right>$ measured from the optical composite spectrum by the conversion factor determined from the best-fit \texttt{SPS+Neb} model. $\left<{\Sigma_\sfrha}\right>$ is then computed using $\left<{\sfrha}\right>$ and mean $R_{\rm e}$ of individual galaxies in each ensemble. $\left<\sfrha\right>$ and $\left<\Sigma_{\sfrha}\right>$ increase significantly with increasing $\left<L(\ha)/L(\rm {UV})\right>$ between the two subsamples. While the instantaneous SFR (i.e., $\sfrha$) differs significantly between the two subsamples, $\sfrsed$ does not change significantly within the measurement uncertainties. By design, galaxies with higher $L(\ha)/L(\rm {UV})$ have on average higher $\ha$ luminosities. However, this does not necessarily imply that these galaxies have higher $\ha$-based SFRs than UV-based SFRs.  The conversion factor that relates the dust-corrected $L(\ha)$ with SFR\footnote{The conversion factors between the dust-corrected $\ha$ luminosity into $\sfrha$ are $2.675$, $4.888$, $2.102$, and $4.236\times10^{-42},M_{\odot}\rm{yr^{-1}erg^{-1}s}$ for the \enquote{$0.07\,Z_{\odot},$100sin}, \enquote{$1.41\,Z_{\odot},$100sin}, \enquote{$0.07\,Z_{\odot},$100bin}, and \enquote{$1.41\,Z_{\odot},$100bin} models, respectively. This is based on the \cite{Chabrier} IMF.} depends on stellar age, metallicity, and the hardness of the ionizing spectrum.  As we show below, there is evidence that galaxies with higher $L(\ha)/L(\rm {UV})$ have a harder ionizing spectrum than those with lower $L(\ha)/L(\rm {UV})$ and, as such, they are likely to have a higher $\ha$ flux per unit SFR (see discussion in Section~\ref{sec:fuvfeature}, and Section~\ref{sec:Sum}). The difference between the nebular and stellar reddening in the \textit{high}-$L(\ha)/L(\rm {UV})$ bin is $\simeq2.1$ times larger when compared to the \textit{low}-$L(\ha)/L(\rm {UV})$. The higher nebular reddening measured for the \textit{high}-$L(\ha)/L(\rm UV)$ bin is not surprising given that galaxies with larger $L(\ha)$ (i.e., higher SFRs) tend to be dustier \citep{Foster_2009,Reddy_2010,Kashino_2013,Reddy_2015,Reddy_2020}. 
\begin{table}
\centering
\begin{threeparttable}[b]
\caption{Average stellar population properties}
\label{tab:eqpcygni}
\def\arraystretch{1.3}%
\begin{tabular*}{\columnwidth}{@{\extracolsep{\fill}} lcc}
\hline
Properties& \textit{low-$\frac{L(\ha)}{L(\rm UV)}$} & \textit{high-$\frac{L(\ha)}{L(\rm UV)}$}\\
\hline
$\left<L(\ha)/L(\rm {UV})\right>$\tnote{a}& $0.007\pm0.002$ & $0.035\pm0.005$ \\
$\left<z\right>$\tnote{b}& $2.132\pm0.031$ & $2.185\pm0.029$ \\
$\left<\log[M_{\ast}/M_{\odot}]\right>$\tnote{c} & $9.88\pm 0.05$ & $9.96\pm0.06$ \\
$\left<R_{\rm e}\right>(\rm{kpc})$\tnote{d} & $2.94\pm0.23$&$2.33\pm0.14$ \\
$\left<12+\log(\rm O/H)\right>$\tnote{e}&$8.52\pm0.02$&$8.39\pm0.02$\\
\hline
$\left<Z_{\ast}/Z_{\odot}\right>$\tnote{f} & $0.099\pm 0.010$& $0.085\pm0.015$ \\
$\left<\log[\rm {Age/yr}]\right>$\tnote{g}& $8.4\pm 0.1$ &$8.0\pm0.2$ \\
$\left<E(B-V)_{\rm cont}\right>$\tnote{h}&$0.090\pm0.010$&$0.074\pm0.008$\\
$\rm{\left<\sfrsed\right>}(M_{\odot}\,yr^{-1})$\tnote{i}& $9.61\pm2.73$&$10.64\pm3.35$ \\
$\rm{\left<\sfrha\right>}(M_{\odot}\,yr^{-1})$\tnote{j}&$8.57\pm1.96$&$22.12\pm2.04$\\
$\rm{\left<\Sigma_{\sfrha}\right>}(M_{\odot}\,\rm yr^{-1} kpc^{-2})$\tnote{k} & $0.16\pm0.04$&$0.65\pm0.10$ \\
$\left<E(B-V)_{\rm neb}\right>$\tnote{l}&$0.29\pm0.03$&$0.49\pm0.06$\\
\hline
$\big<W_{\lambda}$(\rm{\ion{Si}{iv})}$\big>\,(\angstrom)$\tnote{m} & $0.103\pm 0.018$ & $0.146\pm0.015$ \\
$\big<W_{\lambda}$(\rm{\ion{C}{iv})}$\big>\,(\angstrom)$\tnote{n} & $0.206\pm 0.034$ & $0.113\pm 0.024$ \\
$\big<W_{\lambda}$(\rm{\ion{He}{ii})}$\big>\,(\angstrom)$\tnote{o} & $0.428\pm 0.032$ & $0.684\pm 0.033$ \\
\hline
\end{tabular*}
\begin{tablenotes}
\item[a] Mean dust-corrected $\ha$-to-UV luminosity ratio.
\item[b] Mean redshift.
\item[c] Mean stellar mass.
\item[d] Mean effective radius.
\item[e] Mean gas-phase abundances
\item[f] Stellar metallicity  ($Z_{\odot}=0.0142$ from \citealt{Asplund_2009}). 
\item[g] Stellar population age.
\item[h] Stellar continuum reddening. 
\item[i] SED star-formation rate measured from the FUV composite spectrum.
\item[j] $\ha$ star-formation rate measured from the optical composite spectrum.
\item[k] $\ha$ star-formation-rate surface density.
\item[l] Nebular reddening measured from the optical composite spectrum.
\item[m] Equivalent width of \ion{Si}{IV} $\,\lambda\lambda1393,1403$.
\item[n] Equivalent width of \ion{C}{IV} $\,\lambda\lambda1548,1550$.
\item[0] Equivalent width of \ion{He}{II} $\,\lambda1640$.
\end{tablenotes}
\end{threeparttable}
\end{table}

\subsection{Photospheric and stellar wind FUV spectral features vs. $L(\ha)/L(\rm UV)$ }
\label{sec:modelpredic}
Some FUV spectral features are strongly correlated with starburst age, metallicity, and IMF properties, making them excellent proxies for constraining the physical properties of the massive star population \citep{Lamers_1999,Pettini_2000,Leitherer_2001,Mehlert_2002,Smith_2002,Shapley_2003,Keel_2004,Rix_2004,Steidel_2004,Leitherer_2010,Cassata_2013,Grafener_2015,Chisholm_2019,Reddy_2022}. 

In continuous star formation, stars form at a relatively constant rate over time. As a result, the galaxy maintains a steady population of young, massive stars. This leads to a relatively stable presence of FUV P-Cygni features. In contrast, bursty star formation involves periods of intense star formation activity followed by periods of relative quiescence. During a starburst episode, the galaxy produces a large number of massive stars in a short period, which can lead to stronger FUV P-Cygni features as a result of the increased population of massive stars. And, during a post-burst episode, the equivalent widths of the features are expected to weaken \citep{Walborn_1985,Pellerin_2002,Leitherer_2005,Garcia_2017,Calabr_2021}. The FUV spectral features discussed in this work are the P-Cygni component of \ion{Si}{iv}$\,\lambda\lambda 1393,1402$, \ion{C}{iv}$\,\lambda\lambda 1548,1550$, and the stellar component of \textrm{\rm\ion{He}{ii}} $\lambda 1640$. The presence of \ion{C}{iv} and \ion{Si}{iv} P-Cygni features in a galaxy's spectrum suggests the existence of massive stars with $M_{\ast}\geqslant30M_{\odot}$ and short main-sequence lifetime of $\sim 2-5\,$Myr, and therefore is an indicator of the early stages of star formation \citep{Leitherer_1995,Pettini_2000,Leitherer_2001,Shapley_2003,Quider_2009}. The origin of the broad \textrm{\rm\ion{He}{ii}} $\lambda 1640$ stellar wind emission observed in the spectra of local galaxies is the massive short-lived and extremely hot Wolf-Rayet stars \citep{Schaerer_1996,demello_1998,Crowther_2007,Shirazi_2012,Cassata_2013,Visbal_2015,Crowther_2016,Nanayakkara_2019}. The fraction of WR stars declines with decreasing stellar metallicity. Therefore, another mechanism is needed to explain the observation of \textrm{\rm\ion{He}{ii}} $\lambda 1640$ at high redshift galaxies where the metallicity is lower compared to local galaxies. One possible explanation for such observation is the abundance of binary systems at high redshifts that can result in an increase in the fraction of WR stars in low metallicity environments \citep{Shapley_2003,Cantiello_2007,de_Mink_2013}. In fact, according to previous studies, when single evolution stellar population synthesis models are compared to the models including binary evolution in low stellar metallicity, the \ion{He}{ii} stellar feature is best reproduced by the latter \citep{Shirazi_2012,Steidel_2016,Gutkin_2016,Stanway_2016,Senchyna_2017,Eldridge_2017,Smith_2018,Chisholm_2019,Saxena_2020,Reddy_2022}. Therefore, fitting the observed rest-FUV composite spectra with the SPS models that include binary stellar evolution is necessary in order to study the variations in the strength of stellar \textrm{\rm\ion{He}{ii}} $\lambda 1640$ emission. 
\begin{figure*}
\centering
   \subfloat{%
      \includegraphics[width=\textwidth]{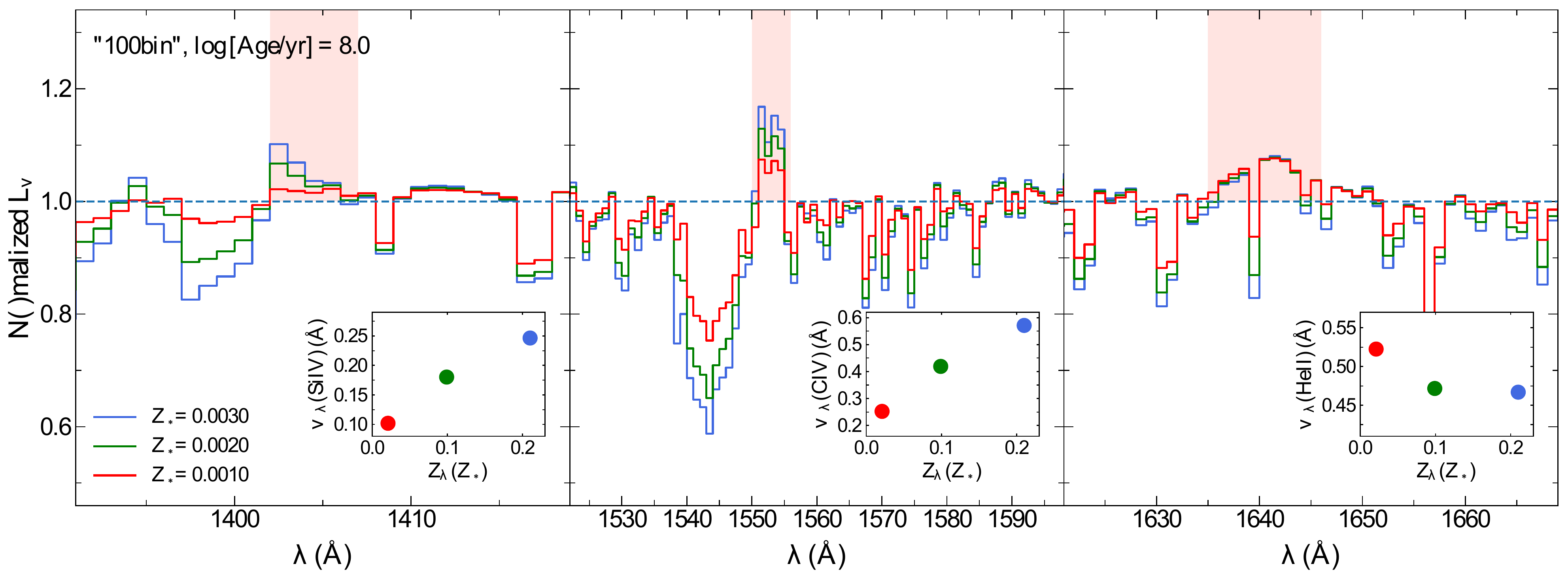}}
      
   \subfloat{%
      \includegraphics[width=\textwidth]{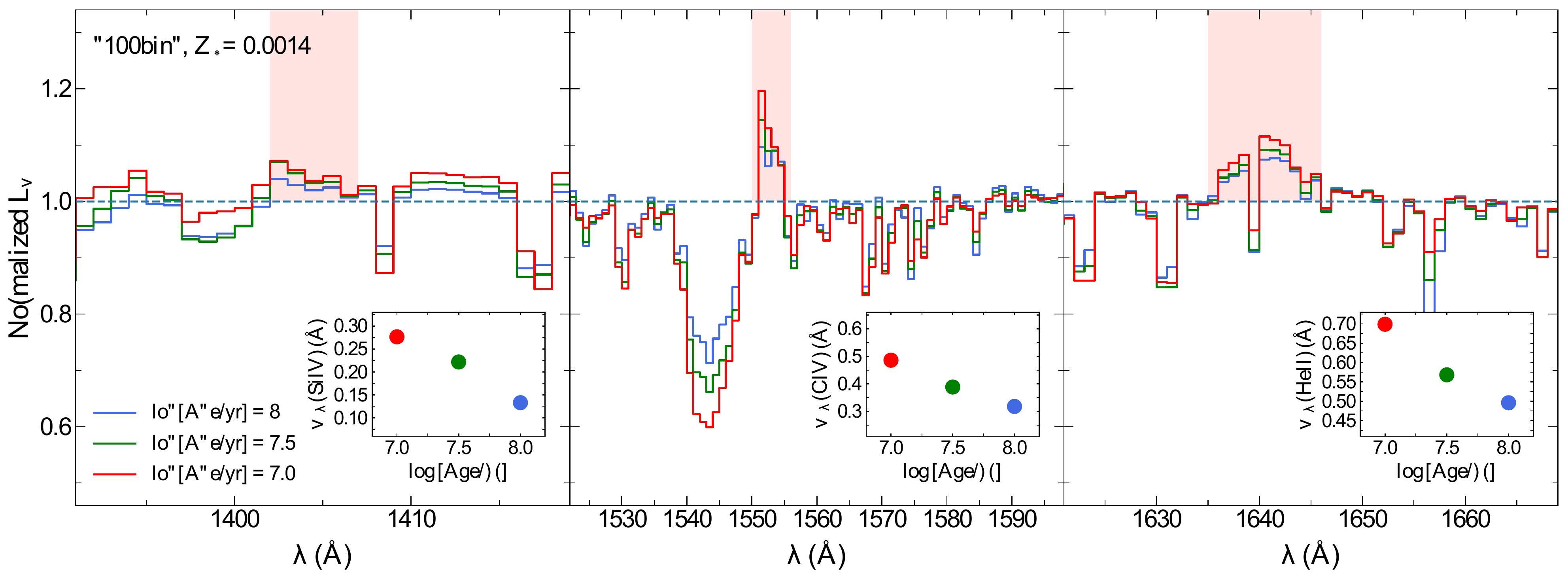}}
      
    \subfloat{
      \includegraphics[width=\textwidth]{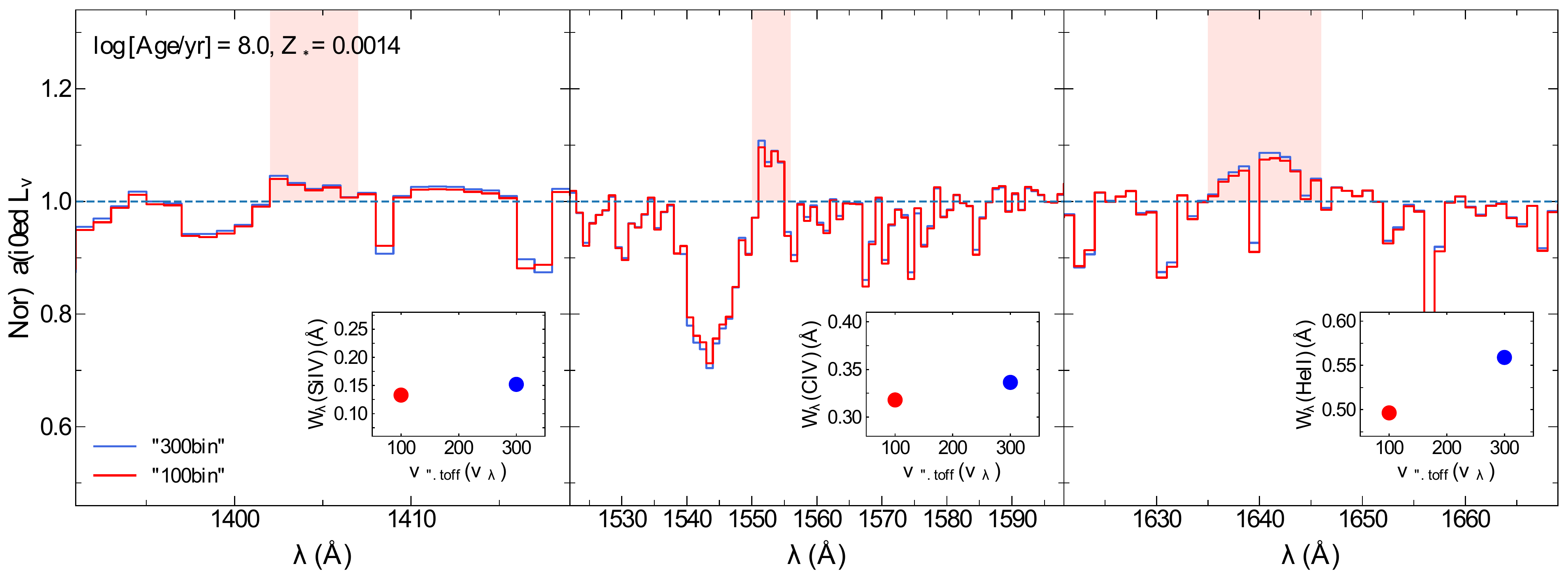}}\\
\caption{\label{fig:spsdependency} Variation of the continuum-normalized \texttt{SPS+Neb} models with stellar metallicity (\textit{top}), stellar age (\textit{middle}), and upper-mass cutoff of the IMF (\textit{bottom}). In each panel only the specified parameter in the lower left is relaxed to change, while the parameters indicated in the upper left are held fixed. In all panels the ionization parameter and nebular metallicity are held fixed to the average values of the MOSDEF-LRIS sample ($\log U=-3.0,\ \log[Z_{\rm neb}/Z_{\odot}]=-0.4$; from \citealt{Reddy_2022}). Each of the three columns presented in this figure corresponds to a particular feature: \ion{Si}{iv} P-Cygni, \ion{C}{iv} P-Cygni, and \ion{He}{ii} stellar winds. The inset panels indicate the equivalent width of each line in each model. The shaded pink indicates the regions by which the width measurements are performed for each feature.}
\end{figure*}
\subsubsection{\texttt{SPS+Neb} Model Predictions of FUV Spectral Features}
\label{sec:varfuv}
Based on the \texttt{SPS+Neb} models, we show an example of the sensitivity of \ion{Si}{iv}, \ion{C}{iv}, and \ion{He}{ii} stellar features to the stellar age, metallicity, and $M_{\rm cutoff}$ of the IMF in Figure~\ref{fig:spsdependency}. The equivalent widths ($W_{\lambda}$) of these features are also shown in the inset panels. These equivalent widths are measured by directly integrating across each line (above the line of unity) in the continuum-normalized models. In each panel, we only adjust one physical parameter at a time and keep the other two unchanged. The fixed values are chosen based on the average parameters derived from the composite spectra of all galaxies in the MOSDEF/LRIS sample.

The top panel of Figure~\ref{fig:spsdependency} compares three constant SFH models with fixed stellar population age of $\log[\rm Age/yr]=8.0$, fixed upper-mass cutoff of $M_{\rm cutoff}=100\,M_{\odot}$, and varying metallicities of $Z_{\ast}=\{0.0010,0.0020,0.0030\}$. As depicted by the inset panels, as the metallicity increases from $Z_{\ast}=0.0010$ to $Z_{\ast}=0.0030$, the equivalent widths of \ion{C}{iv} and \ion{Si}{iv} P-Cygni emission become $\sim2.3$ and $\sim2.5$ times larger, respectively. This is due to the fact that these P-Cygni features are sensitive to mass-loss rate, which increases as metallicity increases. In the case of \ion{He}{ii}, the model with lowest metallicity ($Z_{\ast}=0.0010$) exhibits the largest equivalent width compare to the higher metallicity models. This is due to the fact that stars with lower metallicity at given ages have harder ionizing spectra. 

The middle panel of Figure~\ref{fig:spsdependency} shows three models with fixed metallicity of $Z_{\ast}=0.0014$, fixed mass cutoff of $M_{\rm cutoff}=100\,M_{\odot}$, and varying stellar ages of $\log[\rm Age/yr]=\{7.0,7.5,8.0\}$. The inset panels demonstrate that the younger stellar population model ($\log[\rm Age/yr]=7.0$) show a larger equivalent width of \ion{Si}{iv}, \ion{C}{iv}, and \ion{He}{ii} by a factor of $\sim2.1$, $\sim1.6$, and $\sim1.5$, respectively, when compared to the model with a higher age ($\log[\rm Age/yr]=8.0$). This prediction again demonstrates that the photospheric and stellar wind spectral features are strong at the early stages of star-formation. 

The bottom panel of Figure~\ref{fig:spsdependency} depicts two \texttt{SPS+Neb} models with fixed stellar age of $\log[\rm Age/yr]=8.0$ and stellar metallicity of $Z_{\ast}=0.0014$ and varying upper-mass cutoff of $M_{\rm cutoff}/M_{\odot}=\{100,300\}$. The inset panels indicate that changing the mass cutoff of the IMF from $100\,M_{\odot}$ to $300\,M_{\odot}$ causes the equivalent widths of \ion{Si}{iv}, \ion{C}{iv}, and \ion{He}{ii} to grow $\sim1.1$, $\sim1.1$, and $\sim1.2$ times larger. 

\subsubsection{Observed FUV spectral features in bins of $L(\ha)/L(\rm UV)$}
\label{sec:fuvfeature}
As shown in Section~\ref{sec:varfuv}, the model-predicted equivalent widths of \ion{Si}{iv}, \ion{C}{iv}, and \ion{He}{ii} are sensitive to stellar age, metallicity, and less sensitive to the high-mass cutoff of the IMF. In this section, we examine the variations in the observed equivalent widths of those FUV spectral features from the composite spectra of the two $L(\ha)/L(\rm UV)$ subsamples. The advantage of analyzing equivalent widths of the observed features is that they are unaffected by dust or aperture uncertainties. In addition, the observed equivalent widths are insensitive to the model assumptions (e.g., constant vs. instantaneous burst SFH). 

\begin{figure}
\centering
     \subfloat{
        \includegraphics[width=\columnwidth]{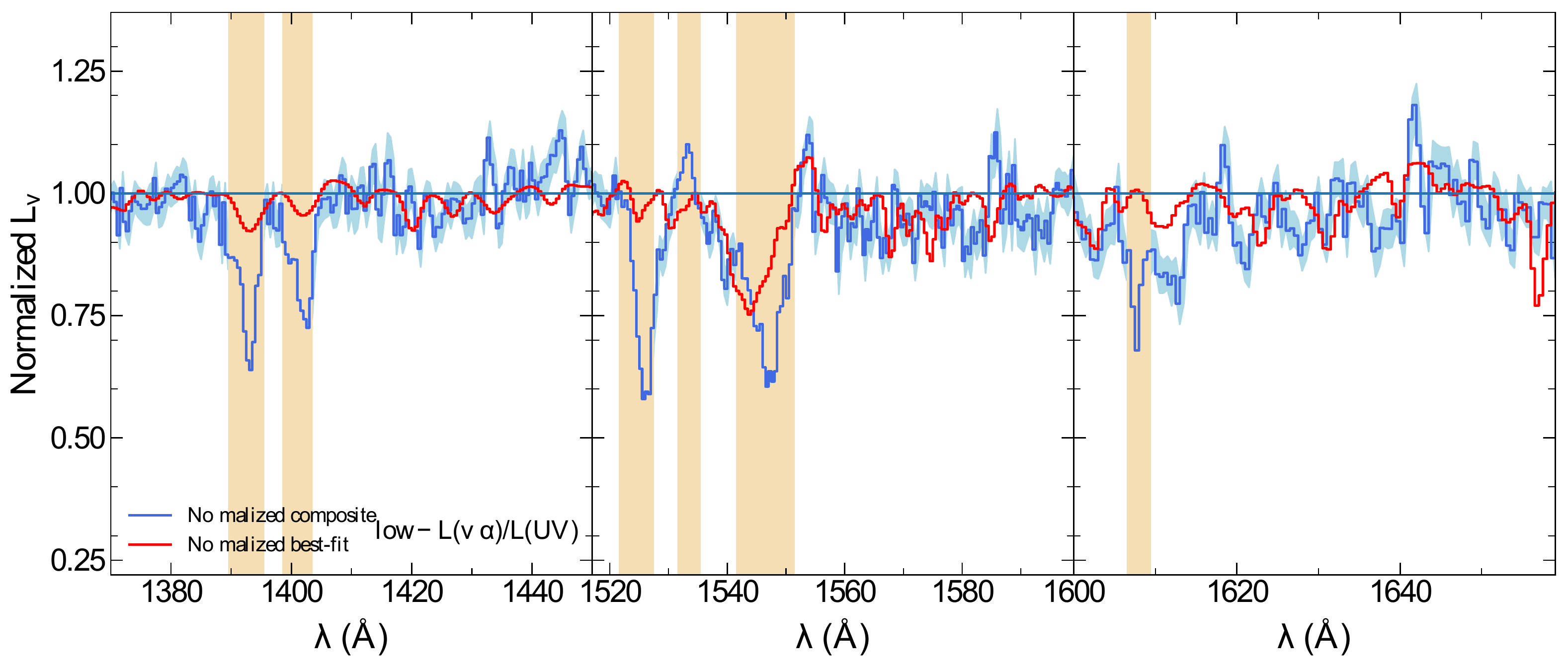}}
        
     \subfloat{
        \includegraphics[width=\columnwidth]{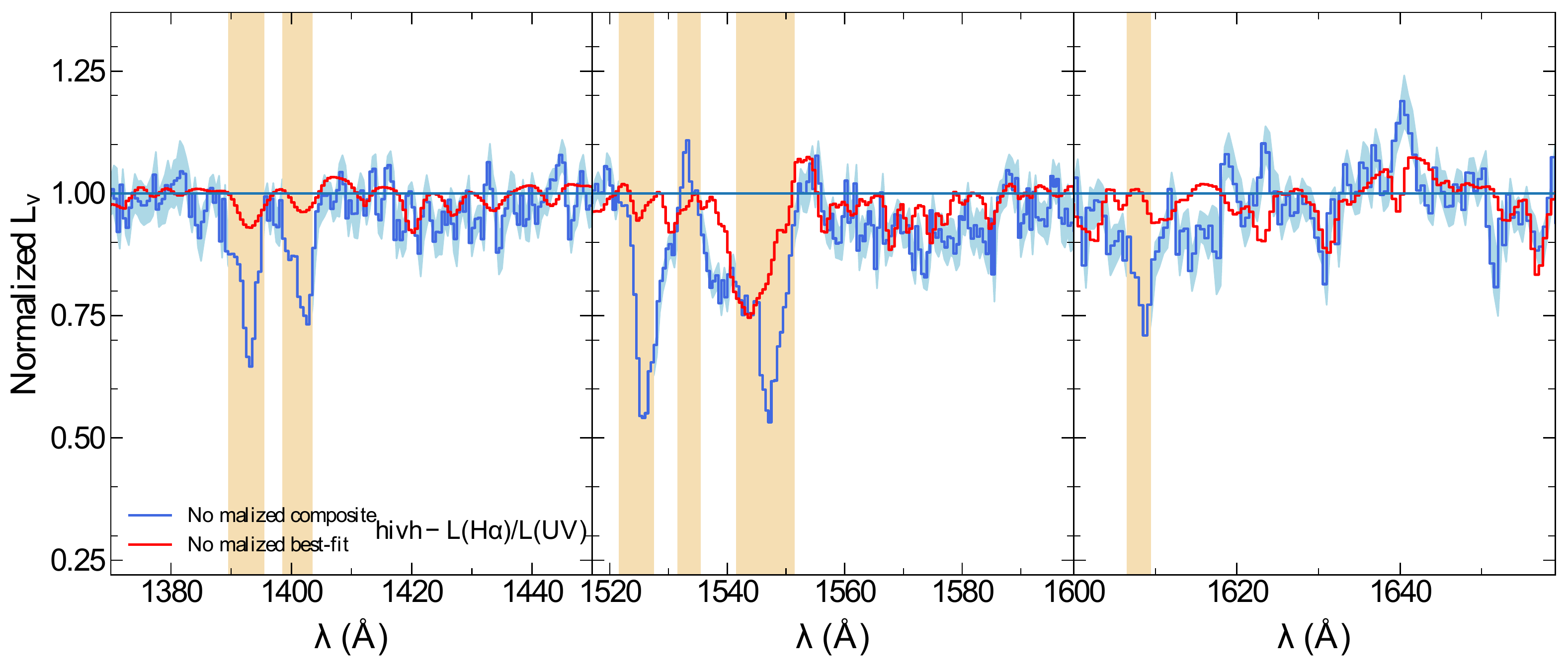}}
\caption{\label{fig:complratio} Continuum-normalized composite spectra ($blue$) of the two $L(\ha)/L(\rm UV)$ subsamples from which the equivalent width measurements are performed. The physical properties of each of the bins, as well as, $W_{\lambda}$(\ion{Si}{iv}), $W_{\lambda}$(\ion{C}{iv}), and $W_{\lambda}$(\ion{He}{ii}) measurements are listed in Table~\ref{tab:eqpcygni}. Those regions that are not included in the fitting process are shaded in orange.}
\end{figure}

The average rest-frame equivalent widths ($\left<W_{\lambda}\right>$) for each of the above-mentioned FUV spectral features are measured by directly integrating across each line in each of the continuum-normalized composite spectra shown in Figure~\ref{fig:complratio} and are reported in Table~\ref{tab:eqpcygni}. To ensure unbiased measurements, we utilize identical wavelength intervals for each bin. These wavelength intervals are derived based on the regions that the lines occupy in the \texttt{SPS+Neb} models. These regions are highlighted in Figure~\ref{fig:spsdependency}. The errors in $W_{\lambda}$ are measured by perturbing the continuum-normalized spectra according to the error in spectra and repeating the measurements many times. The uncertainty is determined by the standard deviation of these perturbations. The final reported uncertainties include the error associated with the normalization process. 

\begin{figure}
\centering
    \includegraphics[width=\columnwidth]{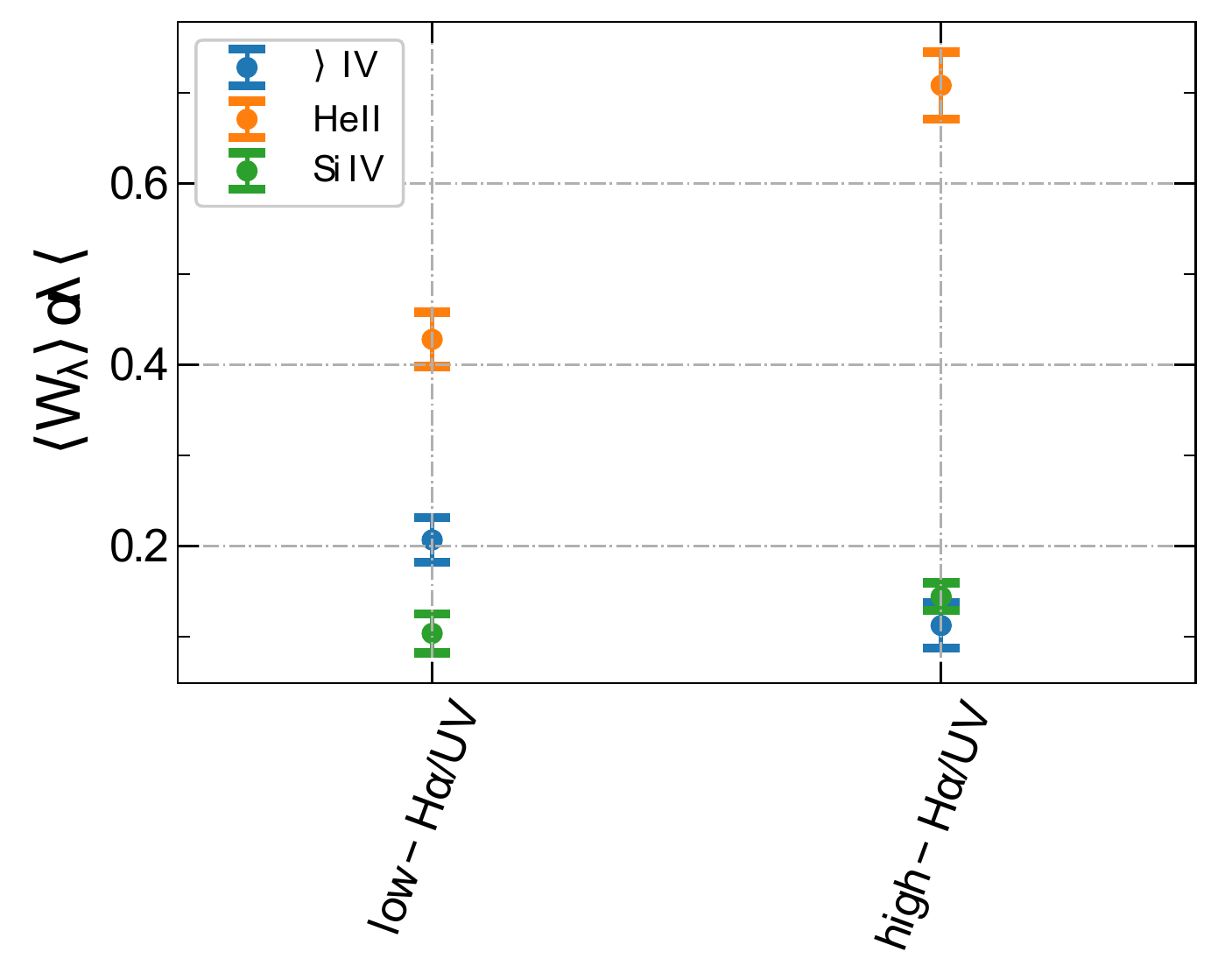}
    \caption{Comparison of the average equivalent widths of  \ion{Si}{iv}$\,\lambda\lambda1393,1402$, \ion{C}{iv}$\,\lambda\lambda1548,1550$, and \textrm{\rm\ion{He}{ii}} $\lambda 1640$ 
    \label{fig:obswidth} stellar emission lines measured from the continuum-normalized spectra of the two bins reported in Table~\ref{tab:eqpcygni}.}
\end{figure}

Figure~\ref{fig:obswidth} shows the comparison between the average rest-frame equivalent widths of \ion{Si}{iv}, \ion{C}{iv}, and \ion{He}{ii} in the $L(\ha)/L(\rm UV)$ subsamples. No significant differences are found 
in $\left<W_{\lambda}(\textrm{\ion{Si}{iv}})\right>$, and $\left<W_{\lambda}(\textrm{\ion{C}{iv}})\right>$ between the \textit{low}- and \textit{high}-$L(\ha)/L(\rm UV)$ bins within the measurement uncertainties. However, $\left<W_{\lambda}(\textrm{\ion{He}{ii}})\right>$ grows by a factor of $\sim1.7$ from the \textit{low}- to \textit{high}-$L(\ha)/L(\rm UV)$ bin. If galaxies with higher $L(\ha)/L(\rm UV)$ are undergoing a burst of star formation, then we would expect them to have higher \ion{C}{iv} and \ion{Si}{iv} P-Cygni emission equivalent widths relative to galaxies with lower $L(\ha)/L(\rm UV)$. 
 
While \ion{Si}{iv} and \ion{C}{iv} P-Cygni emissions are prominently stellar in origin, this is not the case for \ion{He}{ii}. The extremely hot sources that produce stellar \ion{He}{ii} emission also generate enough \ion{$\rm{He^{+}}\,$}{} ionizing photons with wavelengths of $\lambda<228\,\angstrom$ to yield nebular \ion{He}{ii} emission due to recombination, which complicates the interpretation of the \ion{He}{ii} emission. Based on the previous studies (e.g., \citealt{Steidel_2016,Reddy_2022}), we adopt the following procedure to disentangle the stellar and nebular components. 
\begin{figure}
\centering
    
    \includegraphics[width=0.9\columnwidth]{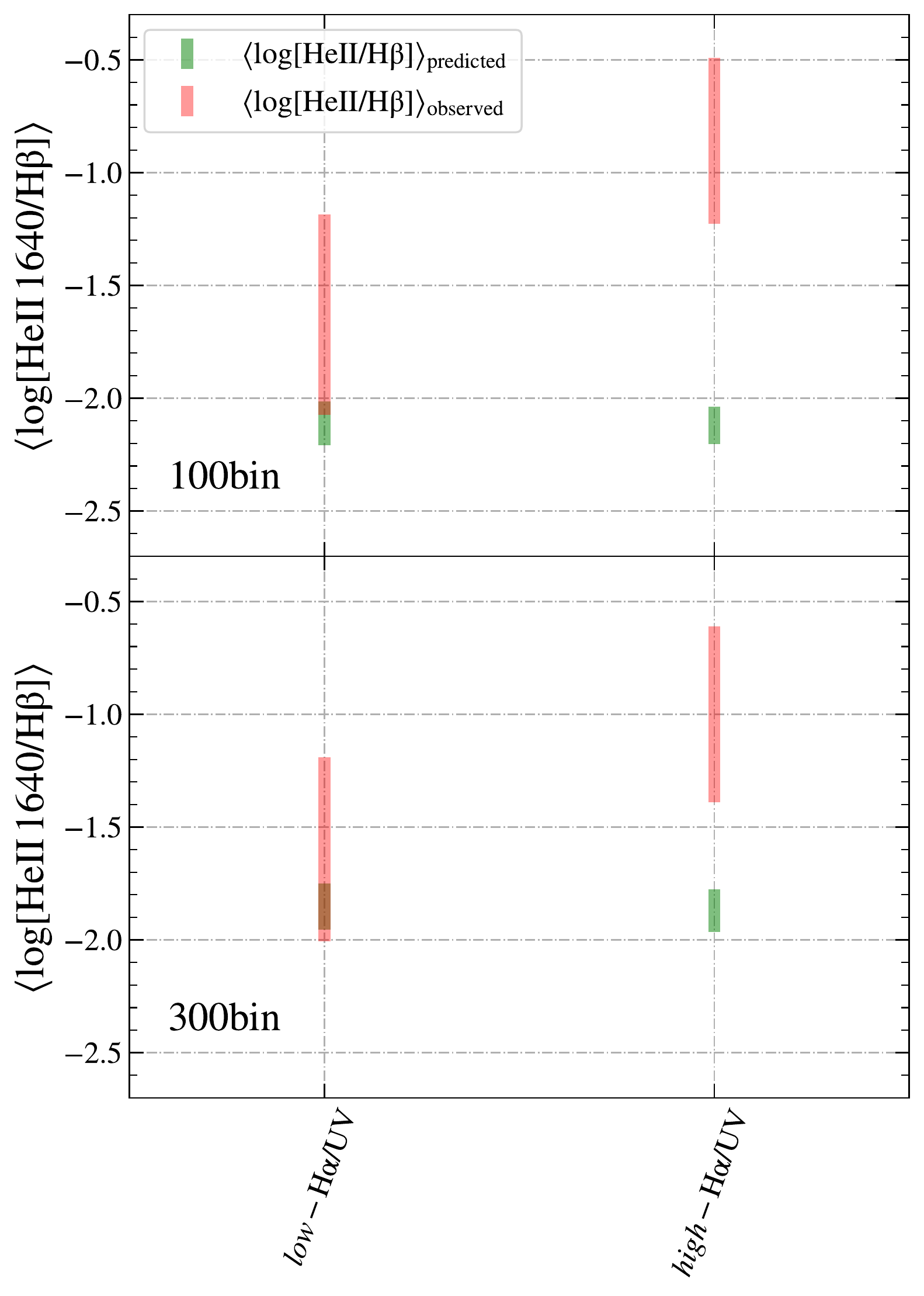}
    \caption{Comparison of the model-predicted nebular \textrm{\rm\ion{He}{ii}} $\lambda 1640$ relative intensity, $\left<\text{\ion{He}{ii}}/\hb\right>$, derived from the \texttt{Cloudy} code and the observed dust-corrected relative intensity measured by subtraction of the best-fit \texttt{SPS+Neb} model from the composite spectrum for each $L(\ha)/L(\rm UV)$ subsample and different model assumptions. The colored bars show the $\pm3 \sigma$ range of the measurement uncertainties. \label{fig:neb}}
\end{figure}
We measure the observed nebular \ion{He}{ii} intensity by subtracting the best-fit \texttt{SPS+Neb} model from the composite spectrum of each bin using the "100bin" and "300bin"~\footnote{When fitting the FUV composite spectra with the "300bin" \texttt{SPS+Neb} models, they can still reproduce all the FUV features discussed in this work. The best-fit stellar population age, metallicity, and continuum reddening of the \textit{high}-$L(\ha)/L(\rm UV)$ subsample obtained with the "300bin" model are $\left<\log[\rm Age/yr]\right>=8.0\pm0.2$, $\left<Z_{\ast}\right>=0.084\pm0.011$, and $\left<E(B-V)_{\rm{cont}}\right>=0.067\pm0.005$.} model assumptions. Because the best-fit model identifies the stellar component, the subtraction of the best-fit model from the observed spectrum is assumed to be purely nebular. The observed nebular \ion{He}{ii} intensity is then dust-corrected assuming $\left<\ebmvgas\right>$ and the \cite{Cardelli} extinction curve, where $\left<\ebmvgas\right>$ is measured directly from the optical composite spectrum. The model-predicted nebular \ion{He}{ii} intensity is derived by using the best-fit \texttt{SPS} model of each bin as an input to the \texttt{Cloudy} photoionization code. The comparison between the model-predicted and observed nebular \ion{He}{ii} emission in terms of relative intensity, $\left<\text{\ion{He}{ii}}/\hb\right>$, is shown in Figure~\ref{fig:neb} for the $L(\ha)/L(\rm UV)$ subsamples. The model-predicted and observed nebular \ion{He}{ii} intensities measured for the \textit{low}-$L(\ha)/L(\rm UV)$  bin agree within the $3\sigma$ uncertainty for both of the mass cutoff assumptions. However, the model prediction of the nebular \ion{He}{ii} intensity does not fully account for the observed nebular \ion{He}{ii} intensity in the \textit{high}-$L(\ha)/L(\rm UV)$ bin even with an increase in the upper-mass cutoff of the IMF.    

Our results indicate that recent SF activity, and low metallicity cannot explain the difference in the \ion{He}{ii} emission of galaxies in the two $L(\ha)/L(\rm UV)$ bins because the stellar age and metallicity derived for the two bins are similar within their respective uncertainties. Next, we investigate whether a top heavy IMF can account for such a difference. First, we separate the nebular and stellar components of the \ion{He}{ii} emission. We then compare the observed nebular \ion{He}{ii} intensity to that predicted by the \texttt{Cloudy} photoionization model using various assumptions on the upper-mass limit of the IMF. We find that even a top heavy IMF model ($M_{\rm cutoff}=300M_{\odot}$) is unable to accurately predict the observed nebular \ion{He}{ii} intensity of the \textit{high}-$L(\ha)/L(\rm UV)$ bin. Another potential contributor that gives rise to the \ion{He}{}$^{+}$ ionizing photons budget in low-metallicity star-forming galaxies is discussed below. 

\cite{Schaerer_2019} suggested that high mass X-ray binaries (HMXBs) are a primary source for producing \ion{He}{}$^{+}$ ionizing photons in low-metallicity star-forming galaxies. They found that only SPS models that include HMXBs are able to reproduce the observed relative intensity of nebular \ion{He}{ii} emission ($\text{\ion{He}{ii}}/\hb$). Studies of both the local and high-redshift universe have suggested that the X-ray luminosity ($L_{\rm X}$) of HMXBs in star-forming galaxies increases with SFR \citep{Nandra_2002,Bauer_2002,Seibert_2002,Grimm_2003,Reddy_2004,Persic_2004,Gilfanov_2004,Persic_2007,Lehmer_2008,Lehmer_2010}, which is expected owing to the young ages of HXMBs ($\sim 10\,$Myr). Several studies have indicated that $L_{\rm X}$ per unit SFR in star-forming galaxies elevates at high redshift (e.g., \citealt{Basu-Zych_2013a,Lehmer_2016,Aird_2017}). This enhancement in $L_{\rm X}$/SFR with redshift may be due to the lower metallicities of high-redshift galaxies, which results in more luminous (and possibly more numerous) HMXBs \citep{Brorby_2016,Douna_2015}. In fact, observational studies have shown evidence of several ultraluminous X-ray sources in nearby galaxies with low metallicities (e.g., \citealt{Mineo_2012,Prestwich_2013,Basu-Zych_2013b}). Following the idea that $L_{\rm X}$/SFR is metallicity-dependent, \cite{Brorby_2016} parameterized the $L_{\rm X}$-SFR-$Z$ relationship, where $Z$ is the gas-phase metallicity, for a sample of local star-forming galaxies as:

\begin{eqnarray}
    \log\left(\frac{L_{\rm X}/\rm{SFR}}{\rm{erg\, s^{-1}}/(M_{\odot}\,\rm{yr^{-1}})}\right)&=&-0.59\times(12+\log(\rm O/H)-8.69)\nonumber \\ & & +39.49.
\label{eq5}
\end{eqnarray}
\cite{Fornasini_2019} studied a sample of MOSDEF galaxies with available X-ray data to investigate the $L_{\rm X}$/SFR and $Z$ relationship at redshift $z\sim2$. They found that for both $\ha-$ and SED-based SFRs, the $L_{\rm X}$-SFR-$Z$ relation is in good agreement with that of the \cite{Brorby_2016} relation for local galaxies (Equation~\ref{eq5}). The results obtained were not affected by the assumed $L(\ha)$ to $\sfrha$ conversion factor. We incorporate $\sfrha$ and $\sfrsed$ into Equation~\ref{eq5} to calculate the average X-ray luminosities for the $L(\ha)/L(\rm UV)$ subsets. We find that the average X-ray luminosity of the \textit{high}-$L(\ha)/L(\rm UV)$ subset is $3\times$ greater with $\sfrha$, and $1.4\times$ greater with $\sfrsed$ when compared to the average X-ray luminosity of the \textit{low}-$L(\ha)/L(\rm UV)$ subset. The statistical differences between the average X-ray luminosities derived using $\sfrha$ and $\sfrsed$ are at $5\sigma$ and $2\sigma$ levels, respectively. The increase in $\left<L_{\rm X}\right>$ with increasing $L(\ha)/L(\rm UV)$ may indicate the presence of luminous HMXBs, which in turn could explain the high nebular \ion{He}{ii} emission observed for galaxies with high $L(\ha)/L(\rm UV)$ ratios. Considering that an increase in $\left<L_{\rm X}\right>$ is also observed when using $\sfrsed$, the conclusion about the existence of HMXBs may be reliably drawn. However, this conclusion is only robust at a $2\sigma$ statistical significance level owing to the large uncertainties of the average SED-based SFRs.

%\end{figure}
\section{Summary and Conclusion}
\label{sec:Sum}
We examine the effectiveness of the dust-corrected globally measured $\ha$-to-UV luminosity ratio in tracing burstiness for typical star-forming galaxies at $z\sim2$. We use the MOSDEF survey to explore stellar population properties differences in bins of $L(\ha)/L(\rm{UV})$. 

In the first part of this analysis, we employ the HST imaging of 310 star-forming galaxies (MOSDEF/MORPH sample) drawn from the MOSDEF survey to construct the star-formation-rate surface density and stellar age maps. We use a Voronoi binning technique to group the pixels based on their $S/N$. We then study the distribution of $\Sigma_{\sfrsed}$ and stellar age of Voronoi bins within each galaxy using a morphological metric called patchiness ($P$). Patchiness is sensitive to deviations from average, therefore galaxies that are undergoing a burst of star-formation contain regions with higher $\Sigma_{\sfrsed}$ and younger stellar age than the mean value for the entire galaxy and are expected to exhibit higher $P(\Sigma_{\sfrsed})$ and/or $P(\rm Age)$. We find no correlation between $L(\ha)/L(\rm UV)$ and $P(\Sigma_{\sfrsed})$, as well as between $L(\ha)/L(\rm UV)$ and $P(\rm Age)$. We suggest that the globally-averaged $L(\ha)/L(\rm UV)$ does not trace stochastic SFH over a time-scale of $\sim 10\,$Myr, which is the typical dynamical timescale probed by the Voronoi bins. We suggest that this lack of correlation may be because of the uncertainties related to the variations in the stellar dust attenuation curve, limited dynamical time scale and spatial resolution probed by the Voronoi bins.

In the second part of this analysis, we use a rest-FUV spectroscopic sample obtained by LRIS (MOSDEF/LRIS sample) to study the average physical properties of $z\sim2$ star-forming galaxies in bins of $L(\ha)/L(\rm UV)$. We use the \texttt{BPASS} constant SFH models combined with the nebular continuum emission generated by the \texttt{Cloudy} radiative transfer code (\texttt{SPS+Neb} models) as our theoretical basis to address the effect of different physical assumptions on the model-predicted $L(\ha)/L(\rm UV)$. As suggested by other studies, the $\ha$-to-UV ratio predicted by \texttt{SPS+Neb} models increases for younger stellar populations, or when the upper end of the IMF increases. The inclusion of binary stellar evolution or lowering the stellar metallicity of the models also cause a rise in the predicted ratio. We divide the 124 galaxies in the MOSDEF/LRIS sample into two bins of $L(\ha)/L(\rm UV)$ with an equal number of galaxies in each to investigate whether the variation observed in the dust-corrected $\ha$-to-UV ratio is related to differences in stellar age, metallicity, and/or upper-end mass of the IMF as suggested by the SPS theoretical models. The main conclusions of the second part of the paper are as follows:

\begin{itemize}

    \item The average stellar population age estimated for the \textit{high-}$L(\ha)/L(\rm UV)$ bin is $\log[\rm{Age/yr}]=8.0\pm0.2$, compared to $\log[\rm{Age/yr}]=8.4\pm0.1$ for the \textit{low-}$L(\ha)/L(\rm UV)$ bin. We find no significant variation in the stellar metallicity between the \textit{low}- and \textit{high}-$L(\ha)/L(\rm UV)$ bin within the measurement uncertainties. The stellar population age of $100\,$Myr derived for the \textit{high-}$L(\ha)/L(\rm UV)$ bin is longer than the dynamical timescale of a few tens of Myr, implying that the \textit{high-}$L(\ha)/L(\rm UV)$ galaxies are not necessarily undergoing a burst of star formation.

    \item Galaxies with higher $L(\ha)/L(\rm UV)$ have on average higher $\ha$ luminosities. Such galaxies also have strong \textrm{\rm\ion{He}{ii}} $\lambda 1640$ emission, but do not exhibit significantly different EWs of \ion{Si}{iv}$\,\lambda\lambda 1393,1402$, \ion{C}{iv}$\,\lambda\lambda 1548,1550$ P-Cygni emissions. These results suggest that the higher Ha/UV ratio of these galaxies is due to their harder ionizing spectra, rather than a higher $\sfrha$ relative to SFR[UV]. The lack of variations in the strength of the P-Cygni emissions between the two $L(\ha)/L(\rm UV)$ subsamples is expected given the insignificant differences found between the stellar age and metallicity of the two subsamples.
    
    \item The difference between the strength of the observed \textrm{\rm\ion{He}{ii}} emission of the \textit{low}- and \textit{high}-$L(\ha)/L(\rm UV)$ subsamples can be further investigated when the nebular and stellar components of the \textrm{\rm\ion{He}{ii}} line are disentangled. We find that the model-predicted nebular \textrm{\rm\ion{He}{ii}} intensity cannot accurately predict the observed amount for the \textit{high}-$L(\ha)/L(\rm UV)$ bin even if the upper-mass limit of the IMF is increased from $M_{\rm cutoff}=100\,M_{\odot}$ to $300\,M_{\odot}$.  According to recent studies, low metallicity star-forming galaxies get the majority of their \ion{He}{}$^{+}$ ionizing photons from high mass X-ray binaries \citep{Schaerer_2019}. We measure the X-ray luminosity of each bin using the $L_{\rm X}$-SFR-$Z$ relation found by \cite{Brorby_2016} for local galaxies. \cite{Fornasini_2019} found that the $L_{\rm X}$-SFR-$Z$ relation for $z\sim2$ galaxies is in good agreement with the \cite{Brorby_2016} local relation for both $\ha$ and SED-based SFRs. We find that the X-ray luminosity is on average larger for galaxies with higher $L(\ha)/L(\rm UV)$ regardless of the assumed SFR indicator. This result may suggest the presence of luminous HMXBs, which could explain the high nebular \ion{He}{ii} emission observed for galaxies with high $\ha$-to-UV ratios. As HMXBs have short lifespans (a few Myr), they effectively indicate recent star formation. Nonetheless, the potential abundance of HMXBs in the \textit{high}-$L(\ha)/L(\rm UV)$ group does not directly imply that galaxies in this subset primarily experience a bursty SFH, since massive stars and HMXBs are being continuously formed even in a constant SFH. In addition, the idea that the \textit{high}-$L(\ha)/L(\rm UV)$ sample is in a burst mode of star formation is not supported by the fact that the equivalent width of the CIV and SiIV P-Cygni features do not vary between the two subsamples. The presence of a more intense ionizing spectrum in galaxies with higher $L(\ha)/L(\rm UV)$ ratios can be backed by the abundance of HMXBs in these galaxies.

\end{itemize}    

Our results cast doubt upon the reliability of $\ha$-to-UV luminosity ratio in tracing bursty SFH of typical star-forming galaxies at $z\sim2$. This is due to the absence of evidence suggesting that galaxies with higher $L(\ha)/L(\rm UV)$ are experiencing a burst, based on their average stellar population age and the lack of variation in P-Cygni features compared to galaxies with lower $L(\ha)/L(\rm UV)$. There is one important implication of this work. It is proposed that star-forming galaxies may be in a bursty phase of star formation at the beginning of the reionization epoch, producing enough ionizing photons to reionize the intergalactic medium. If such is the case, it is important to validate the tracer of burstiness which is commonly used in the literature, and whether there are other phenomena that can affect the $\ha$-to-UV ratio. Using the next-generation telescopes, we will have access to even more high quality data to study the $\ha$-to-UV ratio variations in more detail. For example, James Webb Space Telescope can provide high-resolution rest-FUV spectra of high redshift galaxies to aid in constraining the mode of star formation history and/or hardness of the ionizing spectrum and the causes of the variations in the $\ha$-to-UV ratio. 

Several studies of high redshift galaxies have found evidence of bursty SFHs by comparing $\ha$ and UV SFRs (e.g., \citealt{Atek2022} at $z\sim1.1$, and \citealt{Faisst_2019} at $z\sim4.5$). These works suggest that the excess found in the $\ha$ SFR relative to the UV SFR can only be explained by additional bursts of star formation on top of an underlying smooth star formation. The aforementioned studies have found that $\sfrha$/SFR[UV] ratio is preferentially higher for lower mass galaxies. Galaxies of lower masses, which are also likely to have a lower metallicity, may be conducive to an IMF that is top-heavy \citep{Tremonti_2004,Dalcanton_2007,Lara_2010,Peeples_2011,Lilly_2013,Andrews_2013,Zahid_2014,Chisholm_2018}, and therefore one must be very careful in interpreting the $\ha$-to-UV ratio for such galaxies. The equivalent width of the stellar photosphere features investigated in this work (Section~\ref{sec:modelpredic}) is less affected by the uncertainties associated with the $L(\ha)/L(\rm UV)$ ratio, such as variations in ionizing escape fraction, stochastic IMF, and dust reddening. Hence, the equivalent width of FUV stellar features (e.g., \ion{c}{iv}, and \ion{Si}{iv}) may
more reliably trace recent bursts of star formation than the $\ha$-to-UV ratio. Nonetheless, it is suggested that future studies use higher $S/N$ rest-frame FUV spectra and increased sample sizes to more effectively compare the effectiveness of P-Cygni features in tracing star formation burstiness against the $\ha$-to-UV ratio.

\section*{Acknowledgements}
This work is based on observations taken by the 3D-HST Treasury Program (GO $12177$ and $12328$) with the NASA/ESA HST, which is operated by the Association of Universities for Research in Astronomy, Inc., under NASA contract NAS5-26555. In addition, we would like to acknowledge that this work made use of v2.2.1 of the Binary Population and Spectral Synthesis (\texttt{BPASS}) models as described in \cite{Eldridge_2017}, and \cite{Stanway_2018}. The MOSDEF team acknowledges support from an NSF AAG collaborative grant (AST-1312780, 1312547, 1312764, and 1313171), grant AR-13907 from the Space Telescope Science Institute, grant NNX16AF54G from the NASA ADAP program, and Chandra archival award AR6-17011X. This work would not have been possible without the generous contributions from the 3D-HST collaboration. The MOSDEF data were obtained at the W.M. Keck Observatory, which is operated as a scientific partnership among the California Institute of Technology, the University of California and the National Aeronautics and Space Administration. The Observatory was made possible by the generous financial support of the W.M. Keck Foundation. We recognize and acknowledge the very significant cultural role and reverence that the summit of Mauna Kea has always had within the indigenous Hawaiian community. We are most fortunate to have the opportunity to conduct observations from this mountain.

%%%%%%%%%%%%%%%%%%%%%%%%%%%%%%%%%%%%%%%%%%%%%%%%%%
\section*{Data Availability}

In this work, we use spectroscopic redshifts and rest-frame optical line measurements obtained from the MOSFIRE Deep Evolution Field (MOSDEF) survey \citep{Kriek_2015}. This is publicly available at \url{https://mosdef.astro.berkeley.edu/}.

We also use photometry obtained from the CANDELS \citep{Grogin_2011,Koekemoer_2011} and reprocessed by the 3D-HST grism survey team \citep{Brammer_2012,Skelton_2014,Momcheva2016}. The data sets and catalogs can be found at \url{https://archive.stsci.edu/prepds/3d-hst/}.

We analyze the Far-UV spectra obtained by the Low Resolution Imagerand Spectrometer (LRIS; \citealt{Oke,Steidel_2004}). \cite{Topping} and \cite{Reddy_2022} contain information about the MOSDEF/LRIS data reduction. MOSDEF/LRIS data sets are available upon request. 

\bibliographystyle{mnras}
\bibliography{ms}

\begin{thebibliography}{}
\makeatletter
\relax
\def\mn@urlcharsother{\let\do\@makeother \do\$\do\&\do\#\do\^\do\_\do\%\do\~}
\def\mn@doi{\begingroup\mn@urlcharsother \@ifnextchar [ {\mn@doi@} {\mn@doi@[]}}
\def\mn@doi@[#1]#2{\def\@tempa{#1}\ifx\@tempa\@empty \href {http://dx.doi.org/#2} {doi:#2}\else \href {http://dx.doi.org/#2} {#1}\fi \endgroup}
\def\mn@eprint#1#2{\mn@eprint@#1:#2::\@nil}
\def\mn@eprint@arXiv#1{\href {http://arxiv.org/abs/#1} {{\tt arXiv:#1}}}
\def\mn@eprint@dblp#1{\href {http://dblp.uni-trier.de/rec/bibtex/#1.xml} {dblp:#1}}
\def\mn@eprint@#1:#2:#3:#4\@nil{\def\@tempa {#1}\def\@tempb {#2}\def\@tempc {#3}\ifx \@tempc \@empty \let \@tempc \@tempb \let \@tempb \@tempa \fi \ifx \@tempb \@empty \def\@tempb {arXiv}\fi \@ifundefined {mn@eprint@\@tempb}{\@tempb:\@tempc}{\expandafter \expandafter \csname mn@eprint@\@tempb\endcsname \expandafter{\@tempc}}}

\bibitem[\protect\citeauthoryear{{Aird}, {Coil}  \& {Georgakakis}}{{Aird} et~al.}{2017}]{Aird_2017}
{Aird} J.,  {Coil} A.~L.,   {Georgakakis} A.,  2017, \mn@doi [\mnras] {10.1093/mnras/stw2932}, \href {https://ui.adsabs.harvard.edu/abs/2017MNRAS.465.3390A} {465, 3390}

\bibitem[\protect\citeauthoryear{{Andrews} \& {Martini}}{{Andrews} \& {Martini}}{2013}]{Andrews_2013}
{Andrews} B.~H.,  {Martini} P.,  2013, \mn@doi [\apj] {10.1088/0004-637X/765/2/140}, \href {https://ui.adsabs.harvard.edu/abs/2013ApJ...765..140A} {765, 140}

\bibitem[\protect\citeauthoryear{Asplund, Grevesse, Sauval  \& Scott}{Asplund et~al.}{2009}]{Asplund_2009}
Asplund M.,  Grevesse N.,  Sauval A.~J.,   Scott P.,  2009, \mn@doi [Annual Review of Astronomy and Astrophysics] {10.1146/annurev.astro.46.060407.145222}, 47, 481

\bibitem[\protect\citeauthoryear{{Asquith} et~al.,}{{Asquith} et~al.}{2018}]{Asquith_2018}
{Asquith} R.,  et~al., 2018, \mn@doi [\mnras] {10.1093/mnras/sty1870}, \href {https://ui.adsabs.harvard.edu/abs/2018MNRAS.480.1197A} {480, 1197}

\bibitem[\protect\citeauthoryear{{Atek}, {Furtak}, {Oesch}, {van Dokkum}, {Reddy}, {Contini}, {Illingworth}  \& {Wilkins}}{{Atek} et~al.}{2022}]{Atek2022}
{Atek} H.,  {Furtak} L.~J.,  {Oesch} P.,  {van Dokkum} P.,  {Reddy} N.,  {Contini} T.,  {Illingworth} G.,   {Wilkins} S.,  2022, \mn@doi [\mnras] {10.1093/mnras/stac360}, \href {https://ui.adsabs.harvard.edu/abs/2022MNRAS.511.4464A} {511, 4464}

\bibitem[\protect\citeauthoryear{{Azadi} et~al.,}{{Azadi} et~al.}{2017}]{Azadi-2017}
{Azadi} M.,  et~al., 2017, \mn@doi [\apj] {10.3847/1538-4357/835/1/27}, \href {https://ui.adsabs.harvard.edu/abs/2017ApJ...835...27A} {835, 27}

\bibitem[\protect\citeauthoryear{{Azadi} et~al.,}{{Azadi} et~al.}{2018}]{Azadi_2018}
{Azadi} M.,  et~al., 2018, \mn@doi [\apj] {10.3847/1538-4357/aad3c8}, \href {https://ui.adsabs.harvard.edu/abs/2018ApJ...866...63A} {866, 63}

\bibitem[\protect\citeauthoryear{{Basu-Zych} et~al.,}{{Basu-Zych} et~al.}{2013a}]{Basu-Zych_2013a}
{Basu-Zych} A.~R.,  et~al., 2013a, \mn@doi [\apj] {10.1088/0004-637X/762/1/45}, \href {https://ui.adsabs.harvard.edu/abs/2013ApJ...762...45B} {762, 45}

\bibitem[\protect\citeauthoryear{{Basu-Zych} et~al.,}{{Basu-Zych} et~al.}{2013b}]{Basu-Zych_2013b}
{Basu-Zych} A.~R.,  et~al., 2013b, \mn@doi [\apj] {10.1088/0004-637X/774/2/152}, \href {https://ui.adsabs.harvard.edu/abs/2013ApJ...774..152B} {774, 152}

\bibitem[\protect\citeauthoryear{{Bauer}, {Alexander}, {Brandt}, {Hornschemeier}, {Vignali}, {Garmire}  \& {Schneider}}{{Bauer} et~al.}{2002}]{Bauer_2002}
{Bauer} F.~E.,  {Alexander} D.~M.,  {Brandt} W.~N.,  {Hornschemeier} A.~E.,  {Vignali} C.,  {Garmire} G.~P.,   {Schneider} D.~P.,  2002, \mn@doi [\aj] {10.1086/343778}, \href {https://ui.adsabs.harvard.edu/abs/2002AJ....124.2351B} {124, 2351}

\bibitem[\protect\citeauthoryear{{Bertin, E.} \& {Arnouts, S.}}{{Bertin, E.} \& {Arnouts, S.}}{1996}]{Bertin}
{Bertin, E.} {Arnouts, S.} 1996, \mn@doi [Astron. Astrophys. Suppl. Ser.] {10.1051/aas:1996164}, 117, 393

\bibitem[\protect\citeauthoryear{{Bicker} \& {Fritze-v. Alvensleben}}{{Bicker} \& {Fritze-v. Alvensleben}}{2005}]{Bicker_2005}
{Bicker} J.,  {Fritze-v. Alvensleben} U.,  2005, \mn@doi [\aap] {10.1051/0004-6361:200500194}, \href {https://ui.adsabs.harvard.edu/abs/2005A&A...443L..19B} {443, L19}

\bibitem[\protect\citeauthoryear{{Boselli}, {Boissier}, {Cortese}, {Buat}, {Hughes}  \& {Gavazzi}}{{Boselli} et~al.}{2009}]{Boselli_2009}
{Boselli} A.,  {Boissier} S.,  {Cortese} L.,  {Buat} V.,  {Hughes} T.~M.,   {Gavazzi} G.,  2009, \mn@doi [\apj] {10.1088/0004-637X/706/2/1527}, \href {https://ui.adsabs.harvard.edu/abs/2009ApJ...706.1527B} {706, 1527}

\bibitem[\protect\citeauthoryear{{Brammer} et~al.,}{{Brammer} et~al.}{2012}]{Brammer_2012}
{Brammer} G.~B.,  et~al., 2012, \mn@doi [\apjs] {10.1088/0067-0049/200/2/13}, \href {https://ui.adsabs.harvard.edu/abs/2012ApJS..200...13B} {200, 13}

\bibitem[\protect\citeauthoryear{{Brinchmann}, {Charlot}, {White}, {Tremonti}, {Kauffmann}, {Heckman}  \& {Brinkmann}}{{Brinchmann} et~al.}{2004}]{Brinchmann_2004}
{Brinchmann} J.,  {Charlot} S.,  {White} S.~D.~M.,  {Tremonti} C.,  {Kauffmann} G.,  {Heckman} T.,   {Brinkmann} J.,  2004, \mn@doi [\mnras] {10.1111/j.1365-2966.2004.07881.x}, \href {https://ui.adsabs.harvard.edu/abs/2004MNRAS.351.1151B} {351, 1151}

\bibitem[\protect\citeauthoryear{{Brorby}, {Kaaret}, {Prestwich}  \& {Mirabel}}{{Brorby} et~al.}{2016}]{Brorby_2016}
{Brorby} M.,  {Kaaret} P.,  {Prestwich} A.,   {Mirabel} I.~F.,  2016, \mn@doi [\mnras] {10.1093/mnras/stw284}, \href {https://ui.adsabs.harvard.edu/abs/2016MNRAS.457.4081B} {457, 4081}

\bibitem[\protect\citeauthoryear{Broussard et~al.,}{Broussard et~al.}{2019}]{Broussard_2019}
Broussard A.,  et~al., 2019, \mn@doi [The Astrophysical Journal] {10.3847/1538-4357/ab04ad}, 873, 74

\bibitem[\protect\citeauthoryear{Broussard, Gawiser  \& Iyer}{Broussard et~al.}{2022}]{Broussard_2022}
Broussard A.,  Gawiser E.,   Iyer K.,  2022, \mn@doi [The Astrophysical Journal] {10.3847/1538-4357/ac94c2}, 939, 35

\bibitem[\protect\citeauthoryear{{Calabr{\`o}} et~al.,}{{Calabr{\`o}} et~al.}{2021}]{Calabr_2021}
{Calabr{\`o}} A.,  et~al., 2021, \mn@doi [\aap] {10.1051/0004-6361/202039244}, \href {https://ui.adsabs.harvard.edu/abs/2021A&A...646A..39C} {646, A39}

\bibitem[\protect\citeauthoryear{{Calzetti}, {Armus}, {Bohlin}, {Kinney}, {Koornneef}  \& {Storchi-Bergmann}}{{Calzetti} et~al.}{2000}]{Calzetti_2000}
{Calzetti} D.,  {Armus} L.,  {Bohlin} R.~C.,  {Kinney} A.~L.,  {Koornneef} J.,   {Storchi-Bergmann} T.,  2000, \mn@doi [\apj] {10.1086/308692}, \href {https://ui.adsabs.harvard.edu/abs/2000ApJ...533..682C} {533, 682}

\bibitem[\protect\citeauthoryear{{Cantiello, M.}, {Yoon, S.-C.}, {Langer, N.}  \& {Livio, M.}}{{Cantiello, M.} et~al.}{2007}]{Cantiello_2007}
{Cantiello, M.} {Yoon, S.-C.} {Langer, N.}  {Livio, M.} 2007, \mn@doi [A\&A] {10.1051/0004-6361:20077115}, 465, L29

\bibitem[\protect\citeauthoryear{{Caplar} \& {Tacchella}}{{Caplar} \& {Tacchella}}{2019}]{Caplar_2019}
{Caplar} N.,  {Tacchella} S.,  2019, \mn@doi [\mnras] {10.1093/mnras/stz1449}, \href {https://ui.adsabs.harvard.edu/abs/2019MNRAS.487.3845C} {487, 3845}

\bibitem[\protect\citeauthoryear{Cappellari \& Copin}{Cappellari \& Copin}{2003}]{Cappellari}
Cappellari M.,  Copin Y.,  2003, \mn@doi [Monthly Notices of the Royal Astronomical Society] {10.1046/j.1365-8711.2003.06541.x}, 342, 345

\bibitem[\protect\citeauthoryear{{Cardelli}, {Clayton}  \& {Mathis}}{{Cardelli} et~al.}{1989}]{Cardelli}
{Cardelli} J.~A.,  {Clayton} G.~C.,   {Mathis} J.~S.,  1989, \mn@doi [\apj] {10.1086/167900}, \href {https://ui.adsabs.harvard.edu/abs/1989ApJ...345..245C} {345, 245}

\bibitem[\protect\citeauthoryear{{Cassata} et~al.,}{{Cassata} et~al.}{2013}]{Cassata_2013}
{Cassata} P.,  et~al., 2013, \mn@doi [\aap] {10.1051/0004-6361/201220969}, \href {https://ui.adsabs.harvard.edu/abs/2013A&A...556A..68C} {556, A68}

\bibitem[\protect\citeauthoryear{{Chabrier}}{{Chabrier}}{2003}]{Chabrier}
{Chabrier} G.,  2003, \mn@doi [\pasp] {10.1086/376392}, \href {https://ui.adsabs.harvard.edu/abs/2003PASP..115..763C} {115, 763}

\bibitem[\protect\citeauthoryear{{Chisholm}, {Tremonti}  \& {Leitherer}}{{Chisholm} et~al.}{2018}]{Chisholm_2018}
{Chisholm} J.,  {Tremonti} C.,   {Leitherer} C.,  2018, \mn@doi [\mnras] {10.1093/mnras/sty2380}, \href {https://ui.adsabs.harvard.edu/abs/2018MNRAS.481.1690C} {481, 1690}

\bibitem[\protect\citeauthoryear{Chisholm, Rigby, Bayliss, Berg, Dahle, Gladders  \& Sharon}{Chisholm et~al.}{2019}]{Chisholm_2019}
Chisholm J.,  Rigby J.~R.,  Bayliss M.,  Berg D.~A.,  Dahle H.,  Gladders M.,   Sharon K.,  2019, \mn@doi [The Astrophysical Journal] {10.3847/1538-4357/ab3104}, 882, 182

\bibitem[\protect\citeauthoryear{{Choi}, {Conroy}  \& {Byler}}{{Choi} et~al.}{2017}]{Choi_2017}
{Choi} J.,  {Conroy} C.,   {Byler} N.,  2017, \mn@doi [\apj] {10.3847/1538-4357/aa679f}, \href {https://ui.adsabs.harvard.edu/abs/2017ApJ...838..159C} {838, 159}

\bibitem[\protect\citeauthoryear{{Coil} et~al.,}{{Coil} et~al.}{2015}]{Coil-2015}
{Coil} A.~L.,  et~al., 2015, \mn@doi [\apj] {10.1088/0004-637X/801/1/35}, \href {https://ui.adsabs.harvard.edu/abs/2015ApJ...801...35C} {801, 35}

\bibitem[\protect\citeauthoryear{{Crowther}}{{Crowther}}{2007}]{Crowther_2007}
{Crowther} P.~A.,  2007, \mn@doi [\araa] {10.1146/annurev.astro.45.051806.110615}, \href {https://ui.adsabs.harvard.edu/abs/2007ARA&A..45..177C} {45, 177}

\bibitem[\protect\citeauthoryear{{Crowther} et~al.,}{{Crowther} et~al.}{2016}]{Crowther_2016}
{Crowther} P.~A.,  et~al., 2016, \mn@doi [\mnras] {10.1093/mnras/stw273}, \href {https://ui.adsabs.harvard.edu/abs/2016MNRAS.458..624C} {458, 624}

\bibitem[\protect\citeauthoryear{Dalcanton}{Dalcanton}{2007}]{Dalcanton_2007}
Dalcanton J.~J.,  2007, \mn@doi [The Astrophysical Journal] {10.1086/508913}, 658, 941

\bibitem[\protect\citeauthoryear{{Dale} et~al.,}{{Dale} et~al.}{2016}]{Dale_2016}
{Dale} D.~A.,  et~al., 2016, \mn@doi [\aj] {10.3847/0004-6256/151/1/4}, \href {https://ui.adsabs.harvard.edu/abs/2016AJ....151....4D} {151, 4}

\bibitem[\protect\citeauthoryear{{Dale} et~al.,}{{Dale} et~al.}{2020}]{Dale_2020}
{Dale} D.~A.,  et~al., 2020, \mn@doi [\aj] {10.3847/1538-3881/ab7eb2}, \href {https://ui.adsabs.harvard.edu/abs/2020AJ....159..195D} {159, 195}

\bibitem[\protect\citeauthoryear{{Dickey} et~al.,}{{Dickey} et~al.}{2021}]{Dickey_2021}
{Dickey} C.~M.,  et~al., 2021, \mn@doi [\apj] {10.3847/1538-4357/abc014}, \href {https://ui.adsabs.harvard.edu/abs/2021ApJ...915...53D} {915, 53}

\bibitem[\protect\citeauthoryear{{Dobbs} \& {Pringle}}{{Dobbs} \& {Pringle}}{2009}]{Dobbs_2009}
{Dobbs} C.~L.,  {Pringle} J.~E.,  2009, \mn@doi [\mnras] {10.1111/j.1365-2966.2009.14815.x}, \href {https://ui.adsabs.harvard.edu/abs/2009MNRAS.396.1579D} {396, 1579}

\bibitem[\protect\citeauthoryear{{Dom{\'\i}nguez}, {Siana}, {Brooks}, {Christensen}, {Bruzual}, {Stark}  \& {Alavi}}{{Dom{\'\i}nguez} et~al.}{2015}]{Dominguez}
{Dom{\'\i}nguez} A.,  {Siana} B.,  {Brooks} A.~M.,  {Christensen} C.~R.,  {Bruzual} G.,  {Stark} D.~P.,   {Alavi} A.,  2015, \mn@doi [\mnras] {10.1093/mnras/stv1001}, \href {https://ui.adsabs.harvard.edu/abs/2015MNRAS.451..839D} {451, 839}

\bibitem[\protect\citeauthoryear{{Douna}, {Pellizza}, {Mirabel}  \& {Pedrosa}}{{Douna} et~al.}{2015}]{Douna_2015}
{Douna} V.~M.,  {Pellizza} L.~J.,  {Mirabel} I.~F.,   {Pedrosa} S.~E.,  2015, \mn@doi [\aap] {10.1051/0004-6361/201525617}, \href {https://ui.adsabs.harvard.edu/abs/2015A&A...579A..44D} {579, A44}

\bibitem[\protect\citeauthoryear{{Du} et~al.,}{{Du} et~al.}{2018}]{Du_2018}
{Du} X.,  et~al., 2018, \mn@doi [\apj] {10.3847/1538-4357/aabfcf}, \href {https://ui.adsabs.harvard.edu/abs/2018ApJ...860...75D} {860, 75}

\bibitem[\protect\citeauthoryear{{Eldridge}}{{Eldridge}}{2012}]{Eldridge_2012}
{Eldridge} J.~J.,  2012, \mn@doi [\mnras] {10.1111/j.1365-2966.2012.20662.x}, \href {https://ui.adsabs.harvard.edu/abs/2012MNRAS.422..794E} {422, 794}

\bibitem[\protect\citeauthoryear{{Eldridge}, {Stanway}, {Xiao}, {McClelland}, {Taylor}, {Ng}, {Greis}  \& {Bray}}{{Eldridge} et~al.}{2017}]{Eldridge_2017}
{Eldridge} J.~J.,  {Stanway} E.~R.,  {Xiao} L.,  {McClelland} L.~A.~S.,  {Taylor} G.,  {Ng} M.,  {Greis} S.~M.~L.,   {Bray} J.~C.,  2017, \mn@doi [\pasa] {10.1017/pasa.2017.51}, \href {https://ui.adsabs.harvard.edu/abs/2017PASA...34...58E} {34, e058}

\bibitem[\protect\citeauthoryear{{Elmegreen}}{{Elmegreen}}{2006}]{Elmegreen_2006}
{Elmegreen} B.~G.,  2006, \mn@doi [\apj] {10.1086/505785}, \href {https://ui.adsabs.harvard.edu/abs/2006ApJ...648..572E} {648, 572}

\bibitem[\protect\citeauthoryear{Emami, Siana, Weisz, Johnson, Ma  \& El-Badry}{Emami et~al.}{2019}]{Emami_2019}
Emami N.,  Siana B.,  Weisz D.~R.,  Johnson B.~D.,  Ma X.,   El-Badry K.,  2019, \mn@doi [The Astrophysical Journal] {10.3847/1538-4357/ab211a}, 881, 71

\bibitem[\protect\citeauthoryear{Erb, Steidel, Shapley, Pettini, Reddy  \& Adelberger}{Erb et~al.}{2006}]{Erb_2006}
Erb D.~K.,  Steidel C.~C.,  Shapley A.~E.,  Pettini M.,  Reddy N.~A.,   Adelberger K.~L.,  2006, \mn@doi [The Astrophysical Journal] {10.1086/505341}, 647, 128

\bibitem[\protect\citeauthoryear{{Faisst}, {Capak}, {Emami}, {Tacchella}  \& {Larson}}{{Faisst} et~al.}{2019}]{Faisst_2019}
{Faisst} A.~L.,  {Capak} P.~L.,  {Emami} N.,  {Tacchella} S.,   {Larson} K.~L.,  2019, \mn@doi [\apj] {10.3847/1538-4357/ab425b}, \href {https://ui.adsabs.harvard.edu/abs/2019ApJ...884..133F} {884, 133}

\bibitem[\protect\citeauthoryear{{Feldmann}, {Quataert}, {Hopkins}, {Faucher-Gigu{\`e}re}  \& {Kere{\v{s}}}}{{Feldmann} et~al.}{2017}]{Feldmann_2017}
{Feldmann} R.,  {Quataert} E.,  {Hopkins} P.~F.,  {Faucher-Gigu{\`e}re} C.-A.,   {Kere{\v{s}}} D.,  2017, \mn@doi [\mnras] {10.1093/mnras/stx1120}, \href {https://ui.adsabs.harvard.edu/abs/2017MNRAS.470.1050F} {470, 1050}

\bibitem[\protect\citeauthoryear{{Ferland} et~al.,}{{Ferland} et~al.}{2017}]{Ferland17}
{Ferland} G.~J.,  et~al., 2017, \rmxaa, \href {https://ui.adsabs.harvard.edu/abs/2017RMxAA..53..385F} {53, 385}

\bibitem[\protect\citeauthoryear{{Fetherolf} et~al.,}{{Fetherolf} et~al.}{2020}]{TARA_Voro}
{Fetherolf} T.,  et~al., 2020, \mn@doi [\mnras] {10.1093/mnras/staa2775}, \href {https://ui.adsabs.harvard.edu/abs/2020MNRAS.498.5009F} {498, 5009}

\bibitem[\protect\citeauthoryear{{Fetherolf} et~al.,}{{Fetherolf} et~al.}{2021}]{Tara_2021a}
{Fetherolf} T.,  et~al., 2021, \mn@doi [\mnras] {10.1093/mnras/stab2570}, \href {https://ui.adsabs.harvard.edu/abs/2021MNRAS.508.1431F} {508, 1431}

\bibitem[\protect\citeauthoryear{{Fetherolf} et~al.,}{{Fetherolf} et~al.}{2023}]{Tara_2022}
{Fetherolf} T.,  et~al., 2023, \mn@doi [\mnras] {10.1093/mnras/stac3362}, \href {https://ui.adsabs.harvard.edu/abs/2023MNRAS.518.4214F} {518, 4214}

\bibitem[\protect\citeauthoryear{{Fornasini} et~al.,}{{Fornasini} et~al.}{2019}]{Fornasini_2019}
{Fornasini} F.~M.,  et~al., 2019, \mn@doi [\apj] {10.3847/1538-4357/ab4653}, \href {https://ui.adsabs.harvard.edu/abs/2019ApJ...885...65F} {885, 65}

\bibitem[\protect\citeauthoryear{{F{\"o}rster Schreiber} et~al.,}{{F{\"o}rster Schreiber} et~al.}{2009}]{Foster_2009}
{F{\"o}rster Schreiber} N.~M.,  et~al., 2009, \mn@doi [\apj] {10.1088/0004-637X/706/2/1364}, \href {https://ui.adsabs.harvard.edu/abs/2009ApJ...706.1364F} {706, 1364}

\bibitem[\protect\citeauthoryear{{Freeman} et~al.,}{{Freeman} et~al.}{2017}]{Freeman_2019}
{Freeman} W.~R.,  et~al., 2017, arXiv e-prints, \href {https://ui.adsabs.harvard.edu/abs/2017arXiv171003230F} {p. arXiv:1710.03230}

\bibitem[\protect\citeauthoryear{{Fudamoto} et~al.,}{{Fudamoto} et~al.}{2020}]{Fudamoto_2020}
{Fudamoto} Y.,  et~al., 2020, \mn@doi [\aap] {10.1051/0004-6361/202038163}, \href {https://ui.adsabs.harvard.edu/abs/2020A&A...643A...4F} {643, A4}

\bibitem[\protect\citeauthoryear{{Fujimoto}, {Chevance}, {Haydon}, {Krumholz}  \& {Kruijssen}}{{Fujimoto} et~al.}{2019}]{Fujimoto_2019}
{Fujimoto} Y.,  {Chevance} M.,  {Haydon} D.~T.,  {Krumholz} M.~R.,   {Kruijssen} J.~M.~D.,  2019, \mn@doi [\mnras] {10.1093/mnras/stz641}, \href {https://ui.adsabs.harvard.edu/abs/2019MNRAS.487.1717F} {487, 1717}

\bibitem[\protect\citeauthoryear{{Fumagalli}, {da Silva}  \& {Krumholz}}{{Fumagalli} et~al.}{2011}]{Fumagalli_2011}
{Fumagalli} M.,  {da Silva} R.~L.,   {Krumholz} M.~R.,  2011, \mn@doi [\apjl] {10.1088/2041-8205/741/2/L26}, \href {https://ui.adsabs.harvard.edu/abs/2011ApJ...741L..26F} {741, L26}

\bibitem[\protect\citeauthoryear{{Genzel} et~al.,}{{Genzel} et~al.}{2010}]{Genzel_2010}
{Genzel} R.,  et~al., 2010, \mn@doi [\mnras] {10.1111/j.1365-2966.2010.16969.x}, \href {https://ui.adsabs.harvard.edu/abs/2010MNRAS.407.2091G} {407, 2091}

\bibitem[\protect\citeauthoryear{{Gilfanov}, {Grimm}  \& {Sunyaev}}{{Gilfanov} et~al.}{2004}]{Gilfanov_2004}
{Gilfanov} M.,  {Grimm} H.~J.,   {Sunyaev} R.,  2004, \mn@doi [\mnras] {10.1111/j.1365-2966.2004.07450.x}, \href {https://ui.adsabs.harvard.edu/abs/2004MNRAS.347L..57G} {347, L57}

\bibitem[\protect\citeauthoryear{{Glazebrook}, {Blake}, {Economou}, {Lilly}  \& {Colless}}{{Glazebrook} et~al.}{1999}]{Glazebrook_1999}
{Glazebrook} K.,  {Blake} C.,  {Economou} F.,  {Lilly} S.,   {Colless} M.,  1999, \mn@doi [\mnras] {10.1046/j.1365-8711.1999.02576.x}, \href {https://ui.adsabs.harvard.edu/abs/1999MNRAS.306..843G} {306, 843}

\bibitem[\protect\citeauthoryear{{Gordon}, {Clayton}, {Misselt}, {Landolt}  \& {Wolff}}{{Gordon} et~al.}{2003}]{Gordon}
{Gordon} K.~D.,  {Clayton} G.~C.,  {Misselt} K.~A.,  {Landolt} A.~U.,   {Wolff} M.~J.,  2003, \mn@doi [\apj] {10.1086/376774}, \href {https://ui.adsabs.harvard.edu/abs/2003ApJ...594..279G} {594, 279}

\bibitem[\protect\citeauthoryear{{Governato} et~al.,}{{Governato} et~al.}{2010}]{Governato_2010}
{Governato} F.,  et~al., 2010, \mn@doi [\nat] {10.1038/nature08640}, \href {https://ui.adsabs.harvard.edu/abs/2010Natur.463..203G} {463, 203}

\bibitem[\protect\citeauthoryear{{Gr{\"a}fener} \& {Vink}}{{Gr{\"a}fener} \& {Vink}}{2015}]{Grafener_2015}
{Gr{\"a}fener} G.,  {Vink} J.~S.,  2015, \mn@doi [\aap] {10.1051/0004-6361/201425287}, \href {https://ui.adsabs.harvard.edu/abs/2015A&A...578L...2G} {578, L2}

\bibitem[\protect\citeauthoryear{{Green}, {Glazebrook}, {Gilbank}, {McGregor}, {Damjanov}, {Abraham}  \& {Sharp}}{{Green} et~al.}{2017}]{green_2017}
{Green} A.~W.,  {Glazebrook} K.,  {Gilbank} D.~G.,  {McGregor} P.~J.,  {Damjanov} I.,  {Abraham} R.~G.,   {Sharp} R.,  2017, \mn@doi [\mnras] {10.1093/mnras/stx1119}, \href {https://ui.adsabs.harvard.edu/abs/2017MNRAS.470..639G} {470, 639}

\bibitem[\protect\citeauthoryear{{Grimm}, {Gilfanov}  \& {Sunyaev}}{{Grimm} et~al.}{2003}]{Grimm_2003}
{Grimm} H.~J.,  {Gilfanov} M.,   {Sunyaev} R.,  2003, \mn@doi [\mnras] {10.1046/j.1365-8711.2003.06224.x}, \href {https://ui.adsabs.harvard.edu/abs/2003MNRAS.339..793G} {339, 793}

\bibitem[\protect\citeauthoryear{{Grogin} et~al.,}{{Grogin} et~al.}{2011}]{Grogin_2011}
{Grogin} N.~A.,  et~al., 2011, \mn@doi [\apjs] {10.1088/0067-0049/197/2/35}, \href {https://ui.adsabs.harvard.edu/abs/2011ApJS..197...35G} {197, 35}

\bibitem[\protect\citeauthoryear{Guo et~al.,}{Guo et~al.}{2016}]{Guo_2016}
Guo Y.,  et~al., 2016, \mn@doi [The Astrophysical Journal] {10.3847/1538-4357/833/1/37}, 833, 37

\bibitem[\protect\citeauthoryear{{Gutkin}, {Charlot}  \& {Bruzual}}{{Gutkin} et~al.}{2016}]{Gutkin_2016}
{Gutkin} J.,  {Charlot} S.,   {Bruzual} G.,  2016, \mn@doi [\mnras] {10.1093/mnras/stw1716}, \href {https://ui.adsabs.harvard.edu/abs/2016MNRAS.462.1757G} {462, 1757}

\bibitem[\protect\citeauthoryear{{Hayward} \& {Hopkins}}{{Hayward} \& {Hopkins}}{2017}]{Hayward_2017}
{Hayward} C.~C.,  {Hopkins} P.~F.,  2017, \mn@doi [\mnras] {10.1093/mnras/stw2888}, \href {https://ui.adsabs.harvard.edu/abs/2017MNRAS.465.1682H} {465, 1682}

\bibitem[\protect\citeauthoryear{Hemmati et~al.,}{Hemmati et~al.}{2014}]{Hemmati_2014}
Hemmati S.,  et~al., 2014, \mn@doi [The Astrophysical Journal] {10.1088/0004-637x/797/2/108}, 797, 108

\bibitem[\protect\citeauthoryear{{Hopkins} \& {Beacom}}{{Hopkins} \& {Beacom}}{2006}]{Hopkins_2006}
{Hopkins} A.~M.,  {Beacom} J.~F.,  2006, \mn@doi [\apj] {10.1086/506610}, \href {https://ui.adsabs.harvard.edu/abs/2006ApJ...651..142H} {651, 142}

\bibitem[\protect\citeauthoryear{{Hopkins}, {Kere{\v{s}}}, {O{\~n}orbe}, {Faucher-Gigu{\`e}re}, {Quataert}, {Murray}  \& {Bullock}}{{Hopkins} et~al.}{2014}]{Hopkins_2014}
{Hopkins} P.~F.,  {Kere{\v{s}}} D.,  {O{\~n}orbe} J.,  {Faucher-Gigu{\`e}re} C.-A.,  {Quataert} E.,  {Murray} N.,   {Bullock} J.~S.,  2014, \mn@doi [\mnras] {10.1093/mnras/stu1738}, \href {https://ui.adsabs.harvard.edu/abs/2014MNRAS.445..581H} {445, 581}

\bibitem[\protect\citeauthoryear{{Hoversten} \& {Glazebrook}}{{Hoversten} \& {Glazebrook}}{2008}]{Hoversten_2008}
{Hoversten} E.~A.,  {Glazebrook} K.,  2008, \mn@doi [\apj] {10.1086/524095}, \href {https://ui.adsabs.harvard.edu/abs/2008ApJ...675..163H} {675, 163}

\bibitem[\protect\citeauthoryear{{Hunter}, {Elmegreen}  \& {Ludka}}{{Hunter} et~al.}{2010}]{Hunter_2010}
{Hunter} D.~A.,  {Elmegreen} B.~G.,   {Ludka} B.~C.,  2010, \mn@doi [\aj] {10.1088/0004-6256/139/2/447}, \href {https://ui.adsabs.harvard.edu/abs/2010AJ....139..447H} {139, 447}

\bibitem[\protect\citeauthoryear{{Iglesias-P{\'a}ramo}, {Boselli}, {Gavazzi}  \& {Zaccardo}}{{Iglesias-P{\'a}ramo} et~al.}{2004}]{Iglesias_2004}
{Iglesias-P{\'a}ramo} J.,  {Boselli} A.,  {Gavazzi} G.,   {Zaccardo} A.,  2004, \mn@doi [\aap] {10.1051/0004-6361:20034572}, \href {https://ui.adsabs.harvard.edu/abs/2004A&A...421..887I} {421, 887}

\bibitem[\protect\citeauthoryear{{Jafariyazani}, {Mobasher}, {Hemmati}, {Fetherolf}, {Khostovan}  \& {Chartab}}{{Jafariyazani} et~al.}{2019}]{Jafariyazani_2019}
{Jafariyazani} M.,  {Mobasher} B.,  {Hemmati} S.,  {Fetherolf} T.,  {Khostovan} A.~A.,   {Chartab} N.,  2019, \mn@doi [\apj] {10.3847/1538-4357/ab5526}, \href {https://ui.adsabs.harvard.edu/abs/2019ApJ...887..204J} {887, 204}

\bibitem[\protect\citeauthoryear{{Kashino} et~al.,}{{Kashino} et~al.}{2013}]{Kashino_2013}
{Kashino} D.,  et~al., 2013, \mn@doi [\apjl] {10.1088/2041-8205/777/1/L8}, \href {https://ui.adsabs.harvard.edu/abs/2013ApJ...777L...8K} {777, L8}

\bibitem[\protect\citeauthoryear{{Keel}, {Holberg}  \& {Treuthardt}}{{Keel} et~al.}{2004}]{Keel_2004}
{Keel} W.~C.,  {Holberg} J.~B.,   {Treuthardt} P.~M.,  2004, \mn@doi [\aj] {10.1086/421367}, \href {https://ui.adsabs.harvard.edu/abs/2004AJ....128..211K} {128, 211}

\bibitem[\protect\citeauthoryear{{Kennicutt}}{{Kennicutt}}{1989}]{Kennicutt_1998}
{Kennicutt} Robert~C. J.,  1989, \mn@doi [\apj] {10.1086/167834}, \href {https://ui.adsabs.harvard.edu/abs/1989ApJ...344..685K} {344, 685}

\bibitem[\protect\citeauthoryear{{Kennicutt} \& {Evans}}{{Kennicutt} \& {Evans}}{2012}]{Kennicutt_2012}
{Kennicutt} R.~C.,  {Evans} N.~J.,  2012, \mn@doi [\araa] {10.1146/annurev-astro-081811-125610}, \href {https://ui.adsabs.harvard.edu/abs/2012ARA&A..50..531K} {50, 531}

\bibitem[\protect\citeauthoryear{{Kere{\v{s}}}, {Katz}, {Dav{\'e}}, {Fardal}  \& {Weinberg}}{{Kere{\v{s}}} et~al.}{2009}]{Kere_2009}
{Kere{\v{s}}} D.,  {Katz} N.,  {Dav{\'e}} R.,  {Fardal} M.,   {Weinberg} D.~H.,  2009, \mn@doi [\mnras] {10.1111/j.1365-2966.2009.14924.x}, \href {https://ui.adsabs.harvard.edu/abs/2009MNRAS.396.2332K} {396, 2332}

\bibitem[\protect\citeauthoryear{{Kewley}, {Geller}, {Jansen}  \& {Dopita}}{{Kewley} et~al.}{2002}]{Kewley_2002}
{Kewley} L.~J.,  {Geller} M.~J.,  {Jansen} R.~A.,   {Dopita} M.~A.,  2002, \mn@doi [\aj] {10.1086/344487}, \href {https://ui.adsabs.harvard.edu/abs/2002AJ....124.3135K} {124, 3135}

\bibitem[\protect\citeauthoryear{{Kewley}, {Jansen}  \& {Geller}}{{Kewley} et~al.}{2005}]{Kewley_2005}
{Kewley} L.~J.,  {Jansen} R.~A.,   {Geller} M.~J.,  2005, \mn@doi [\pasp] {10.1086/428303}, \href {https://ui.adsabs.harvard.edu/abs/2005PASP..117..227K} {117, 227}

\bibitem[\protect\citeauthoryear{Knapen \& James}{Knapen \& James}{2009}]{Knapen_2009}
Knapen J.~H.,  James P.~A.,  2009, \mn@doi [The Astrophysical Journal] {10.1088/0004-637x/698/2/1437}, 698, 1437

\bibitem[\protect\citeauthoryear{Koekemoer et~al.,}{Koekemoer et~al.}{2011}]{Koekemoer_2011}
Koekemoer A.~M.,  et~al., 2011, \mn@doi [The Astrophysical Journal Supplement Series] {10.1088/0067-0049/197/2/36}, 197, 36

\bibitem[\protect\citeauthoryear{Kriek et~al.,}{Kriek et~al.}{2015}]{Kriek_2015}
Kriek M.,  et~al., 2015, \mn@doi [The Astrophysical Journal Supplement Series] {10.1088/0067-0049/218/2/15}, 218, 15

\bibitem[\protect\citeauthoryear{{Lamers}, {Haser}, {de Koter}  \& {Leitherer}}{{Lamers} et~al.}{1999}]{Lamers_1999}
{Lamers} H. J.~G.~L.~M.,  {Haser} S.,  {de Koter} A.,   {Leitherer} C.,  1999, \mn@doi [\apj] {10.1086/307127}, \href {https://ui.adsabs.harvard.edu/abs/1999ApJ...516..872L} {516, 872}

\bibitem[\protect\citeauthoryear{{Lang} et~al.,}{{Lang} et~al.}{2014}]{Lang_2014}
{Lang} P.,  et~al., 2014, \mn@doi [\apj] {10.1088/0004-637X/788/1/11}, \href {https://ui.adsabs.harvard.edu/abs/2014ApJ...788...11L} {788, 11}

\bibitem[\protect\citeauthoryear{{Lara-L{\'o}pez} et~al.,}{{Lara-L{\'o}pez} et~al.}{2010}]{Lara_2010}
{Lara-L{\'o}pez} M.~A.,  et~al., 2010, \mn@doi [\aap] {10.1051/0004-6361/201014803}, \href {https://ui.adsabs.harvard.edu/abs/2010A&A...521L..53L} {521, L53}

\bibitem[\protect\citeauthoryear{Law, Steidel, Erb, Larkin, Pettini, Shapley  \& Wright}{Law et~al.}{2009}]{Law_2009}
Law D.~R.,  Steidel C.~C.,  Erb D.~K.,  Larkin J.~E.,  Pettini M.,  Shapley A.~E.,   Wright S.~A.,  2009, \mn@doi [The Astrophysical Journal] {10.1088/0004-637X/697/2/2057}, 697, 2057

\bibitem[\protect\citeauthoryear{Law, Steidel, Shapley, Nagy, Reddy  \& Erb}{Law et~al.}{2012}]{Law_2012}
Law D.~R.,  Steidel C.~C.,  Shapley A.~E.,  Nagy S.~R.,  Reddy N.~A.,   Erb D.~K.,  2012, \mn@doi [The Astrophysical Journal] {10.1088/0004-637X/745/1/85}, 745, 85

\bibitem[\protect\citeauthoryear{Lee et~al.,}{Lee et~al.}{2009}]{Lee_2009}
Lee J.~C.,  et~al., 2009, \mn@doi [The Astrophysical Journal] {10.1088/0004-637x/706/1/599}, 706, 599

\bibitem[\protect\citeauthoryear{{Lee} et~al.,}{{Lee} et~al.}{2011}]{Lee_2011}
{Lee} J.~C.,  et~al., 2011, \mn@doi [\apjs] {10.1088/0067-0049/192/1/6}, \href {https://ui.adsabs.harvard.edu/abs/2011ApJS..192....6L} {192, 6}

\bibitem[\protect\citeauthoryear{{Lehmer} et~al.,}{{Lehmer} et~al.}{2008}]{Lehmer_2008}
{Lehmer} B.~D.,  et~al., 2008, \mn@doi [\apj] {10.1086/588459}, \href {https://ui.adsabs.harvard.edu/abs/2008ApJ...681.1163L} {681, 1163}

\bibitem[\protect\citeauthoryear{{Lehmer}, {Alexander}, {Bauer}, {Brandt}, {Goulding}, {Jenkins}, {Ptak}  \& {Roberts}}{{Lehmer} et~al.}{2010}]{Lehmer_2010}
{Lehmer} B.~D.,  {Alexander} D.~M.,  {Bauer} F.~E.,  {Brandt} W.~N.,  {Goulding} A.~D.,  {Jenkins} L.~P.,  {Ptak} A.,   {Roberts} T.~P.,  2010, \mn@doi [\apj] {10.1088/0004-637X/724/1/559}, \href {https://ui.adsabs.harvard.edu/abs/2010ApJ...724..559L} {724, 559}

\bibitem[\protect\citeauthoryear{{Lehmer} et~al.,}{{Lehmer} et~al.}{2016}]{Lehmer_2016}
{Lehmer} B.~D.,  et~al., 2016, \mn@doi [\apj] {10.3847/0004-637X/825/1/7}, \href {https://ui.adsabs.harvard.edu/abs/2016ApJ...825....7L} {825, 7}

\bibitem[\protect\citeauthoryear{{Leitherer}}{{Leitherer}}{2005}]{Leitherer_2005}
{Leitherer} C.,  2005, {A Far-Ultraviolet View of Starburst Galaxies}.
p.~89, \mn@doi{10.1007/1-4020-3539-X\_16}

\bibitem[\protect\citeauthoryear{{Leitherer} \& {Heckman}}{{Leitherer} \& {Heckman}}{1995}]{Leitherer_1995}
{Leitherer} C.,  {Heckman} T.~M.,  1995, \mn@doi [\apjs] {10.1086/192112}, \href {https://ui.adsabs.harvard.edu/abs/1995ApJS...96....9L} {96, 9}

\bibitem[\protect\citeauthoryear{{Leitherer}, {Le{\~a}o}, {Heckman}, {Lennon}, {Pettini}  \& {Robert}}{{Leitherer} et~al.}{2001}]{Leitherer_2001}
{Leitherer} C.,  {Le{\~a}o} J. R.~S.,  {Heckman} T.~M.,  {Lennon} D.~J.,  {Pettini} M.,   {Robert} C.,  2001, \mn@doi [\apj] {10.1086/319814}, \href {https://ui.adsabs.harvard.edu/abs/2001ApJ...550..724L} {550, 724}

\bibitem[\protect\citeauthoryear{{Leitherer}, {Ortiz Ot{\'a}lvaro}, {Bresolin}, {Kudritzki}, {Lo Faro}, {Pauldrach}, {Pettini}  \& {Rix}}{{Leitherer} et~al.}{2010}]{Leitherer_2010}
{Leitherer} C.,  {Ortiz Ot{\'a}lvaro} P.~A.,  {Bresolin} F.,  {Kudritzki} R.-P.,  {Lo Faro} B.,  {Pauldrach} A. W.~A.,  {Pettini} M.,   {Rix} S.~A.,  2010, \mn@doi [\apjs] {10.1088/0067-0049/189/2/309}, \href {https://ui.adsabs.harvard.edu/abs/2010ApJS..189..309L} {189, 309}

\bibitem[\protect\citeauthoryear{{Leung} et~al.,}{{Leung} et~al.}{2019}]{Leung-2019}
{Leung} G. C.~K.,  et~al., 2019, \mn@doi [\apj] {10.3847/1538-4357/ab4a7c}, \href {https://ui.adsabs.harvard.edu/abs/2019ApJ...886...11L} {886, 11}

\bibitem[\protect\citeauthoryear{{Lilly}, {Carollo}, {Pipino}, {Renzini}  \& {Peng}}{{Lilly} et~al.}{2013}]{Lilly_2013}
{Lilly} S.~J.,  {Carollo} C.~M.,  {Pipino} A.,  {Renzini} A.,   {Peng} Y.,  2013, \mn@doi [\apj] {10.1088/0004-637X/772/2/119}, \href {https://ui.adsabs.harvard.edu/abs/2013ApJ...772..119L} {772, 119}

\bibitem[\protect\citeauthoryear{{Ma} et~al.,}{{Ma} et~al.}{2018}]{Ma_2018}
{Ma} X.,  et~al., 2018, \mn@doi [\mnras] {10.1093/mnras/sty1024}, \href {https://ui.adsabs.harvard.edu/abs/2018MNRAS.478.1694M} {478, 1694}

\bibitem[\protect\citeauthoryear{Madau \& Dickinson}{Madau \& Dickinson}{2014}]{Madau_2014}
Madau P.,  Dickinson M.,  2014, \mn@doi [Annual Review of Astronomy and Astrophysics] {10.1146/annurev-astro-081811-125615}, 52, 415

\bibitem[\protect\citeauthoryear{{Madau}, {Ferguson}, {Dickinson}, {Giavalisco}, {Steidel}  \& {Fruchter}}{{Madau} et~al.}{1996}]{Madau_1996}
{Madau} P.,  {Ferguson} H.~C.,  {Dickinson} M.~E.,  {Giavalisco} M.,  {Steidel} C.~C.,   {Fruchter} A.,  1996, \mn@doi [\mnras] {10.1093/mnras/283.4.1388}, \href {https://ui.adsabs.harvard.edu/abs/1996MNRAS.283.1388M} {283, 1388}

\bibitem[\protect\citeauthoryear{Mas-Ribas, Dijkstra  \& Forero-Romero}{Mas-Ribas et~al.}{2016}]{Mas_Ribas_2016}
Mas-Ribas L.,  Dijkstra M.,   Forero-Romero J.~E.,  2016, \mn@doi [The Astrophysical Journal] {10.3847/1538-4357/833/1/65}, 833, 65

\bibitem[\protect\citeauthoryear{{McLean} et~al.,}{{McLean} et~al.}{2012}]{Mclean}
{McLean} I.~S.,  et~al., 2012, in {McLean} I.~S.,  {Ramsay} S.~K.,   {Takami} H.,  eds,  Society of Photo-Optical Instrumentation Engineers (SPIE) Conference Series Vol. 8446, Ground-based and Airborne Instrumentation for Astronomy IV. p. 84460J, \mn@doi{10.1117/12.924794}

\bibitem[\protect\citeauthoryear{{Mehlert} et~al.,}{{Mehlert} et~al.}{2002}]{Mehlert_2002}
{Mehlert} D.,  et~al., 2002, \mn@doi [\aap] {10.1051/0004-6361:20021052}, \href {https://ui.adsabs.harvard.edu/abs/2002A&A...393..809M} {393, 809}

\bibitem[\protect\citeauthoryear{Meurer et~al.,}{Meurer et~al.}{2009}]{Meurer_2009}
Meurer G.~R.,  et~al., 2009, \mn@doi [The Astrophysical Journal] {10.1088/0004-637x/695/1/765}, 695, 765

\bibitem[\protect\citeauthoryear{{Mineo}, {Gilfanov}  \& {Sunyaev}}{{Mineo} et~al.}{2012}]{Mineo_2012}
{Mineo} S.,  {Gilfanov} M.,   {Sunyaev} R.,  2012, \mn@doi [\mnras] {10.1111/j.1365-2966.2011.19862.x}, \href {https://ui.adsabs.harvard.edu/abs/2012MNRAS.419.2095M} {419, 2095}

\bibitem[\protect\citeauthoryear{{Momcheva} et~al.,}{{Momcheva} et~al.}{2016}]{Momcheva2016}
{Momcheva} I.~G.,  et~al., 2016, \mn@doi [\apjs] {10.3847/0067-0049/225/2/27}, \href {https://ui.adsabs.harvard.edu/abs/2016ApJS..225...27M} {225, 27}

\bibitem[\protect\citeauthoryear{{Nanayakkara} et~al.,}{{Nanayakkara} et~al.}{2019}]{Nanayakkara_2019}
{Nanayakkara} T.,  et~al., 2019, \mn@doi [\aap] {10.1051/0004-6361/201834565}, \href {https://ui.adsabs.harvard.edu/abs/2019A&A...624A..89N} {624, A89}

\bibitem[\protect\citeauthoryear{Nandra, Mushotzky, Arnaud, Steidel, Adelberger, Gardner, Teplitz  \& Windhorst}{Nandra et~al.}{2002}]{Nandra_2002}
Nandra K.,  Mushotzky R.~F.,  Arnaud K.,  Steidel C.~C.,  Adelberger K.~L.,  Gardner J.~P.,  Teplitz H.~I.,   Windhorst R.~A.,  2002, \mn@doi [The Astrophysical Journal] {10.1086/341888}, 576, 625

\bibitem[\protect\citeauthoryear{{Noeske} et~al.,}{{Noeske} et~al.}{2007}]{Noeske_2007}
{Noeske} K.~G.,  et~al., 2007, \mn@doi [\apjl] {10.1086/517926}, \href {https://ui.adsabs.harvard.edu/abs/2007ApJ...660L..43N} {660, L43}

\bibitem[\protect\citeauthoryear{{Oke} et~al.,}{{Oke} et~al.}{1995}]{Oke}
{Oke} J.~B.,  et~al., 1995, \mn@doi [\pasp] {10.1086/133562}, \href {https://ui.adsabs.harvard.edu/abs/1995PASP..107..375O} {107, 375}

\bibitem[\protect\citeauthoryear{{Papovich}, {Dickinson}  \& {Ferguson}}{{Papovich} et~al.}{2001}]{Papovich_2001}
{Papovich} C.,  {Dickinson} M.,   {Ferguson} H.~C.,  2001, \mn@doi [\apj] {10.1086/322412}, \href {https://ui.adsabs.harvard.edu/abs/2001ApJ...559..620P} {559, 620}

\bibitem[\protect\citeauthoryear{{Peeples} \& {Shankar}}{{Peeples} \& {Shankar}}{2011}]{Peeples_2011}
{Peeples} M.~S.,  {Shankar} F.,  2011, \mn@doi [\mnras] {10.1111/j.1365-2966.2011.19456.x}, \href {https://ui.adsabs.harvard.edu/abs/2011MNRAS.417.2962P} {417, 2962}

\bibitem[\protect\citeauthoryear{{Pellerin} et~al.,}{{Pellerin} et~al.}{2002}]{Pellerin_2002}
{Pellerin} A.,  et~al., 2002, \mn@doi [\apjs] {10.1086/342268}, \href {https://ui.adsabs.harvard.edu/abs/2002ApJS..143..159P} {143, 159}

\bibitem[\protect\citeauthoryear{{Persic} \& {Rephaeli}}{{Persic} \& {Rephaeli}}{2007}]{Persic_2007}
{Persic} M.,  {Rephaeli} Y.,  2007, \mn@doi [\aap] {10.1051/0004-6361:20054146}, \href {https://ui.adsabs.harvard.edu/abs/2007A&A...463..481P} {463, 481}

\bibitem[\protect\citeauthoryear{{Persic}, {Rephaeli}, {Braito}, {Cappi}, {Della Ceca}, {Franceschini}  \& {Gruber}}{{Persic} et~al.}{2004}]{Persic_2004}
{Persic} M.,  {Rephaeli} Y.,  {Braito} V.,  {Cappi} M.,  {Della Ceca} R.,  {Franceschini} A.,   {Gruber} D.~E.,  2004, \mn@doi [\aap] {10.1051/0004-6361:20034500}, \href {https://ui.adsabs.harvard.edu/abs/2004A&A...419..849P} {419, 849}

\bibitem[\protect\citeauthoryear{{Pettini}, {Steidel}, {Adelberger}, {Dickinson}  \& {Giavalisco}}{{Pettini} et~al.}{2000}]{Pettini_2000}
{Pettini} M.,  {Steidel} C.~C.,  {Adelberger} K.~L.,  {Dickinson} M.,   {Giavalisco} M.,  2000, \mn@doi [\apj] {10.1086/308176}, \href {https://ui.adsabs.harvard.edu/abs/2000ApJ...528...96P} {528, 96}

\bibitem[\protect\citeauthoryear{{Pflamm-Altenburg}, {Weidner}  \& {Kroupa}}{{Pflamm-Altenburg} et~al.}{2007}]{Pflamm_2007}
{Pflamm-Altenburg} J.,  {Weidner} C.,   {Kroupa} P.,  2007, \mn@doi [\apj] {10.1086/523033}, \href {https://ui.adsabs.harvard.edu/abs/2007ApJ...671.1550P} {671, 1550}

\bibitem[\protect\citeauthoryear{{Pflamm-Altenburg}, {Weidner}  \& {Kroupa}}{{Pflamm-Altenburg} et~al.}{2009}]{Pflamm_2009}
{Pflamm-Altenburg} J.,  {Weidner} C.,   {Kroupa} P.,  2009, \mn@doi [\mnras] {10.1111/j.1365-2966.2009.14522.x}, \href {https://ui.adsabs.harvard.edu/abs/2009MNRAS.395..394P} {395, 394}

\bibitem[\protect\citeauthoryear{{Prestwich}, {Tsantaki}, {Zezas}, {Jackson}, {Roberts}, {Foltz}, {Linden}  \& {Kalogera}}{{Prestwich} et~al.}{2013}]{Prestwich_2013}
{Prestwich} A.~H.,  {Tsantaki} M.,  {Zezas} A.,  {Jackson} F.,  {Roberts} T.~P.,  {Foltz} R.,  {Linden} T.,   {Kalogera} V.,  2013, \mn@doi [\apj] {10.1088/0004-637X/769/2/92}, \href {https://ui.adsabs.harvard.edu/abs/2013ApJ...769...92P} {769, 92}

\bibitem[\protect\citeauthoryear{{Quider}, {Pettini}, {Shapley}  \& {Steidel}}{{Quider} et~al.}{2009}]{Quider_2009}
{Quider} A.~M.,  {Pettini} M.,  {Shapley} A.~E.,   {Steidel} C.~C.,  2009, \mn@doi [\mnras] {10.1111/j.1365-2966.2009.15234.x}, \href {https://ui.adsabs.harvard.edu/abs/2009MNRAS.398.1263Q} {398, 1263}

\bibitem[\protect\citeauthoryear{{Reddy} \& {Steidel}}{{Reddy} \& {Steidel}}{2004}]{Reddy_2004}
{Reddy} N.~A.,  {Steidel} C.~C.,  2004, \mn@doi [\apjl] {10.1086/383087}, \href {https://ui.adsabs.harvard.edu/abs/2004ApJ...603L..13R} {603, L13}

\bibitem[\protect\citeauthoryear{{Reddy}, {Erb}, {Pettini}, {Steidel}  \& {Shapley}}{{Reddy} et~al.}{2010}]{Reddy_2010}
{Reddy} N.~A.,  {Erb} D.~K.,  {Pettini} M.,  {Steidel} C.~C.,   {Shapley} A.~E.,  2010, \mn@doi [\apj] {10.1088/0004-637X/712/2/1070}, \href {https://ui.adsabs.harvard.edu/abs/2010ApJ...712.1070R} {712, 1070}

\bibitem[\protect\citeauthoryear{Reddy, Pettini, Steidel, Shapley, Erb  \& Law}{Reddy et~al.}{2012}]{Reddy_2012}
Reddy N.~A.,  Pettini M.,  Steidel C.~C.,  Shapley A.~E.,  Erb D.~K.,   Law D.~R.,  2012, \mn@doi [The Astrophysical Journal] {10.1088/0004-637x/754/1/25}, 754, 25

\bibitem[\protect\citeauthoryear{{Reddy} et~al.,}{{Reddy} et~al.}{2015}]{Reddy_2015}
{Reddy} N.~A.,  et~al., 2015, \mn@doi [\apj] {10.1088/0004-637X/806/2/259}, \href {https://ui.adsabs.harvard.edu/abs/2015ApJ...806..259R} {806, 259}

\bibitem[\protect\citeauthoryear{{Reddy}, {Steidel}, {Pettini}  \& {Bogosavljevi{\'c}}}{{Reddy} et~al.}{2016}]{Reddy_2016}
{Reddy} N.~A.,  {Steidel} C.~C.,  {Pettini} M.,   {Bogosavljevi{\'c}} M.,  2016, \mn@doi [\apj] {10.3847/0004-637X/828/2/107}, \href {https://ui.adsabs.harvard.edu/abs/2016ApJ...828..107R} {828, 107}

\bibitem[\protect\citeauthoryear{{Reddy} et~al.,}{{Reddy} et~al.}{2018}]{Reddy_2018}
{Reddy} N.~A.,  et~al., 2018, \mn@doi [\apj] {10.3847/1538-4357/aaa3e7}, \href {https://ui.adsabs.harvard.edu/abs/2018ApJ...853...56R} {853, 56}

\bibitem[\protect\citeauthoryear{{Reddy} et~al.,}{{Reddy} et~al.}{2020}]{Reddy_2020}
{Reddy} N.~A.,  et~al., 2020, \mn@doi [\apj] {10.3847/1538-4357/abb674}, \href {https://ui.adsabs.harvard.edu/abs/2020ApJ...902..123R} {902, 123}

\bibitem[\protect\citeauthoryear{{Reddy} et~al.,}{{Reddy} et~al.}{2022}]{Reddy_2022}
{Reddy} N.~A.,  et~al., 2022, \mn@doi [\apj] {10.3847/1538-4357/ac3b4c}, \href {https://ui.adsabs.harvard.edu/abs/2022ApJ...926...31R} {926, 31}

\bibitem[\protect\citeauthoryear{Rezaee, Reddy, Shivaei, Fetherolf, Emami  \& Khostovan}{Rezaee et~al.}{2021}]{Rezaee_2021}
Rezaee S.,  Reddy N.,  Shivaei I.,  Fetherolf T.,  Emami N.,   Khostovan A.~A.,  2021, \mn@doi [Monthly Notices of the Royal Astronomical Society] {10.1093/mnras/stab1885}, 506, 3588

\bibitem[\protect\citeauthoryear{{Richards} et~al.,}{{Richards} et~al.}{2016}]{Richards_2016}
{Richards} S.~N.,  et~al., 2016, \mn@doi [\mnras] {10.1093/mnras/stv2453}, \href {https://ui.adsabs.harvard.edu/abs/2016MNRAS.455.2826R} {455, 2826}

\bibitem[\protect\citeauthoryear{{Rix}, {Pettini}, {Leitherer}, {Bresolin}, {Kudritzki}  \& {Steidel}}{{Rix} et~al.}{2004}]{Rix_2004}
{Rix} S.~A.,  {Pettini} M.,  {Leitherer} C.,  {Bresolin} F.,  {Kudritzki} R.-P.,   {Steidel} C.~C.,  2004, \mn@doi [\apj] {10.1086/424031}, \href {https://ui.adsabs.harvard.edu/abs/2004ApJ...615...98R} {615, 98}

\bibitem[\protect\citeauthoryear{{Rodighiero} et~al.,}{{Rodighiero} et~al.}{2014}]{Rodighiero_2014}
{Rodighiero} G.,  et~al., 2014, \mn@doi [\mnras] {10.1093/mnras/stu1110}, \href {https://ui.adsabs.harvard.edu/abs/2014MNRAS.443...19R} {443, 19}

\bibitem[\protect\citeauthoryear{{Salim} et~al.,}{{Salim} et~al.}{2007}]{Salim_2007}
{Salim} S.,  et~al., 2007, \mn@doi [\apjs] {10.1086/519218}, \href {https://ui.adsabs.harvard.edu/abs/2007ApJS..173..267S} {173, 267}

\bibitem[\protect\citeauthoryear{Sanders et~al.,}{Sanders et~al.}{2018}]{Sanders_2018}
Sanders R.~L.,  et~al., 2018, \mn@doi [The Astrophysical Journal] {10.3847/1538-4357/aabcbd}, 858, 99

\bibitem[\protect\citeauthoryear{{Saxena} et~al.,}{{Saxena} et~al.}{2020}]{Saxena_2020}
{Saxena} A.,  et~al., 2020, \mn@doi [\aap] {10.1051/0004-6361/201937170}, \href {https://ui.adsabs.harvard.edu/abs/2020A&A...636A..47S} {636, A47}

\bibitem[\protect\citeauthoryear{{Schaerer}}{{Schaerer}}{1996}]{Schaerer_1996}
{Schaerer} D.,  1996, \mn@doi [\apjl] {10.1086/310193}, \href {https://ui.adsabs.harvard.edu/abs/1996ApJ...467L..17S} {467, L17}

\bibitem[\protect\citeauthoryear{{Schaerer}, {Fragos}  \& {Izotov}}{{Schaerer} et~al.}{2019}]{Schaerer_2019}
{Schaerer} D.,  {Fragos} T.,   {Izotov} Y.~I.,  2019, \mn@doi [\aap] {10.1051/0004-6361/201935005}, \href {https://ui.adsabs.harvard.edu/abs/2019A&A...622L..10S} {622, L10}

\bibitem[\protect\citeauthoryear{{Schmidt}}{{Schmidt}}{1959}]{Schmidt_1959}
{Schmidt} M.,  1959, \mn@doi [\apj] {10.1086/146614}, \href {https://ui.adsabs.harvard.edu/abs/1959ApJ...129..243S} {129, 243}

\bibitem[\protect\citeauthoryear{Seibert, Heckman  \& Meurer}{Seibert et~al.}{2002}]{Seibert_2002}
Seibert M.,  Heckman T.~M.,   Meurer G.~R.,  2002, \mn@doi [The Astronomical Journal] {10.1086/341043}, 124, 46

\bibitem[\protect\citeauthoryear{{Senchyna} et~al.,}{{Senchyna} et~al.}{2017}]{Senchyna_2017}
{Senchyna} P.,  et~al., 2017, \mn@doi [\mnras] {10.1093/mnras/stx2059}, \href {https://ui.adsabs.harvard.edu/abs/2017MNRAS.472.2608S} {472, 2608}

\bibitem[\protect\citeauthoryear{{Shapley}, {Steidel}, {Adelberger}, {Dickinson}, {Giavalisco}  \& {Pettini}}{{Shapley} et~al.}{2001}]{Shapley_2001}
{Shapley} A.~E.,  {Steidel} C.~C.,  {Adelberger} K.~L.,  {Dickinson} M.,  {Giavalisco} M.,   {Pettini} M.,  2001, \mn@doi [\apj] {10.1086/323432}, \href {https://ui.adsabs.harvard.edu/abs/2001ApJ...562...95S} {562, 95}

\bibitem[\protect\citeauthoryear{{Shapley}, {Steidel}, {Pettini}  \& {Adelberger}}{{Shapley} et~al.}{2003}]{Shapley_2003}
{Shapley} A.~E.,  {Steidel} C.~C.,  {Pettini} M.,   {Adelberger} K.~L.,  2003, \mn@doi [\apj] {10.1086/373922}, \href {https://ui.adsabs.harvard.edu/abs/2003ApJ...588...65S} {588, 65}

\bibitem[\protect\citeauthoryear{{Shapley}, {Steidel}, {Pettini}, {Adelberger}  \& {Erb}}{{Shapley} et~al.}{2006}]{Shapley_2006}
{Shapley} A.~E.,  {Steidel} C.~C.,  {Pettini} M.,  {Adelberger} K.~L.,   {Erb} D.~K.,  2006, \mn@doi [\apj] {10.1086/507511}, \href {https://ui.adsabs.harvard.edu/abs/2006ApJ...651..688S} {651, 688}

\bibitem[\protect\citeauthoryear{{Shen}, {Madau}, {Conroy}, {Governato}  \& {Mayer}}{{Shen} et~al.}{2014}]{Shen_2014}
{Shen} S.,  {Madau} P.,  {Conroy} C.,  {Governato} F.,   {Mayer} L.,  2014, \mn@doi [\apj] {10.1088/0004-637X/792/2/99}, \href {https://ui.adsabs.harvard.edu/abs/2014ApJ...792...99S} {792, 99}

\bibitem[\protect\citeauthoryear{{Shirazi} \& {Brinchmann}}{{Shirazi} \& {Brinchmann}}{2012}]{Shirazi_2012}
{Shirazi} M.,  {Brinchmann} J.,  2012, \mn@doi [\mnras] {10.1111/j.1365-2966.2012.20439.x}, \href {https://ui.adsabs.harvard.edu/abs/2012MNRAS.421.1043S} {421, 1043}

\bibitem[\protect\citeauthoryear{{Shivaei} et~al.,}{{Shivaei} et~al.}{2015}]{Shivaei_2015}
{Shivaei} I.,  et~al., 2015, \mn@doi [\apj] {10.1088/0004-637X/815/2/98}, \href {https://ui.adsabs.harvard.edu/abs/2015ApJ...815...98S} {815, 98}

\bibitem[\protect\citeauthoryear{{Shivaei} et~al.,}{{Shivaei} et~al.}{2018}]{Shivaei_2018}
{Shivaei} I.,  et~al., 2018, \mn@doi [\apj] {10.3847/1538-4357/aaad62}, \href {https://ui.adsabs.harvard.edu/abs/2018ApJ...855...42S} {855, 42}

\bibitem[\protect\citeauthoryear{{Shivaei} et~al.,}{{Shivaei} et~al.}{2020}]{Shivaei_2020}
{Shivaei} I.,  et~al., 2020, \mn@doi [\apj] {10.3847/1538-4357/aba35e}, \href {https://ui.adsabs.harvard.edu/abs/2020ApJ...899..117S} {899, 117}

\bibitem[\protect\citeauthoryear{{Siana} et~al.,}{{Siana} et~al.}{2007}]{Siana_2007}
{Siana} B.,  et~al., 2007, \mn@doi [\apj] {10.1086/521185}, \href {https://ui.adsabs.harvard.edu/abs/2007ApJ...668...62S} {668, 62}

\bibitem[\protect\citeauthoryear{Skelton et~al.,}{Skelton et~al.}{2014}]{Skelton_2014}
Skelton R.~E.,  et~al., 2014, \mn@doi [The Astrophysical Journal Supplement Series] {10.1088/0067-0049/214/2/24}, 214, 24

\bibitem[\protect\citeauthoryear{{Smith}, {Norris}  \& {Crowther}}{{Smith} et~al.}{2002}]{Smith_2002}
{Smith} L.~J.,  {Norris} R. P.~F.,   {Crowther} P.~A.,  2002, \mn@doi [\mnras] {10.1046/j.1365-8711.2002.06042.x}, \href {https://ui.adsabs.harvard.edu/abs/2002MNRAS.337.1309S} {337, 1309}

\bibitem[\protect\citeauthoryear{{Smith}, {G{\"o}tberg}  \& {de Mink}}{{Smith} et~al.}{2018}]{Smith_2018}
{Smith} N.,  {G{\"o}tberg} Y.,   {de Mink} S.~E.,  2018, \mn@doi [\mnras] {10.1093/mnras/stx3181}, \href {https://ui.adsabs.harvard.edu/abs/2018MNRAS.475..772S} {475, 772}

\bibitem[\protect\citeauthoryear{{Smith}, {van Zee}, {Salim}, {Dale}, {Staudaher}, {Wrock}  \& {Maben}}{{Smith} et~al.}{2021}]{Smith_2021}
{Smith} M.~V.,  {van Zee} L.,  {Salim} S.,  {Dale} D.,  {Staudaher} S.,  {Wrock} T.,   {Maben} A.,  2021, \mn@doi [\mnras] {10.1093/mnras/stab1530}, \href {https://ui.adsabs.harvard.edu/abs/2021MNRAS.505.3998S} {505, 3998}

\bibitem[\protect\citeauthoryear{{Somerville} \& {Primack}}{{Somerville} \& {Primack}}{1999}]{Somerville_1999}
{Somerville} R.~S.,  {Primack} J.~R.,  1999, \mn@doi [\mnras] {10.1046/j.1365-8711.1999.03032.x}, \href {https://ui.adsabs.harvard.edu/abs/1999MNRAS.310.1087S} {310, 1087}

\bibitem[\protect\citeauthoryear{Sparre, Hayward, Feldmann, Faucher-Giguère, Muratov, Kereš  \& Hopkins}{Sparre et~al.}{2016}]{Sparre_2016}
Sparre M.,  Hayward C.~C.,  Feldmann R.,  Faucher-Giguère C.-A.,  Muratov A.~L.,  Kereš D.,   Hopkins P.~F.,  2016, \mn@doi [Monthly Notices of the Royal Astronomical Society] {10.1093/mnras/stw3011}, 466, 88

\bibitem[\protect\citeauthoryear{{Sparre}, {Hayward}, {Feldmann}, {Faucher-Gigu{\`e}re}, {Muratov}, {Kere{\v{s}}}  \& {Hopkins}}{{Sparre} et~al.}{2017}]{Sparre_2017}
{Sparre} M.,  {Hayward} C.~C.,  {Feldmann} R.,  {Faucher-Gigu{\`e}re} C.-A.,  {Muratov} A.~L.,  {Kere{\v{s}}} D.,   {Hopkins} P.~F.,  2017, \mn@doi [\mnras] {10.1093/mnras/stw3011}, \href {https://ui.adsabs.harvard.edu/abs/2017MNRAS.466...88S} {466, 88}

\bibitem[\protect\citeauthoryear{{Springel}}{{Springel}}{2000}]{Springel_2000}
{Springel} V.,  2000, \mn@doi [\mnras] {10.1046/j.1365-8711.2000.03187.x}, \href {https://ui.adsabs.harvard.edu/abs/2000MNRAS.312..859S} {312, 859}

\bibitem[\protect\citeauthoryear{{Springel}, {Di Matteo}  \& {Hernquist}}{{Springel} et~al.}{2005}]{Springel_2005}
{Springel} V.,  {Di Matteo} T.,   {Hernquist} L.,  2005, \mn@doi [\mnras] {10.1111/j.1365-2966.2005.09238.x}, \href {https://ui.adsabs.harvard.edu/abs/2005MNRAS.361..776S} {361, 776}

\bibitem[\protect\citeauthoryear{{Stanway} \& {Eldridge}}{{Stanway} \& {Eldridge}}{2018}]{Stanway_2018}
{Stanway} E.~R.,  {Eldridge} J.~J.,  2018, \mn@doi [\mnras] {10.1093/mnras/sty1353}, \href {https://ui.adsabs.harvard.edu/abs/2018MNRAS.479...75S} {479, 75}

\bibitem[\protect\citeauthoryear{{Stanway}, {Eldridge}  \& {Becker}}{{Stanway} et~al.}{2016}]{Stanway_2016}
{Stanway} E.~R.,  {Eldridge} J.~J.,   {Becker} G.~D.,  2016, \mn@doi [\mnras] {10.1093/mnras/stv2661}, \href {https://ui.adsabs.harvard.edu/abs/2016MNRAS.456..485S} {456, 485}

\bibitem[\protect\citeauthoryear{{Steidel}, {Pettini}  \& {Adelberger}}{{Steidel} et~al.}{2001}]{Steidel_2001}
{Steidel} C.~C.,  {Pettini} M.,   {Adelberger} K.~L.,  2001, \mn@doi [\apj] {10.1086/318323}, \href {https://ui.adsabs.harvard.edu/abs/2001ApJ...546..665S} {546, 665}

\bibitem[\protect\citeauthoryear{Steidel, Shapley, Pettini, Adelberger, Erb, Reddy  \& Hunt}{Steidel et~al.}{2004}]{Steidel_2004}
Steidel C.~C.,  Shapley A.~E.,  Pettini M.,  Adelberger K.~L.,  Erb D.~K.,  Reddy N.~A.,   Hunt M.~P.,  2004, \mn@doi [The Astrophysical Journal] {10.1086/381960}, 604, 534

\bibitem[\protect\citeauthoryear{{Steidel}, {Strom}, {Pettini}, {Rudie}, {Reddy}  \& {Trainor}}{{Steidel} et~al.}{2016}]{Steidel_2016}
{Steidel} C.~C.,  {Strom} A.~L.,  {Pettini} M.,  {Rudie} G.~C.,  {Reddy} N.~A.,   {Trainor} R.~F.,  2016, \mn@doi [\apj] {10.3847/0004-637X/826/2/159}, \href {https://ui.adsabs.harvard.edu/abs/2016ApJ...826..159S} {826, 159}

\bibitem[\protect\citeauthoryear{Theios, Steidel, Strom, Rudie, Trainor  \& Reddy}{Theios et~al.}{2019}]{Theios_2019}
Theios R.~L.,  Steidel C.~C.,  Strom A.~L.,  Rudie G.~C.,  Trainor R.~F.,   Reddy N.~A.,  2019, \mn@doi [The Astrophysical Journal] {10.3847/1538-4357/aaf386}, 871, 128

\bibitem[\protect\citeauthoryear{{Topping}, {Shapley}, {Reddy}, {Sanders}, {Coil}, {Kriek}, {Mobasher}  \& {Siana}}{{Topping} et~al.}{2020}]{Topping}
{Topping} M.~W.,  {Shapley} A.~E.,  {Reddy} N.~A.,  {Sanders} R.~L.,  {Coil} A.~L.,  {Kriek} M.,  {Mobasher} B.,   {Siana} B.,  2020, \mn@doi [\mnras] {10.1093/mnras/staa1410}, \href {https://ui.adsabs.harvard.edu/abs/2020MNRAS.495.4430T} {495, 4430}

\bibitem[\protect\citeauthoryear{{Tremonti} et~al.,}{{Tremonti} et~al.}{2004}]{Tremonti_2004}
{Tremonti} C.~A.,  et~al., 2004, \mn@doi [\apj] {10.1086/423264}, \href {https://ui.adsabs.harvard.edu/abs/2004ApJ...613..898T} {613, 898}

\bibitem[\protect\citeauthoryear{{Vidal-Garc{\'\i}a}, {Charlot}, {Bruzual}  \& {Hubeny}}{{Vidal-Garc{\'\i}a} et~al.}{2017}]{Garcia_2017}
{Vidal-Garc{\'\i}a} A.,  {Charlot} S.,  {Bruzual} G.,   {Hubeny} I.,  2017, \mn@doi [\mnras] {10.1093/mnras/stx1324}, \href {https://ui.adsabs.harvard.edu/abs/2017MNRAS.470.3532V} {470, 3532}

\bibitem[\protect\citeauthoryear{{Visbal}, {Haiman}  \& {Bryan}}{{Visbal} et~al.}{2015}]{Visbal_2015}
{Visbal} E.,  {Haiman} Z.,   {Bryan} G.~L.,  2015, \mn@doi [\mnras] {10.1093/mnras/stv785}, \href {https://ui.adsabs.harvard.edu/abs/2015MNRAS.450.2506V} {450, 2506}

\bibitem[\protect\citeauthoryear{{Walborn}, {Nichols-Bohlin}  \& {Panek}}{{Walborn} et~al.}{1985}]{Walborn_1985}
{Walborn} N.~R.,  {Nichols-Bohlin} J.,   {Panek} R.~J.,  1985, NASA Reference Publication, \href {https://ui.adsabs.harvard.edu/abs/1985NASRP1155.....W} {1155}

\bibitem[\protect\citeauthoryear{{Wang} \& {Lilly}}{{Wang} \& {Lilly}}{2020}]{Wang_2020}
{Wang} E.,  {Lilly} S.~J.,  2020, \mn@doi [\apj] {10.3847/1538-4357/ab7b7d}, \href {https://ui.adsabs.harvard.edu/abs/2020ApJ...892...87W} {892, 87}

\bibitem[\protect\citeauthoryear{{Weisz} et~al.,}{{Weisz} et~al.}{2012}]{Weisz_2012}
{Weisz} D.~R.,  et~al., 2012, \mn@doi [\apj] {10.1088/0004-637X/744/1/44}, \href {https://ui.adsabs.harvard.edu/abs/2012ApJ...744...44W} {744, 44}

\bibitem[\protect\citeauthoryear{Wuyts et~al.,}{Wuyts et~al.}{2011}]{Wuyts_2011}
Wuyts S.,  et~al., 2011, \mn@doi [The Astrophysical Journal] {10.1088/0004-637x/738/1/106}, 738, 106

\bibitem[\protect\citeauthoryear{Wuyts et~al.,}{Wuyts et~al.}{2012}]{Wuyts_2012}
Wuyts S.,  et~al., 2012, \mn@doi [The Astrophysical Journal] {10.1088/0004-637x/753/2/114}, 753, 114

\bibitem[\protect\citeauthoryear{Wuyts et~al.,}{Wuyts et~al.}{2013}]{Wuyts_2013}
Wuyts S.,  et~al., 2013, \mn@doi [The Astrophysical Journal] {10.1088/0004-637X/779/2/135}, 779, 135

\bibitem[\protect\citeauthoryear{{Zahid}, {Dima}, {Kudritzki}, {Kewley}, {Geller}, {Hwang}, {Silverman}  \& {Kashino}}{{Zahid} et~al.}{2014}]{Zahid_2014}
{Zahid} H.~J.,  {Dima} G.~I.,  {Kudritzki} R.-P.,  {Kewley} L.~J.,  {Geller} M.~J.,  {Hwang} H.~S.,  {Silverman} J.~D.,   {Kashino} D.,  2014, \mn@doi [\apj] {10.1088/0004-637X/791/2/130}, \href {https://ui.adsabs.harvard.edu/abs/2014ApJ...791..130Z} {791, 130}

\bibitem[\protect\citeauthoryear{{da Silva}, {Fumagalli}  \& {Krumholz}}{{da Silva} et~al.}{2012}]{dasilva_2012}
{da Silva} R.~L.,  {Fumagalli} M.,   {Krumholz} M.,  2012, \mn@doi [\apj] {10.1088/0004-637X/745/2/145}, \href {https://ui.adsabs.harvard.edu/abs/2012ApJ...745..145D} {745, 145}

\bibitem[\protect\citeauthoryear{{da Silva}, {Fumagalli}  \& {Krumholz}}{{da Silva} et~al.}{2014}]{dasilva_2014}
{da Silva} R.~L.,  {Fumagalli} M.,   {Krumholz} M.~R.,  2014, \mn@doi [\mnras] {10.1093/mnras/stu1688}, \href {https://ui.adsabs.harvard.edu/abs/2014MNRAS.444.3275D} {444, 3275}

\bibitem[\protect\citeauthoryear{{de Mello}, {Schaerer}, {Heldmann}  \& {Leitherer}}{{de Mello} et~al.}{1998}]{demello_1998}
{de Mello} D.~F.,  {Schaerer} D.,  {Heldmann} J.,   {Leitherer} C.,  1998, \mn@doi [\apj] {10.1086/306317}, \href {https://ui.adsabs.harvard.edu/abs/1998ApJ...507..199D} {507, 199}

\bibitem[\protect\citeauthoryear{de Mink, Langer, Izzard, Sana  \& de Koter}{de~Mink et~al.}{2013}]{de_Mink_2013}
de Mink S.~E.,  Langer N.,  Izzard R.~G.,  Sana H.,   de Koter A.,  2013, \mn@doi [The Astrophysical Journal] {10.1088/0004-637x/764/2/166}, 764, 166

\bibitem[\protect\citeauthoryear{{van der Wel} et~al.,}{{van der Wel} et~al.}{2014}]{van_der_Wel2014}
{van der Wel} A.,  et~al., 2014, \mn@doi [\apjl] {10.1088/2041-8205/792/1/L6}, \href {https://ui.adsabs.harvard.edu/abs/2014ApJ...792L...6V} {792, L6}

\makeatother
\end{thebibliography}
\appendix

\bsp	% typesetting comment
\label{lastpage}
\end{document}